# A High-Angular Resolution Multiplicity Survey of the Open Clusters $\alpha$ Persei and Praesepe

J. Patience [1], A. M. Ghez,

UCLA Division of Astronomy & Astrophysics, Los Angeles, CA 90095-1562

I. N. Reid [2] & K. Matthews

Palomar Observatory, California Institute of Technology, Pasadena, CA 91125

[1] present address: IGPP, LLNL, 7000 East Ave. L-413, Livermore, CA 94550; patience@igpp.ucllnl.org

[2] present address: Space Telescope Science Institute, 3700 San Martin Dr., Baltimore, MD 21218

Abstract

Two hundred and forty-two members of the Praesepe and α Persei clusters have been surveyed with high angular resolution 2.2 μm speckle imaging on the IRTF 3-m, the Hale 5-m, and the Keck 10-m, along with direct imaging using the near-infrared camera (NICMOS) aboard the Hubble Space Telescope (HST). The observed stars range in spectral type from B (~5 $M_{SUN}$) to early-M (~0.5 $M_{SUN}$), with the majority of the targets more massive than ~0.8 MSUN. The 39 binary and 1 quadruple systems detected encompass separations from 0".053 to 7".28; 28 of the systems are new detections and there are 9 candidate substellar companions. The results of the survey are used to test binary star formation and evolution scenarios and to investigate the effects of companion stars on X-ray emission and stellar rotation. The main results are:

• Over the projected separation range of 26-581 AU and magnitude differences of ΔK<4.0 mag (comparable to mass ratios, $q=m_{sec}/m_{prim}$ > 0.25), the companion star fraction (*CSF*) for α Persei is 0.09 ± 0.03 and for Praesepe is 0.10 ± 0.03. This fraction is consistent with the field G-dwarf value, implying that there is not a systematic decline in multiplicity with age at these separations on timescales of a few x $10^7$yrs. The combination of previous spectroscopic work and the current cluster survey results in a cluster binary separation distribution that peaks at $4\ ^{+1}_{-1.5}$ AU, a significantly smaller value than the peaks of both the field G-dwarf and the nearby T Tauri distributions. If the field G-dwarf distribution represents a superposition of distributions from the populations that contributed to the field, then the data implies that ~30% of field binaries formed in dark clouds like the nearby T Tauri stars and the remaining ~70% formed in denser regions.

• An exploration of the binary star properties reveals a cluster *CSF* that increases with decreasing target mass and a cluster mass ratio distribution that rises more sharply for higher mass stars, but is independent of binary separation. These observational trends are consistent with several models of capture in small clusters and simulations of accretion following fragmentation in a cluster environment. Other types of capture and fragmentation are either inconsistent with these data or currently lack testable predictions.

• Among the cluster A stars, there is a higher fraction of binaries in the subset with X-ray detections, consistent with the hypothesis that lower mass companions are the true source of X-ray emission.

• Finally, in the younger cluster α Persei, the rotational velocities for solar-type binaries with separations less that 60 AU are significantly higher than that of wider systems. This suggests that companions may critically affect the rotational evolution of young stars.

## 1. Introduction

Extensive observations have established that binary stars are very common among field solar-type stars and are approximately twice as prevalent among T Tauri stars in the closest star-forming regions (Abt & Levy 1976; Duquennoy & Mayor 1991; Ghez et al. 1993, 1997a; Leinert et al. 1993; Simon et al. 1995; Köhler & Leinert 1998). The large discrepancy in binary fractions between pre-main sequence stars and main sequence stars at separations of tens to hundreds of AU has provided the impetus for both theoretical studies (e.g., Kroupa 1995; Durisen & Sterzik 1994) and a number of high angular resolution imaging investigations of binary stars in nearby open clusters with ages intermediate between the pre-main sequence stars (~2 Myr Simon et al. 1993) and the solar neighborhood (~5 Gyr for Duquennoy & Mayor 1991 sample). An adaptive optics survey of the Pleiades (~125 Myr, Stauffer et al. 1998) measured a binary fraction comparable to the field over the separation range 11 to 792 AU (Bouvier et al. 1997). In contrast, a speckle imaging survey of the older Hyades (~660 Myr, Torres et al. 1997), covering the 5 to 50 AU separation range, measured a multiplicity intermediate between the T Tauri and field stars (Patience et al. 1998). In this paper, the multiplicity of $\alpha$ Persei (90 Myr, Stauffer et al. 1999) and Praesepe (~660 Myr, Mermilliod 1981) are investigated using speckle imaging with large ground-based telescopes and direct imaging with the Hubble Space Telescope (*HST*). The first aspect of this project addresses the disparity in binary fractions by measuring the evolution of the binary star fraction over the critical separation range of tens to hundreds of AU.

The database of binary star characteristics resolved by large multiplicity surveys can also be used to observationally test several binary star formation scenarios. Binary star formation mechanisms fall into two broad categories, capture and fragmentation (e.g., review in Clarke 1996). Although capture in large clusters does not occur frequently enough to produce a significant fraction of binaries (Aarseth & Hills 1972), capture in small-N clusters with or without the effects of circumstellar disks has been modelled as a viable formation mechanism (McDonald & Clarke 1993,1995; Sterzik & Durisen 1998). Different modes of fragmentation have also been proposed—fragmentation of the protostellar cloud core (e.g., review in Clarke 1996) or of the protostellar disk (e.g., Bonnell & Bate 1994); other simulations follow the acccretion of material after fragmentation (Bonnell & Bate 1997; Bate 2000a, 2000b). Many formation theories have distinct predictions for the resulting binary star properties such as how the companion star fraction and mass ratio distribution depend upon mass. In order to investigate the mass dependence of these properties, the stars surveyed for this work cover a range of masses, from the massive B stars to the sub-solar mass late K and early M stars (~5 $M_{SUN}$ to ~0.5 $M_{SUN}$).

Another topic that can be addressed with multiplicity surveys is the effect, both active and passive, of a companion on the primary. Of particular interest for this study is the possible active role of binary companions in the rotational history of solar-type stars. Although most solar-type stars have low rotational velocities (~15 km/s) with limited spread in values (~6-70 km/s) during the pre-main sequence stage (e.g. Bertout 1989), by the age of the $\alpha$ Persei cluster, 90 Myrs, these stars exhibit a range of rotational velocities of more than 200 km/s (Stauffer et al. 1985, Prosser 1992). While the subsequent spin-down due to mass loss in a magnetized wind is well understood (Schatzman 1962), the mechanism which makes some stars rapid rotators, while most remain slow rotators at the age of $\alpha$ Persei is not known; two opposing ideas, however, involve binary stars. Both scenarios are predicated upon the assumption that stellar rotation is regulated by a star-disk interaction (Konigl 1991, Armitage & Clarke 1996) in which the stellar magnetic field is linked to the slowly rotating outer disk. In the disk-braking scenario, the stellar rotation rate is critically dependent upon the circumstellar disk lifetime, with the rapid rotators associated with short-lived disks. One model suggests that binary stars completely disrupt the circumstellar disks creating rapid rotators (Bouvier et al. 1997), while an alternate theory suggests that, except for tidally locked systems, binary stars should be slow rotators. In the second model, binary stars truncate rather than obliterate disks, leaving a remnant ring of material that actually survives longer and brakes the star over a longer period of time (Armitage & Clarke 1996). Recent observations of rotation periods among Trapezium stars have questioned the idea of disk-regulated rotation (Stassun et al. 1999), but this Orion study focused on lower mass stars than those observed in $\alpha$ Persei, and mass may be an important factor in rotation rates (Herbst et al. 2000). Angular momentum evolution varies with mass and therefore this part of the investigation is limited to the solar-type stars. With the maximum range of rotational velocities, $\alpha$ Persei is the ideal cluster to investigate this possible role of companions.

A potential passive effect of companions relates to X-ray detections; late-type companion stars may be an important factor in explaining the X-ray detections from late-B and A type stars. B6 to A5 stars lack both the strong winds of higher mass stars and the dynamo of lower mass stars believed to generate X-rays along the rest of the main sequence (c.f. Pallavicini 1989). While shearing motions in the coronae of late B to A stars may represent a mechanism to generate X-rays (Tout & Pringle 1994), unresolved companions may be the true source of the observed X-rays and provide a simpler explanation. Multiplicity surveys test this hypothesis, as binaries discovered by spectroscopy or by the current high-resolution imaging survey have separations smaller than typical ~10" ROSAT detection error boxes.

In this paper, the results of high angular resolution multiplicity surveys of α Persei and Praesepe are reported and are used to investigate the formation and evolution of binaries as well as the role of companions in stellar rotation and X-ray emission. The data were obtained using speckle imaging on large ground-based telescopes and direct imaging with the Hubble Space Telescope (*HST*) for an additional set of fainter stars. The samples are defined in §2 and the observations and data analysis are described in §3. The results and a description of the sensitivity are reported in §4, which also details the sensitivities of several comparison surveys. In the discussion, §5-7, the evolution of both the binary fraction and the binary separation distribution are investigated; the companion star fraction and the mass ratio distribution are compared to theoretical models; and the connection between binarity and X-ray emission and stellar rotation is explored. The main conclusions are summarized in §8.

## 2. The Sample

The two clusters selected for this study, α Persei and Praesepe, are ideal samples for multiplicity surveys, as these clusters share two critical properties -- distances comparable to the nearby star-forming regions and ages intermediate between T Tauri stars and field stars. The younger cluster, α Persei, has an age of 50 - 90 Myr (Prosser 1992; Stauffer et al. 1999), with the most recent age assessment based on the Li depletion boundary giving the larger value; the more recent value is used in this study. The distance to α Persei is 176 ± 5 pc (Pinsonneault et al. 1998). The older cluster, Praesepe, has an age of ~660 Myr (Mermilliod 1981) and a distance of 171 ± 4 pc (Pinsonneault et al.). The spatial extent of the clusters, 10-15 pc in diameter, introduce an additional uncertainty in the distance to individual members.

The stars observed for this survey were drawn from cluster membership studies with the additional criteria of a K<~10mag for the speckle data and 10 mag < K < 12.3 mag for the α Persei HST data. The range of masses for the α Persei stars is ~5 $M_{SUN}$ to ~0.5 $M_{SUN}$. The $K_s$ magnitudes are available for almost the entire sample from the 2MASS catalogue; appendix 2 describes the conversion from $K_s$ to K magnitudes. Only three giants in Praesepe exceed the 2MASS magnitude limit, however, these stars are not included in the analysis of binary properties. For the 25 a Persei and 6 Praesepe main sequence stars not covered in the current 2MASS data release, the K magnitudes were estimated from an empirical V mag to K mag relation based on a fit to the observed stars; a separate fit was made for each cluster. Again, the details of the V to K relation are given in appendix 2. Tables 1 and 2 list the observed samples of α Persei and Praesepe. Each table gives the name, 2000 coordinates, V magnitude, B-V color, spectral type, 2MASS K magnitude, and X-ray flux or upper limit; for α Persei, the measured *vsini* is also listed in Table 1.

The stars in the α Persei speckle sample have spectral types from B to G, range in K magnitude from 5.1 to 10.5, and account for 109 of the 112 brightest (based on V magnitudes) stars with definite membership. An additional 33 fainter stars with K magnitudes between 10.0 and 12.3 and a large range of rotational velocities were observed with *HST*. In total, 142 stars comprise the α Persei sample. Because of the lower galactic latitude (-7 degrees) of α Persei, a combination of proper motions, radial velocities, and indications of stellar activity ($H_{\alpha}$, Li features) is required to assess the membership of each star in the younger cluster; results from studies of the brighter stars (Heckman et al. 1956) and the fainter stars (Stauffer et al. 1985, Prosser 1992) are compiled in Prosser (1992).

The Praesepe speckle sample includes most of the brightest stars in the cluster, with spectral types from A to G; the 100 observed stars represent more than one-half of the Klein-Wassink (1927) membership list which is based upon photometry and proper motion. Excluding the giants, the range of K magnitudes is 5.9 mag to 9.2 mag, corresponding to masses for the Praesepe main sequence stars of ~2.4 $M_{SUN}$ to ~1.0 $M_{SUN}$. More recent spectroscopic observations of Praesepe stars have revealed a few Klein-Wassink stars as nonmembers (Mermilliod & Mayor 1999), and two of these non-member stars are part of the speckle sample – KW 258 and KW 553. The results for these two stars are reported, but they are not included in the analysis.

## 3. Observations and Data Analysis

### 3.1 Ground-based Speckle Imaging and Shift-and-add

For this study, speckle imaging data were obtained through the K-band ($\lambda_o$=2.2 μm; Δλ=0.4μm) filter with three different cameras: the facility near-infrared camera on the 5m Hale Telescope at Palomar Observatory, NSFCAM (Rayner et al. 1993; Shure et al. 1994) on the 3m NASA Infrared Telescope Facility (IRTF), and NIRC (Matthews & Soifer 1994; Matthews et al. 1996) on the 10m Keck Telescope. With plate scales of 0".0326/pix for Palomar, 0".0532/pix for NSFCAM, and 0".0206/pix for NIRC, each camera allows for diffraction-limited imaging at 2.2 μm. The diffraction-limit, λ/D, ranges from 0".045 to 0".15 for these observations. All of the cameras include 256x256 arrays, however, only subarrays were recorded at Palomar and the IRTF. Due to demands on the available disk space, the Palomar data are limited to a 64 x 64 subarray, or a 2".14 x 2".14 field-of-view and the IRTF data are restricted to a 128 x 128 subarray, resulting in a wider 6".80 x 6".80 field-of-view.

The observing program was designed to obtain a sample as large as possible, but also to ensure that the survey data covered a large angular separation range. The survey was initiated in 1995 at Palomar and completed in 1999 at the IRTF; a small sample was also observed in 1999 at Keck to begin to probe smaller separations. Speckle analysis is particularly effective for resolving binaries with separations close to the diffraction limit, while shift-and-add processing is most effective for detecting companions outside the seeing halo (radius ~0".5) of the target star. Tables 3a (α Persei) and 3b (Praesepe) summarize the number and type of observations made for the sample. The entire ground-based sample was observed with one of the three cameras for full speckle analysis. Almost all of the targets observed from the ground were also analyzed with the shift-and-add technique.

The speckle observing procedure is the same for each camera and involves recording a total of 3,000 to 4,000 exposures (fewer for Keck) of ~0.1s in stacks of ~500 images. By alternating between observations of the target star and a nearby reference point source, atmospheric effects are measured and removed through analysis developed by Labeyrie (1970) and Lohmann et al. (1983). Details of the data analysis are given in Ghez et al. (1993) and Patience et al. (1998). The results of the speckle processing of binary stars determines the ΔK magnitude difference between primary and secondary, the projected separation, and the position angle, while the parameters measured for single stars are detection limits for unseen companions calculated for several separations.

In addition to speckle analysis, nearly all of the stars in each cluster were also processed with the shift-and-add technique (Bates & Cady 1980, Christou 1991). The speckle data collected with the wider field NSFCAM and NIRC cameras were all reanalyzed with this method to search for wider companions. In order to have uniform coverage at the widest separations, all but 9 α Persei and 11 Praesepe stars observed at Palomar were reobserved at the IRTF for shift-and-add analysis; the shift-and-add datasets require fewer frames and are more efficient to obtain. The shift-and-add analysis provides the same parameters for binary and single stars as the speckle procedure (see Ghez et al. 1998 for details).

3.2 Direct Imaging with Hubble Space Telescope

HST observations with the onboard near-infrared camera NICMOS (Thompson et al. 1998) were taken of 33 members of α Persei fainter than the speckle limiting magnitude (K ~ 10 mag). Using the highest resolution camera of NICMOS—NIC1—which has a pixel scale of 0".043/pix (Thompson et al. 1998), images were recorded through the F140W filter ($\lambda o$ = 1.3 μm, $\Delta\lambda$ = 1.0 μm) in order to maximize sensitivity to companion stars. With the shorter wavelength F140W filter, the diffraction-limited resolution ($\lambda/D$) for HST is 0".11, comparable to ground-based K-band speckle observations which comprise the majority of the survey data set. To minimize the effects of bad pixels, two offset images were obtained for each target. The data were taken in the MULTIACCUM mode, which performs a series of non-destructive readouts designed to increase the dynamic range of the observations. The integration time was at least 2 minutes for each star and as high as 10 minutes for some stars, depending on how efficiently groups of stars could be packed into a single HST orbit. As with the speckle and shift-and-add analysis, the HST data reduction results determine the magnitude difference – in this case ΔF140W -- between primary and secondary, the projected separation, and the position angle for binary stars and the ΔF140W detection limits for unseen companions for the single stars. The IRAF phot package was used to measure the magnitudes and positions of the binaries and the detection limits of the singles. The ΔF140W values are reported, but, in order to express the magnitude differences in terms of mass ratios, the ΔF140W values were converted to ΔJ based on a relation derived from the α Persei single stars given in appendix 2.

## 4 Results

### 4.1 Binary Star Detections and Single Star Detection Limits for α Persei and Praesepe Members

The majority of the detected binaries and multiples are newly resolved by this survey. Of the 209 (100 Praesepe, 109 α Persei) stars observed with the speckle and shift-and-add techniques, 22 are resolved as binaries; 12 of these systems are new detections. The Praesepe survey accounts for 12 of the binaries and the remaining 10 binaries are α Persei members. Among the extreme properties of the detected binaries are a separation as small as 0".053 and a K magnitude difference as large as 4.5. Of the 33 α Persei members observed with HST, 17 binaries and 1 quadruple are imaged, with all except one newly resolved. Many of these potential companions, however, are very faint and are beyond the dynamic range of the ground-based data set. Tables 4a, 4b, and 5 provide the properties of all the detected binaries and notes about previous measurements which are described in the Appendix. The α Persei binaries are divided into two sets, those with stellar companions (Table 4a) and those with candidate substellar companions (Table 4b). Table 5 lists all the Praesepe binaries, since all are in the stellar mass range. The analysis of the binary properties is confined to the systems with stellar companions.

Each of the systems listed in either Tables 4a or Table 5 is assumed to be a physically associated pair, since the small separation coverage and limited dynamic range of the surveys restrict the probability of a chance superposition. Based on the number of stars recorded on the Schmidt plates in the region of α Persei, $7 \times 10^{-4}$/arcsec$^2$ (Prosser 1992), the chance alignment probability is only ~1% for the IRTF field-of-view. Since the Schmidt plates are sensitive to stars fainter than the average detection limit of this IR survey, this is a conservative estimate. Praesepe is located at a higher galactic latitude than α Persei, 32 degrees compared to –7 degrees, making the probability of a chance alignment considerably smaller for binaries in Praesepe.

For the single stars, Tables 6 (α Persei) and 7 (Praesepe) give the detection limits for companions at a separation of 0".15; these tables also include notes about previous measurements. As explained in §3, many stars observed with speckle at Palomar were reobserved at the IRTF with a more limited shift-and-add data set; for these targets, the limit at 0".15 is determined form the Palomar data, but the IRTF observing date is also listed in Tables 6 and 7 to indicate the larger separation range coverage. These detection limits for the unresolved stars are used to quantify the sensitivity of the survey and to define the complete region in the next subsection. The median detection limits for several separations are summarized on Figure 1, which also shows the binaries detected in each cluster.

4.2 α Persei and Praesepe Survey Sensitivity

Since the α Persei and Praesepe members were observed with several telescope systems, the range of separation and magnitude difference is limited before combining the data sets. Because the observations were made with telescopes ranging in diameter from 3m to 10m, the resolution limit is not the same for all targets. The largest value of λ/D is 0".15 for the 3m IRTF at 2.2 μm and this separation is chosen as the inner separation cutoff for the complete region. The widest separation considered is 3".4, one-half the field-of-view of the IRTF camera. Of the 242 star α Persei/Praesepe sample, only the 18 stars observed exclusively with Keck and 20 targets observed only at Palomar do not cover the entire separation range of 0".15 to 3".4. The Keck data are limited to separations less than 2".63, while the Palomar data only extend to 1".07. Given the 5pc difference in cluster distances and the 0".15 to 3".4 angular range, the common projected separation range 26 to 581 AU is considered the complete separation range. The imposition of a separation range excludes 7 binaries, leaving 26 systems with appropriate separations. One Praesepe binary – KW 284 – and 4 α Persei systems –HE 935, AP 60, HE 965, and HE 581 – have separations smaller than the inner cutoff. Two additional α Persei HST binaries with companions bright enough to be above the stellar limit – AP 108 and AP 106 – have separations larger than the outer cutoff.

At a separation of 0".15, the median speckle detection limit is $\Delta K = 3.6$ mag which corresponds to a mass ratio limit of $q$ ~0.3. Although the HST data are not as sensitive at the closest separations, the sensitivity is comparable or higher for most of the separation range. No corrections are applied since the HST observations represent only ~15% of the α Persei/Praesepe sample and only ~5% of the combined cluster sample analyzed in sections 6.2. Two different cutoffs are used in the analysis. The discussion about the frequency of companions (*CSF*) involves a cutoff based on the observed $\Delta K$ and includes the 23 binaries with appropriate sepations and with $\Delta K <= 4$ *mag*, while the discussion of the mass ratio distribution imposes a more conservative cutoff based on the derived mass ratio value and only includes the 19 systems with separations between 26 to 581 AU and $q >= 0.4$. The mass ratio limit is not unique for a given $\Delta K$ since the mass-magnitude relation involves several functions (Henry & McCarthy 1993; Patience et al. 1998); the $\Delta K=4$ *mag* limit is approximately a mass ratio limit of $q$~$0.25$ and the mass ratio limit of $q=0.40$ translates ino a magnitude limit of approximately $\Delta K$~$3$ *mag* (see appendix 2 for details). Since the stars in the samples represent different masses with spectral types from B to K, a mass ratio limit corresponds to companion mass detection limits that scale with the target star mass. Within the complete separation range, one Praesepe binary—KW 212—and 2 α Persei binaries – AP 75 and AP 193—have mass ratios below 0.25. One additional Praesepe pair—KW 282—and 4 additional α Persei pairs—AP 98, AP 17, AP 139, and AP 121—have mass ratios between 0.25 and 0.40.

### 4.3 Comparison Surveys

In addition to the current α Persei and Praesepe surveys, a number of multiplicity surveys have been conducted in regions with different ages. The current results are enhanced by placing them in the context of these previous studies. In order to best compare the current surveys with previous work, this section describes the detections, sensitivity, and separation range coverage of these investigations of nearby star-forming regions, additional clusters, and the solar-neighborhood reported in the literature. The discussion of the α Persei and Praesepe results and how they compare with other samples and with theoretical expectations begins in §5.

#### 4.3.1 Open Clusters

The results of two previous surveys of open clusters—the Pleiades and the Hyades—are included in much of the analysis of binary star properties. An adaptive optics survey of 143 Pleiades G and K dwarfs (mass range ~0.6 $M_{SUN}$ to 1.1 $M_{SUN}$) (Bouvier et al. 1997) provides an excellent comparison sample with an intermediate age between α Persei and Praesepe. Since the Pleiades is slightly closer (D = 132 pc Pinsonneault et al. 1998), the separation range 26 to 581 AU corresponds to 0".20 to 4".40, a subset of the Pleiades survey that spanned 0".08 to 6".0. Within the separation range of 26 AU to 581 AU and $\Delta K < 4$mag, there are 17 Pleiades companions, all of which have colors consistent with cluster membership. Overall, the detection limits of the Pleiades survey are quite similar to the speckle survey presented here. Specifically, the Pleiades observations are less sensitive (by 0.5—1.5 mag) over the small separation range from 0".20 to 0".4 and slightly more sensitive (by 0.5—1.0 mag) at separations wider than ~1".0. Each of the 26 AU to 581 AU and $K < 4.0$ mag range Pleiades binaries would have been detected by the α Persei and Praesepe speckle surveys. Conversely, all but three of the Praesepe and α Persei binaries in the 26 to 581 AU and $\Delta K < 4$ *mag* range would have been detected by the Pleiades survey, and the largest ΔK systems in α Per and Praesepe are only 0.5-1.0 mag beyond the Pleiades detection limit.

Another open cluster, the Hyades, has also been searched with high resolution techniques for companions to 162 A—early K stars (mass range ~0.6 $M_{SUN}$ to 2.4 $M_{SUN}$) (Patience et al. 1998). Its closer distance (D = 46.3 pc  Perryman et al. 1998), however, translates the observed 0".1 to 1".07 angular separation range into 5 to 50 AU. There are 23 binaries in the Hyades sample with $\Delta K < 4.0$ and projected separations from 5 AU to 50 AU. The Hyades data were taken with the same Palomar speckle camera and therefore have a similar companion sensitivity over this smaller projected separation range. Although there is little projected linear separation range overlap with the current survey, the proportion of Hyades binaries in the 5 to 50 AU range is only slightly higher than that of the α Persei, Praesepe, and Pleiades clusters in the 26-581 AU range. To increase the sample size, the distribution of Hyades speckle binary properties is combined with the other cluster data in order to study mass ratio distributions and the mass dependence of the binary properties (§6).

To extend the separation range coverage of the cluster data, systems detected in spectroscopic studies are used to construct an overall binary separation distribution (§5.3). The Hyades has been thoroughly investigated with both speckle imaging (Mason et al. 1993; Patience et al. 1998) and spectroscopy (c.f. Griffin et al. 1988; Stefanik & Latham 1992). Given the proximity of the Hyades and length of time the cluster has been observed, the two techniques provide continuous coverage of a large range of binaries from a separation of 0.02 AU (conversion of the measured 2dy period assuming a system mass of 1.4 $M_{SUN}$ and a conversion factor of 1.26 between separation and semi-major axis) to a separation of 50 AU. Nearly all of the 162-star Hyades speckle sample has been observed spectroscopically and 52 binaries have published periods, of which 16 are also resolved with speckle. Although the mass ratios are not known for most of the spectroscopic systems, the fact that all the closest speckle binaries were seen with spectroscopy suggests that the spectroscopic survey has a similar sensitivity. The few additional spectroscopic systems detected in the overlap range, but not detected with speckle are near the limit of the speckle survey; consequently, orbital motion could easily explain why these binaries were not resolved (Patience et al. 1998). A large sample of Praesepe members has also been investigated with long term spectroscopic surveys (Mermilliod & Mayor 1999, Abt & Willmarth 1999) and these results are also included in the analysis of the larger range of separations (details in appendix 3).

#### 4.3.2 Pre-Main Sequence Stars

The youngest comparison sample of stars is drawn from the many observations of T Tauri stars in the nearby star-forming regions—Taurus, Ophiuchus, Chamaeleon, Corona Australis, and Lupus. Since these star-forming regions are ~140 pc distant, the speckle and direct imaging studies covering 0".19 – 4".15 share the same 26-581 AU projected separation range as the α Persei and Praesepe observations. Over this separation range, the 2 Myr comparison sample consists of the 254 stars that have been observed by both speckle and direct imaging surveys (Ghez et al. 1993, 1997a; Leinert et al. 1993; Simon et al. 1995; and Köhler & Leinert 1998). As with the open clusters, additional surveys exist which are sensitive to binaries with separations outside 26-581 AU. Direct imaging surveys sensitive to systems wider than 4” augment the separation range coverage (Ghez et al. 1997a; Leinert et al. 1993; and Köhler & Leinert 1998)—the direct imaging sample is almost as large, with 240 stars. For binaries with separations less than 25 AU, the data are drawn from the 69 star 5m speckle sample (9-25 AU Ghez et al. 1993), the 82 star (some overlap with 5m speckle) lunar occultation sample (1-25 AU Simon et al. 1995), and the 53 star spectroscopy sample (Mathieu et al. 1989). In addition to the similarity in the observed separation range, the 2Myr sample also covers a comparable mass range to the cluster stars; although more difficult to determine, estimates of the masses range from ~0.2 $M_{SUN}$ to 2.5 $M_{SUN}$, with the nearly all the T Tauri targets >0.5 $M_{SUN}$.

Because of the excess emission and uncertain ages associated with young stars, the ΔK values measured for T Tauri binaries do not uniquely correspond to a mass ratio, making the mass ratio sensitivity level of T Tauri surveys more difficult to quantify. Based on the results of multi-wavelength studies which do determine mass ratios (White 1999, Ghez et al. 1997b, Hartigan et al. 1994) and on the set of theoretical evolutionary models (Baraffe et al.1998) favored by the analysis of the coevality of the GG Tau system (White et al. 1999), a mass ratio of $q$ = 0.25 (similar to the ΔK=4.0 limit of the older samples) is roughly consistent with a cutoff of ΔK~3 for the younger T Tauri stars. In the 26-581 AU range, there are 76 binaries with ΔK of 3.0 or less. An additional 19 binaries satisfy the ΔK cutoff at wider separations extending to 1582 AU (11”.3), while there are 20 binaries in the 1-25 AU range and 5 spectroscopic systems with separations from 0.02-1 AU (periods from ~2dys to ~1 year).

Another group of young stars that would provide interesting comparison samples is the population of young stars in Orion; current surveys, however, overlap only a limited portion of the 26 - 581 AU range considered for the clusters. Although the higher stellar densities associated with giant molecular clouds (GMCs) make these regions more likely progenitors of open clusters, the greater distance to the nearest GMC Orion (D~450pc, Genzel & Stutzki 1989) prevents a complete comparison with the current survey. The results from several large surveys -- an optical HST survey (Prosser et al. 1994) filters F547M (λo=5446 Å) and F875M and two K-band ground-based surveys (Simon et al. 1999, Petr et al. 1998) -- are included in the analysis of the binary separation distribution; the Orion data are compared with the wider cluster systems. The WFPC study of the Trapezium (Prosser et al. 1994) observed 319 targets in the Trapezium, the Simon et al. program covered 292 stars and the Petr et al. dataset includes 45 targets. The Orion sample have considerable overlap, however, and the total number of targets observed by at least one study is 480. Because the data are incomplete in both separation range and sensitivity, no attempt is made to correct the observed values so that they match the detection limits of the cluster surveys. Although the proportion of binaries measured by a study of wide (1000-5000 AU) common proper motion Orion systems (Scally et al. 1999) cannot be directly compared, the conclusions of the wide binary search are discussed in §5.3.2.

4.3.3 Field Stars

The oldest sample is the comprehensive spectroscopic and direct imaging survey of 164 solar neighborhood stars with spectral types ranging from F7-G9 (mass range 0.85 $M_{SUN}$ - 1.3 $M_{SUN}$) (Duquennoy & Mayor 1991). The 26 to 581 AU range corresponds to 4.76 to 6.79 in the log(Period[days]) units used in the G-dwarf study; this conversion assumes that binaries have an average total mass of 1.4 $M_{SUN}$ (measured by the G-dwarf survey) and that the semi-major axis is 1.26 times the projected separation (Fischer & Marcy 1992). This G-dwarf survey lists both detected and corrected companion star fractions; the corrected value accounts for undetected fainter companions down to the bottom of the main-sequence. Since the corrected values correspond to a mass ratio detection limit of 0.10, the G-dwarf results are reduced by 16% in order to restrict the correction to systems with q>0.2, comparable to the cluster limits. The scale factor is based on the G-dwarf mass ratio distribution (Duquennoy & Mayor 1991, their Figure 8 and last line of their Table 7); 16% of the companions have mass ratios undetectable by the α Persei and Praesepe cluster surveys.

Similar in age to the nearby G-dwarfs, but spanning a range in spectral type from A to M, the 106 northern stars within 8 pc provide another comparison sample. Drawing from many sources that enumerate the multiple systems in this well-surveyed sample, the binary census in this sample is estimated to be almost complete for all separations and for companion masses extending to the Hydrogen-burning limit (Reid & Gizis 1997). With the large number of low mass M stars in this sample, the uniform companion mass limit of 0.08 $M_{SUN}$ translates into a mass ratio limit above 0.25 for the ~40% of the sample with masses below 0.35 $M_{SUN}$. Because of the more similar mass range and sensitivity level, the G-dwarf sample is a better comparison for investigations of the age dependence of the overall binary distribution and of the 26-581AU *CSF*, however, the 8 pc sample does provide a data set with which the mass range can be extended.

## 5. Discussion of the Companion Star Fraction (CSF)

### 5.1 Definitions

The multiplicity of each cluster can be determined by counting either the number of multiple systems or the number of companions. The multiple star fraction (*MSF*) compares the number of binaries (*b*) and triples (*t*) to the total sample consisting of singles (*s*), binaries, and triples:

$$MSF = \frac{b+t}{s+b+t}$$

while the companion star fraction (*CSF*) counts the number of companions relative to the sample size:

$$CSF = \frac{b+2t}{s+b+t}$$

The analysis presented in this discussion uses the *CSF* rather than the *MSF*. The calculation of the cluster *CSF* proceeds in two ways: as a single value over the restricted projected separation range of 26-581AU (§ 5.2) and as a distribution function over a large range of projected separations (§ 5.3).

### 5.2 *CSF* over the 26 to 581 AU Range

Over the projected separation range 26 to 581 AU, the multiplicity of the open clusters α Persei, Praesepe, and the Pleiades is well-determined for ΔK less than 4.0 or mass ratios <~ 0.25 (see § 4.2 and 4.3). Based on the binary star detections within these projected separation and mass ratio boundaries, the α Persei $CSF_{\text{26-581AU}}$ is 0.09 ± 0.03, the Praesepe $CSF_{\text{26-581AU}}$ is 0.10 ± 0.03, and the Pleiades $CSF_{\text{26-581AU}}$ is 0.12 ± 0.03. There appears to be no significant difference between the $CSF_{\text{26-581AU}}$ of the three clusters. The combined results from the 385 members of these 3 clusters gives a $CSF_{\text{26-581AU}}$ of 0.10 ± 0.02

Figure 2 plots the $CSF_{\text{26-581AU}}$ as a function of age for *q >~ 0.25* range for 5 different samples. Since both the solar neighborhood G-dwarf survey and the surveys of the nearby (~140pc) star forming regions of Taurus, Ophiuchus, Chamaeleon, Corona Australis, and Lupus cover the entire 26 to 581 AU range, these two age groups are easily compared to the three clusters α Persei, the Pleiades and Praesepe. The criteria used to construct the comparison samples are discussed in § 4.3.2, and the resulting *CSF* for the T Tauri stars is 0.30±0.03 and is 0.16±0.03 for the G-dwarfs. Both α Persei and Praesepe have *CSF*s 3 times lower than the T Tauri stars (Ghez et al. 1993, 1997a; Leinert et al. 1993; Simon et al. 1995; and Köhler & Leinert 1998), but comparable to the the older solar neighborhood G-dwarfs (Duquennoy & Mayor 1991).

One proposed explanation of the factor of two discrepancy between the *CSF* of pre-main sequence stars and solar-aged stars is the disruption of multiple systems over time (c.f. Ghez et al. 1993). Theoretical N-body models by Kroupa (1995a, 1995b) of clusters with an initial total *CSF* (over all separations) of 1.0 predict different trends in total multiplicity depending on the initial stellar density, and these trends can be compared to the observational results. Although the evolutionary models predict the most pronounced effect on the binary fraction within the central 2 pc of a cluster, they also suggest that the overall binary fraction changes with age (Kroupa 1995a). Four cases are considered for the theoretical models: a loose association of binaries comparable to Taurus, a dense region comparable to the Trapezium, and two models with intermediate stellar densities. The typical separation between cluster members in the Hyades of 0.02 pc (Simon 1997) is intermediate between the mean separation of 0.003 pc for the Trapezium (McCaughrean & Stauffer 1994) and the wide spacing of 0.3 pc (Gomez et al. 1993) for Taurus members. The theoretical models predict that the binary fraction of a loose population declines by 10-25% from 2Myrs to the age of α Persei and then falls by an additional ~10% over the age range of the clusters, while the dense populations experience a rapid drop in binary fraction of 60-70% before the age of even the youngest star-forming regions and then remain at a constant low value. For the two intermediate densities more representative of the open clusters, the *CSF* evolution over the ~600 Myrs covered by the clusters also declines by at most ~10%. Under the assumption that the *CSF* evolution for the restricted separation range observed follows the same evolutionary trend as the total *CSF*, the cluster portion of Figure 2 can be compared with these simulations; since the T Tauri star sample is from a lower stellar density region, it cannot be included in the examination of evolution. Although the lack of change in the cluster *CSF* is consistent with the simulations, the predicted effect of ~1% (10% of the cluster *CSF*) is within the uncertainty, limiting the significance of this test. Studies of younger samples with similar densities are required for a more conclusive exploration of evolution as a possible cause of the observed *CSF* differences. Alternatively, the high *CSF* of the nearby dark cloud star-forming regions may be a consequence of different environmental conditions such as stellar density (Kroupa) or temperature (Durisen & Sterzik 1994).

Although previous results from the Hyades speckle survey covering a closer, narrower separation range of 5 to 50 AU were consistent with a decline in the *CSF* with age (Patience et al. 1998), the current data sets—which cover a different, wider separation range and provide more data points—do not support a systematic reduction in the *CSF* with age that occurs on timescales longer than a few x $10^7$ years. Another study (Mason et al. 1998) which suggested a tentative decline in multiplicity with age also focused on a closer separation range—2 to 127 AU—than the current α Persei and Praesepe surveys. The difference in conclusion between surveys probing different separation ranges suggests that examining the *CSF* over a wider range is important. The overall *CSF* distribution plotted in Figure 4a and discussed in the next section may explain the discrepant Hyades result.

### 5.3 Overall *CSF* Distributions

An assessment of the total *CSF* for open clusters covering all separations ≤ 581 AU and all magnitude differences $\Delta K < 4\ mag$ (mass ratios > ~0.25) is estimated by combining the 26 to 581 AU results from the three cluster surveys described above with spectroscopic surveys of the Hyades and Praesepe and with a speckle imaging survey of the Hyades which extend the data to closer separations (see §4.3.1). Merging the results assumes that all four clusters exhibit the same distribution, but this is supported by both the 26-581 AU speckle/AO data for three of the clusters and the 0.05 AU to 3 AU spectroscopic data for the Hyades and Praesepe. A series of K-S tests comparing the imaging portion of each pair of clusters over their common separation range shows no significant differencebetween any pair; Figures 3 a-d show the α Persei, Praesepe, Pleiades, and Hyades distributions individually. The combined cluster *CSF*, plotted in Figure 4a, extends over 4.5 decades of separation (0.02 AU to 581 AU). Summing the *CSF* in each bin over the entire range produces a total observed *CSF* for the clusters of 0.48 ± 0.05. Based on a Gaussian fit to the data, also shown in Figure 4a, the peak of the binary distribution occurs at a value of log(sep[AU]) = 0.6 ± 0.1 ($4^{+1}_{-1.5}$ AU).

The *CSF* distributions of younger and older samples, constructed from the surveys described in §4.3, are shown in Figures 4b-d. The oldest sample is the solar neighborhood (≤ 22 pc) G-dwarf survey, which covers a range larger than the entire separation range of the cluster distribution. Figure 4b shows the G-dwarf distribution, including the incompleteness corrections applied to the entire sample. Over the same 0.02-581 AU range covered by the cluster distribution, the corrected G-dwarf CSF is 0.49 ± 0.05; reducing this value by 16% to 0.41 ± 0.05 provides the best estimate to account for differences in companion detection sensitivity (see §4.3.3). Modelled as a Gaussian, the distribution has a peak at log(sep[AU]) = 1.6 ± 0.2 ($40^{+23}_{-15}$ AU), significantly different from the cluster value. The shifted cluster peak relative to the G-dwarf distribution explains why the Hyades *CSF* measured over the 5 to 50 AU range was larger than the G-dwarf value, while the *CSF* observed in the 26 to 581 AU range is consistent with the solar neighborhood value.

Figure 4c shows the distribution for nearby T Tauri binaries with $\Delta K \leq 3$ mag ($q > \sim 0.25$) in Taurus, Ophiuchus, Chamaeleon, Corona Australis, and Lupus. Compared to the cluster *CSF* histogram, the overall T Tauri CSF distribution has a larger integrated value and the peak is located at larger separations; for the same 0.02-581 AU range as the cluster distribution, the observed *CSF* is 0.69 ± 0.08. A Gaussian fit to the nearby T Tauri data yields a peak location of log(sep) = 1.8 ± 0.2 ($62^{+38}_{-22}$ AU), which is consistent with the G-dwarf peak, but more than 3σ larger than the cluster value of 4 AU. Another set of young binaries, drawn from optical HST images of Orion (Prosser et al.1994) and AO/speckle infrared images of the Trapezium region (Petr et al. 1998, Simon et al. 1999), are plotted in Figure 4d. Since these data sets are limited in their separation range coverage, a Gaussian is not fit to the Orion data. Incompleteness limits the ability to determine whether or not the distributions are different.

Brandner & Köhler (1998) have suggested that the field G-dwarf binary distribution represents a superposition of contributions from different populations, although they did not estimate the relative contributions. Figure 5 compares the fits to the separation distributions of the T Tauri (dotted line), open cluster (thin line) and field (dashed line) samples. Since the G-dwarf data have been corrected to account for companions down to the bottom of the main-sequence, the Gaussian fit to the field has been scaled down (by the same factor of 0.84 explained in §4.3.3) so that the sensitivity is comparable to the cluster and T Tauri distributions. Taking the cluster distribution as representative of stars forming in giant molecular clouds[1], the best fit to the G-dwarf distribution is obtained for a combination of $30^{+15}_{-10}$ % dark cloud binaries and $70^{+10}_{-15}$ % GMC binaries. The best fit is determined by minimizing the chi-squared difference between the Gaussian fits to the scaled G-dwarf distribution and the cluster/T Tauri combination, with the differences measured at increments of 0.25 in log(sep); the uncertainty represents the change in percentage that increases chi-squared by one. In Figure 5, the thick solid curve shows the best fit superposition of T Tauri and cluster binaries.

This analysis implies that a significant fraction of field binaries (and by extension, all field stars) may have formed in lower stellar density regions such as Taurus, rather than in the denser and more populous giant molecular cloud complexes which are the likely progenitors of open clusters. Although based only on the nondetection of wide proper motion systems, Scally et al. (1999) reach a similar conclusion from their observations of the Orion Nebula Cluster. The wide binary analysis suggests an 80%:20% division between formation in a clustered environment and loose association. In contrast, starcount studies have proposed estimates as high as 96% of stars are formed in a clustered environment (not necessarily bound) rather than in a uniform distribution (e.g. Lada et al.1991).

---

[1] The cluster distribution is intended to represent the shape for a dense star-forming region and this analysis is not meant to imply that such a large percentage of stars are members of open clusters at an earlier stage; the population of open clusters is not high enough to account for more than ~10% of all stars (Miller & Scalo 1978, Adams & Myers 2001).

## 6. Discussion of Observational Tests of Binary Formation Models

The binary star properties measured by this survey provide observational tests of several binary formation models. The mass ratio ($q=M_{sec}/M_{prim}$) distribution (§6.1), the mass dependence of both the *CSF* (§6.2.1) and the *q* distribution (§6.2.2), and the separation dependence of the *q* distribution (§6.3) are all important observational constraints on formation mechanisms. Although not all theoretical simulations extend to a point in binary evolution at which main sequence properties are determined, several formation scenarios including capture (§6.4.1) (McDonald & Clarke 1993, 1995; Sterzik & Durisen 1998) and fragmentation (§6.4.2) (Clarke 1996, Bate & Bonnell 1997, Bate 2000a, 2000b) make specific predictions that can be compared to the open cluster data. The survey results are presented first along with comparisons to previous observations and a discussion of the predictions from theoretical models follows.

### 6.1 Mass Ratio (*q*) Distribution

The shape of the *q* distribution presents a means of testing the predictions of capture theories, while the number of peaks in the mass ratio (*q*) distribution may provide an indication of the number of binary star formation mechanisms. The *q* distributions for α Persei, Praesepe, and the Pleiades over 26-581 AU and the Hyades over 5-50 AU all rise monotonically toward smaller mass ratios. Based on a series of K-S tests comparing each pair of these four clusters, none of the q distributions are significantly different. Consequently, all the results are used to construct an overall cluster *q* distribution, shown in Figure 6. This *q* distribution consists of the 54 cluster binaries with mass ratios of 0.40 or greater; this limit was chosen so that incompleteness should not be a problem (see §4.2). The histogram rises slightly toward systems with a smaller mass companion relative to the primary mass (low *q*), but is also consistent with a flat distribution. Since there is no evidence for bimodality in the cluster data, the *q* distribution does not support the idea of binary formation by two processes with distinctly different mass ratio distributions. Comparisons with particular models are made in §6.4.

The slope of the *q* distribution can be compared to the expectations of different types of capture models. The increase toward smaller mass ratios is not steep enough to be consistent with the $q^{-2.35}$ Salpeter power law. The shallower sloped function resulting from random pairing of two stars drawn from a mass function representative of an open cluster population is consistent with the observations; the mass function used for this simulation was a combination of three power laws: $N(m) \sim m^{1.5}$ for $m < 0.1$, $N(m) \sim m^{-1.05}$ for $m = 0.1\text{-}1.0$, and $N(m) \sim m^{-2(1+\log m)}$ for $m > 1.0$ (Reid & Gizis 1997, Meusinger et al. 1996).

### 6.2 Separation Dependence of the *q* Distribution

The combined open cluster data cover a large range of binary star separations (from 5 to 581 AU) and encompass the important size scale associated with circumstellar disks. Two formation scenarios–disk fragmentation and models of accretion following fragmentation –suggest that the binary properties should depend upon the system separation. Dividing the 54 cluster binaries with mass ratios exceeding 0.4 at successively larger binary separations and comparing the close and wide *q* distributions does not reveal distinct mass ratio distributions. In particular, when the sample is split at larger separations comparable to the sizes associated with disks (~100 AU), the mass ratio distribution of the closer systems remains indistinguishable from that of the wider systems. A tentative break at ~200 AU in the separation distribution of mass ratios is seen in studies of nearby T Tauri binaries (White & Ghez 2000; Köhler & Leinert 1998), but these T Tauri surveys extend to larger separations than the cluster observations, and the number of cluster binaries with separations >200 AU is small. Previous results based on the comprehensive G-dwarf survey (Mazeh et al. 1992) suggest that the *q* distribution of the closest binaries with periods less than 3000 dys, or separtions less than 5 AU, is different from the longer period *q* distribution, with a larger proportion of similar mass systems; again the cluster data would not be expected to reproduce this result since the dividing separation is at the boundary of the observed range.

### 6.3 Mass Dependence of the *CSF* and *q* Distribution

With a total of 544 main sequence stars ranging in spectral type from B to early M or masses of ~ 5 to 0.5 $M_{SUN}$, the combined cluster data set presents a unique sample to investigate possible correlations with binary mass. Among the 544 stars are 54 binaries with $q > 0.4$ and 63 binaries with $\Delta K < 4$ mag ($q > \sim 0.25$). Both capture and fragmentation models predict specific trends in the *CSF* and *q* distribution, making the mass dependence of these properties an important discriminant between different formation scenarios. Only the youngest cluster, α Persei still contains the more massive, short-lived B stars. The three speckle surveys of α Persei, Praesepe, and the Hyades include most of the A and F stars in these clusters. Since the AO Pleiades survey concentrated on Solar-type stars, it contributes mainly G and K stars. Additional G and K stars are provided by the α Persei and Hyades surveys. Finally, the Pleiades observations and the α Persei HST data include a small sample of M stars. Figures 7a-d plot spectral type/$(B-V)_0$ color histograms for each sample of stars in the four open clusters.

### 6.3.1 Mass Ratio (*q*) Distribution vs. Mass

The mass ratios of the 54 binaries used to construct the overall mass ratio distribution were sorted by their color and separated into two distributions; the dividing color was varied and K-S tests were performed to determine whether or not the two sets of mass ratios are significantly different. At a dividing color of $(B - V)_0 = 0.53$ (spectral type F8, mass ~1.2 $M_{SUN}$) the two *q* distributions are the most different and have only a 0.2% probability of being drawn from the same population. Figure 8 normalizes the *q* distribution in each sub-sample by the number of binaries and compares the fraction of binaries as a function of *q* for the higher and lower mass primary stars. The *q* distribution for higher mass primaries increases sharply toward smaller *q* values, while the *q* distribution of lower mass primaries is relatively flat. An extension of this trend of fewer small-*q* binaries for lower mass primaries is seen in the 8 pc sample which consists largely of M stars; the *q* distribution for the 8 pc sample increases toward *q~1* systems (Reid & Gizis 1997).

### 6.3.2 *CSF* vs. Mass

In order to investigate the mass dependence of the *CSF*, the 544 star open cluster sample, covering masses from 0.5 to 5 $M_{SUN}$, is divided into 4 roughly equal subsets based on the dereddened target color. The binaries included in this analysis are the 63 cluster binaries with $\Delta K <= 4$mag and separations from 5 to 50 AU (for the Hyades) or 26 to 581 AU (for the more distant open clusters). Although the Hyades data cover a different separation range, the overall *CSF* is comparable to the farther clusters and the Hyades stars constitute a similar fraction of each color group (~0.3). Figure 9 plots the *CSF* calculated for each of the 4 groups as a function of the median color of the subset; the straight line on the plot represents the *CSF* of the entire sample, $0.12 \pm 0.01$. The combined data show a higher *CSF* for the lower mass stars. The best fit line through the combined data has a slope of $0.13 \pm 0.04$, which is 3 $\sigma$ different from a flat line.

The upward trend in multiplicity with decreasing mass over the A-K star range was also suggested by previous results from the 8 pc sample (c.f. Reid & Gizis 1997), but the open cluster sample contains many more early-type stars and extends to the B stars. Over a similar mass range to the cluster dataset, a different trend of increased *CSF* for Herbig AeBe stars relative to T Tauri stars has been reported in an AO survey (Bouvier & Corporon 2000), but these results depend strongly on the large incompleteness corrections applied to the Herbig Ae/Be data -- the cluster data in Figure 9 are not corrected. This sample contains few M stars and it is difficult to directly compare previous results involving lower mass M stars that extend below the masses considered for the cluster surveys, but several M-star survey results are mentioned for completeness. The 8-pc M-dwarf data is not included in Figure 9 because the low masses of these stars makes a $\Delta K=4$ mag ($q<\sim 0.25$) companion below the stellar limit and detecting such faint objects is difficult. When measured differently with a companion mass cutoff rather than a mass ratio cutoff like the open cluster results, a decline in total *CSF* (over all separations) between G-dwarfs and M-dwarfs has been reported (Fischer & Marcy 1992, Reid & Gizis 1997). Over the separation range 14 to 825 AU, a low fraction of binaries amongst Hyades M stars has also been observed (Reid & Gizis 1997). Preliminary results from a continuation of the field M dwarf survey show a similar M-dwarf *CSF* to the G-dwarf value (Udry et al. 2000), but this study only includes spectroscopic systems. With the discovery of an increasing sample of lower mass L dwarfs, the binary fraction of this population is begining to be explored, and initial results suggest that, as with M dwarfs, the proportion of binaries is low (Reid et al. 2001). In conclusion, when measured over the late B to late K star range with a mass ratio limit, the *CSF* increases with decreasing mass, however, studies that measure or correct to a companion mass limit and extend to M stars show different trends of a *CSF* that is higher for Herbig AeBe than T Tauri Pre-Main Sequence stars and that declines over the G to M star range in the field.

### 6.3.3 Summary of Observational Trends

Several trends in the binary properties are evident from the sample of 544 open cluster members with spectral types from B to M and, containing 54 $q > 0.4$ binaries and 63 $q>\sim 0.25$ binaries, and with separations from 5-50 AU (for the Hyades) or 26-581 AU (for the farther clusters); Table 8a summarizes the observations. Considering binaries with $\Delta K < 4.0$ ($q >\sim 0.25$), the *CSF* increases for increasing color (equivalently decreasing mass) as shown in Figure 9. The mass ratio distribution from $q=0.4$-$1.0$, given in Figure 6, is flat or slightly increasing toward smaller mass ratios. The mass ratio distribution has a different shape, however, depending upon the primary mass; the higher mass primaries exhibit a steeper rise toward smaller $q$ systems as shown in Figure 8.

Finally, there is no correlation with separation in the *q* distribution over the 5-581 AU range covered by the cluster surveys.

### 6.4 Comparison with Star Formation Models

#### 6.4.1 Capture Models

Capture occuring in small-N clusters generates binaries with sufficient efficiency to explain the observed total frequency of binary and multiple systems (e.g. McDonald & Clarke 1993, 1995; Sterzik & Durisen 1998). The first model of this class considered -- small-N capture without disks (McDonald & Clarke 1993) -- preferentially produces binaries composed of the two most massive stars, creating a distribution that increases toward unity, inconsistent with the observational data (Figure 6). Because this type of capture preferentially forms binaries with the most massive stars, the resulting *CSF* is expected to decrease with decreasing primary mass, also contrary to the observed positive slope of the cluster data (Figure 9). A further refinement of the small-N capture process includes the effects of circumstellar disks in the calculations. Disk-assisted capture randomizes the companion to the primary – again the most massive star -- and creates more small-*q* binaries, similar to the observed distribution (Figure 6) (McDonald & Clarke 1995). Dynamical decay of few-body systems, another model of capture, predicts that the *CSF* is higher for more massive stars and that the *q* distribution for more massive stars should have a steeper slope (Sterzik & Durisen). Since the predicted trend in CSF as a function of mass involves a secondary mass cutoff rather than a mass ratio limit, it is difficult to test this expectation with the limited dynamic range of the current survey. It is possible to compare the *q* distributions in Figure 8 with the predicted effect of more small *q* systems among the more massive primaries and the observations match the theory. In summary, the observational data from the combined cluster sample are inconsistent with diskless small-N capture, but consistent with disk-assisted small-N capture and dynamical decay of few-body systems. Table 8b lists the comparisons of observations with these capture models.

#### 6.4.2 Fragmentation Models

The mass and separation dependences of the binary properties provide tests of several fragmentation models. One formation mechanism, disk fragmentation, is believed to operate over specific separation scales associated with the size of a circumstellar disk. The *q* distribution does not show a significant difference for separations larger and smaller than a typical disk size (~100AU— 200AU), suggesting that disk fragmentation is not the predominant form of binary formation in this separation range. This conclusion, however, is limited by the separation range of the survey.

Another type of fragmentation -- scale-free fragmentation -- predicts that the properties of binaries should be independent of mass (Clarke 1996). Specifically, the observed positive slope of the *CSF* vs. mass (color) is a problem for this theory which predicts that, for data with a mass ratio cutoff, the *CSF* should be independent of mass. The expected straight line plotted on Figure 9 does not match the data. Additionally, the observation that the *q* distribution for higher mass stars is significantly different from that of the lower mass stars, contradicts a scale-free formation process. Large-scale surveys that have measured a very low frequency of brown dwarf companions (e.g. Oppenheimer 1999, Macintosh et al. 2000) imply that the stellar limit (0.08 $M_{SUN}$) may be an important scale in the binary formation process, while the cluster results, which do not reach this limit, suggest that another more massive scale may influence binary properties.

Numerical models of the fragmentation process cannot currently simulate the full evolution from molecular cloud core to observable binary properties, however, recent simulations of accretion onto protobinary fragments have generated several testable predictions. In an isolated environment, binaries that accrete more material from a rotating cloud core with high specific angular momentum are driven toward similar component masses -- *q~1* systems (Bate 2000a). Since the initial mass of the protobinary increases with separation (Bate 2000a), close binaries need to accrete more material to obtain a given final mass; over a fixed separtation range, more massive systems will also accrete more material. Both closer and more massive binaries should contain more *q~1* binaries based on these simulations. Since the close and wide distributions show no clear difference in the fraction of highest *q* systems, the data do not show the predicted trend, but this may be due to the cutoffs in separation. The limited number of *q~1* systems also do not appear to vary with mass as expected. Accretion simulations modified to represent a clustered rather than a loose environment (Bate 2000b) have a different outcome, since the stellar motion is not correlated with the gas, giving the infalling gas a low specific angular momentum. In this case, massive stars are expected to form more low *q* binaries, in agreement with Figure 8.

It is also important to note that capture and fragmentation may not be separate processes, as fragmentation may produce the small clusters that are the initial conditions of the capture scenarios (e.g. Boss 1996). In summary, the observational trends are inconsistent with disk fragmentation, scale-free fragmentation, and simulations of accretion following fragmentation in a loose environment; a model of accretion following fragmentation in a cluster environment is consistent with the open cluster data. Table 8b summarizes the comparisons of observations and the fragmentation models.

## 7. Discussion of Binary Star Effects on Stellar Properties

The presence of a companion star can both passively skew the measured stellar properties of the primary if it is unresolved and actively alter the distribution of gas and dust around forming stars. One interesting consequence of companion stars is a possible explanation for the unexpected detection of X-ray emission from late-B and A stars. The first subsection, 7.1, investigates this effect. The second subsection, 7.2, discusses the role of companions in stellar rotational evolution.

### 7.1 Binaries and X-ray emission from A stars

In order to investigate the role of lower mass companions in the X-ray detections of late-B and A stars, this section considers a subsample of the cluster dataset -- the B and A stars in α Persei, Praesepe, and the Hyades which have been targeted by both ROSAT and high resolution multiplicity surveys. Since these stars lack the strong winds or dynamos which generate X-rays in O through early-B and F through M stars (c.f. Pallavicini 1989), unresolved companions may be the true source of the ROSAT detections. Binaries detected by speckle and spectroscopic surveys have separations within the error boxes, typically ~10”x10”, of the ROSAT detections.

Combining the results of the α Persei, Praesepe and Hyades surveys, a total of 90 stars in the $(B-V)_0$ color range of –0.13 to 0.30 have been searched for companions with speckle; half of these stars have also been included in spectroscopic multiplicity surveys (Abt 1965, Abt & Levy 1976, Bolte 1991, Burkhart & Coupry 1989, Burkhart & Coupry 1998, Prosser 1992, Morrell & Abt 1992, Stefanik & Latham 1992, Abt & Wilmarth 1999). Of these 90 stars, 78 were observed with either ROSAT raster scans or pointed observations, and 22 have X-ray detections (Randich et al. 1996, Prosser et al. 1996, Randich & Schmitt 1994, Stern et al. 1995). Table 9 lists the number of stars with X-ray detections or X-ray upper limits in each cluster and gives the number of binaries detected by either speckle or spectroscopy for each group.

All the speckle binaries have sufficiently small mass ratios for the companion to have a late enough spectral type to generate the observed X-rays. Some of the detected spectroscopic binaries, however, are SB2 systems consisting of two A stars which would not explain the X-ray emission; the number of SB2s are listed in parentheses after the number of binaries in Table 9 and these binaries are not counted in the following *CSF* calculations. Most of the spectroscopic systems are SB1s, which have lower mass companions. The *CSF* of the subset with X-ray detections is $CSF_{\text{X-ray}} = 0.6 \pm 0.2$ and the corresponding value of the subset without X-ray detections is $CSF_{\text{no X-ray}} = 0.21 \pm 0.06$. Because of the limited dynamic range of the surveys, the $CSF_{\text{X-ray}}$ is not expected to be 1.0 even if each B6 to A star has a companion producing the X-ray emission. For example, M star companions are not detectable. Although the faint companion hypothesis was not the favored interpretation of a high-resolution study of Herbig Ae/Be stars—the younger counterparts to the cluster B and A stars – the Herbig Ae/Be results also reveal a larger number of binaries among stars having X-ray detections (6 of 12) than among stars lacking X-ray emission (0 of 10) (Zinnecker & Preibisch 1994). This supports the cluster results of a $CSF_{\text{X-ray}}$ ~3 times higher than the $CSF_{\text{no X-ray}}$ and both Herbig AeBe and cluster statistics suggest that companions may be responsible for the X-ray detections. Due to the small sample sizes, however, the cluster difference only marginally significant; observations of a larger sample of early type stars will be required to determine more conclusively whether or not low mass companions can account for the apparent emission from late-B to A-type stars.

7.2 *CSF* and Stellar Rotation

Since the solar-type stars (F7 and later $(B-V)_0 \geq 0.50$, Kenyon & Hartmann 1995) in the youngest cluster α Persei are in a unique stage of stellar rotational evolution characterized by a wide range of rotational velocities (Stauffer et al. 1985, Prosser 1992.), the sample for the following analysis is limited to α Persei members in this spectral range. Two competing theories involving binary stars have been proposed to explain the range of observed rotational velocities. As discussed in the Introduction, these models make opposing predictions for the *CSF* of slow and rapid rotators.

Among the 71 solar-type stars in α Persei, the measured *vsini* values cover the entire measurable range from >200 km/s to <10 km/s. A division of the sample at a *vsini* of 20km/s places approximately equal numbers of targets in the slow and rapid category, with 37 slow rotators and 34 rapid rotators. Over the projected separation range of 26 to 581 AU and ΔK < 4 mag (q~ 0.25 to 1.0), the *CSF* of the slow rotators is $CSF_{\text{slow}} = 0.19 \pm 0.07$ and the *CSF* of the rapid rotators is $CSF_{\text{rapid}} = 0.09 \pm 0.05$. The difference between the two values of multiplicity is not statistically significant (~1 σ) and changing the cutoff value for slow and rapid rotation does not alter the result.

Although the *CSF* does not vary with the stellar rotation rate over the complete range of separation and mass ratio; it is possible that only binaries with certain parameters such as close separations have a discernable effect on the primary star rotation rate. Among the 10 Solar-type binaries with separations from 26 AU to 581 AU and $\Delta K \leq 4.0$ mag, there is no difference in the distribution of rotational velocities as a function of either mass ratio or separation. A similar result was found for the Pleiades (Bouvier et al. 1997). The Keck and Palomar observations in this α Persei survey, however, are sensitive to separations closer than the 26 to 581 AU range used in the rest of this study. The sample of stars observed with higher resolution is small, but not biased toward either slow or rapid rotators, since the stars observed with Keck and Palomar have a range of *vsini* values comparable to that of the entire sample. The HST data are sensitive to wider and fainter systems than the 581 AU cutoff. Figure 10 plots the larger set of *all* 16 solar-type binaries detected in α Persei and reveals a trend of increased rotational velocity with smaller separations. Dividing the sample of binaries at 60 AU results in a K-S test probability of only 2% that the close and wide systems have the same *vsini* distribution. The median *vsini* value is 105 ± 69 km/s for the binaries with separations < 60 AU, but only 11 ± 21 km/s for the systems wider than 60 AU. Because of the small sample, this result is preliminary and needs to be confirmed. Among the binaries, separation rather than mass ratio appears to be a more important factor in determining stellar rotation rates. Assuming a connection between disk lifetime rotation rate, the α Persei results suggest that binaries with separations closer than 60 AU have shorter-lived circumstellar disks which do not survive long enough to effectively slow the stellar rotation. The α Persei result is consistent with the observation of reduced millimeter and submillimeter flux from T Tauri binaries with separations less than ~50-100 AU compared to wider binaries in several star forming regions (Jensen et al. 1994, Osterloh & Beckwith 1995, Jensen et al. 1996).

## 8. Summary

With near-infrared speckle and HST NICMOS images of α Persei and Praesepe members, an accounting and analysis of the cluster binaries has been performed. Combining data from the IRTF, Palomar, Keck, and HST, a total of 142 α Persei and 100 Praesepe members with spectral types from B to early-M and masses from ~5 $M_{SUN}$ to ~0.5 $M_{SUN}$ have been observed. Among the Praesepe sample is a total of 12 binaries, while there are 21 binaries and 1 triple in the larger α Persei sample. In α Persei, an additional 6 binaries and 1 quadruple are possible multiples with substellar companions. These potential substellar companions, however, require follow-up confirmation and are not included in the analysis. The detected systems range in separation from 0".053 to ~5".0, with the majority newly resolved; only 7 of the 41 multiples are previously known systems.

Because of the youth of these cluster stars, it is possible to test for correlations between the presence of a companion and signatures of stellar activity such as X-ray emission and rapid rotation. The combination of the data on α Persei and Praesepe early-type stars with previous speckle and spectroscopic data on the Hyades shows that the majority, 0.6±0.2, of late-B and A-type stars with X-ray detections have companions; a *CSF* of only 0.21 ± 0.06 is measured for the stars in this spectral range without X-ray detections. This high *CSF* among X-ray targets suggests that the unexpected X-ray emission from these stars originates from later-type companions, although the small sample of early-type stars limits the statistical significance of the difference between the early-type stars with and without X-ray detections.

The role of companions to solar-type α Persei stars in the wide range of rotational velocities measured for the members of this youngest cluster is also explored. Different theories have suggested that binaries should be associated with either the slow or rapid rotators, but the $CSF_{26\text{-}581AU}$ for the slow rotators is indistinguishable from that of the rapid rotators. With the inclusion of the separations below 26 AU and fainter systems, the distribution of *vsini* values is, however, significantly different for the binaries with separations less than 60 AU compared to the wider systems; the median *vsini* for the close binaries is 105 ± 69 km/s compared to 11 ± 21 km/s for the wide binaries. This suggests that the closest binaries may disrupt the circumstellar disk to an extent that it cannot provide a braking mechanism.

Over the projected separation 26 to 581 AU and magnitude difference range ΔK < 4mag (q >~ 0.25), both α Persei and Praesepe have a $CSF_{26\text{-}581AU}$ of 0.10 ± 0.03. This value is consistent with the $CSF_{26\text{-}581AU}$ for the intermediate-aged Pleiades cluster and the solar-aged G-dwarfs, but significantly lower than the $CSF_{26\text{-}581AU}$ for nearby T Tauri stars. With the similarity in the 26-581 AU multiplicity from ~90 Myr to 5Gyr, there is no evidence for a systematic decline in *CSF* that occurs on timescales less than $10^7$ - $10^8$ yrs. Given the similarity in *CSF* for the clusters, the α Persei, Praesepe, and Pleiades 26 to 581 AU binaries are combined and then merged with the Hyades/Praesepe spectroscopic and Hyades speckle data sets in order to construct an overall cluster *CSF* vs. separation distribution spanning the range -1.75 to 2.75 in log(sep[AU]). This results in an estimate of the total cluster *CSF* for $\Delta K \leq 4.0$ $(q > \sim 0.25)$ of 0.48 ± 0.05. The cluster distribution peaks at 5 AU, a significantly smaller value than the Solar Neighborhood and nearby T Tauri star distribution peaks. Taking the cluster and T Tauri distributions as representative of their initial populations, a simple population synthesis model leads to the suggestion that the nearby G-dwarf binary distribution is a combination of approximately one-third dark cloud T Tauri binaries and two-thirds cluster/GMC binaries.

The observational results from the combined cluster sample involving the *CSF* and *q* distribution are compared with a number of binary formation scenarios. Considering all binaries with separations from 26 to 581 AU and with mass ratios from 0.40 to 1.0, the *q* distribution rises slightly toward lower *q*. This result is consistent with disk-assisted small-N capture, but the slope is too shallow to be explained by random pairing from a Salpeter mass function and too steep to be consistent with diskless small-N capture. The mass and separation dependence of the *CSF* and *q* distribution represent important tests of other simulations. A trend of increasing *CSF* with decreasing mass is seen over the range of the sample, and the *q* distribution of systems with $M_{prim}$> ~1.2 $M_{SUN}$ (B- F stars) is significantly different from that of the systems with $M_{prim}$< ~1.2 $M_{SUN}$ (G and later stars), with a steeper slope (fewer *q~1* systems) associated with the higher mass primaries. The later result is predicted by both dynamical decay of small groups of stars and accretion in clusters, but contradicts accretion simulations in loose associations. The presence of a clear mass dependence in the *q* distribution and a strong suggestion of one in the *CSF* is inconsistent with scale-free fragmentation. Finally, the overall *q* distribution does not vary with separation,which may be difficult to explain with models of disk fragmentation. In summary, the observational results are consistent with disk-assisted small-N capture, dynamical decay of 3-5 body systems, random capture from a cluster mass function, and accretion following fragmentation in a cluster, and contradict the predictions of several other formation scenarios -- random capture from a Salpeter mass function, diskless small-N capture, disk fragmentation, scale-free fragmentation, and accretion following fragmentation in a loose environment.

Acknowledgements

J. P. and A. G. were visiting astronomers at the Infrared Telescope Facility which is operated by the University of Hawaii under contract from the National Aeronautics and Space Administration. Some of the data presented in this paper were obtained at the W. M. Keck Observatory, which is operated as a scientific partnership among the California Institute of Technology, the University of California, and the National Aeronautics and Space Administration. The observatory was made possible by the generous financial support of the W. M. Keck Foundation. The authors wish to extend special thanks to those of Hawaiian ancestry on whose sacred mountain we are priviledged to be guests. Without their generous hospitality, these observations would not have been possible. With great appreciation, we thank the many people who have helped obtain the large amount of data required for this project: Alycia Weinberger at Palomar, Russel White, Lisa Prato, and Angelle Tanner at the IRTF, Al Schultz for HST, and Caer McCabe at Keck. We thank the telescope operators for their assistance with our observing program: Juan Carrasco and Rick Burruss at Palomar, Bill Golisch, Dave Griep, and Charlie Kaminsky at the IRTF, and Barbara Schaeffer and Theresa Chelminiak at Keck. Finally, we thank the referee Rainer Köhler for his helpful suggestions. Support for this work was provided by grant no. NAG5-6975 through the Origins of Solar Systems Project, grant no. GO-07833.01-96A from the Space Telescope Science Institute, and a Packard Foundation fellowship. J.P. also received support from a dissertation year fellowship from the American Association of University Women. Some of this work was performed under the auspices of the U. S. Department of Energy by the University of California, Lawrence Livermore National Laboratory under Contract W-7405-ENG-48. This research required extensive use of the SIMBAD database, operated at CDS, Strasbourg, France.

## Appendix 1: Comparison with Previous α Persei and Praesepe Surveys

Both α Persei and Praesepe have been targeted by a number of previous binary star searches employing a variety of techniques – spectroscopy, lunar occultation, optical speckle, and direct imaging. These surveys provide information about binaries with separations that complement the current IR speckle survey. Since many of these previous surveys include only a subset of the brighter stars, it is not currently possible to obtain a complete census of multiple systems.

### Praesepe

Praesepe has been surveyed by several techniques covering a wide range of separations. Spectroscopic studies of the early-type stars and photometric binaries have detected the closest binaries in the cluster. A survey of Praesepe A stars (Burkhart & Coupry 1998) includes 9 stars in the IR speckle sample. Notes in Tables 5 and 7 indicate the 3 singles and 6 binaries – KW 40, KW 224, KW 276, KW 279, KW 300, and KW 350—detected by this radial velocity survey. One of the spectroscopic binaries – KW 224 – also has a ~1" speckle pair, making it a triple system; the other spectroscopic binaries are speckle singles. Another spectroscopic survey targeting stars located above the main sequence in the H-R diagram includes 8 stars from this sample; 5 are spectroscopic singles and 3 are spectroscopic multiples—KW 365, KW 292, and KW 142 (Bolte 1991). One of the spectroscopic binaries—KW 365—is also a speckle binary; this speckle companion may explain why KW 365 is an SB3 system (Mermilliod et al. 1994), and it is counted as a triple rather than a quadruple. Results from long term radial velocity monitoring surveys of Praesepe (Mermilliod & Mayor 1999; Abt & Willmarth 1999) include 49 of the 98 members (2 of 100 targets are nonmembers) observed in the infrared sample -- 41 speckle singles and 8 speckle binaries. Notes about the individual stars are included in Tables 5 and 7. Of the 41 speckle singles, 1 is a triple -- KW 40 -- and 16 are spectroscopic binaries -- KW 534, KW 16, KW 47, KW 50, KW 142, KW 181, KW 182, KW 229, KW 268, KW 279, KW 300, KW 341, KW 371, KW 416, KW 479, AND KW 496. Among the 8 speckle binaries monitored spectroscopically, 4 are either spectroscopic systems with long periods or they are photometric/visual systems which suggest that they are binary, not triple stars -- KW 284, KW 275, KW 458, and KW 224. One of the speckle binaries is also a spectroscopic system, but with a short period which indicates it is a triple star -- KW 365.

Lunar occultation measurements detect the next closest set of binaries. Peterson and White (1984) observed 27 stars of this Praesepe sample and resolved 3 of them—KW 212, KW 143 and KW 203. Both KW 212 and KW 203 are detected by the IR survey, but KW 143 has a separation below the limit of the IR speckle survey. An optical speckle survey by Mason (1993) has 48 stars in common with this IR study, and 4 of these 48 stars are resolved—KW 265, KW 284, KW 212, and KW 203. Three binaries—KW 212, KW 203 and KW 284—are also detected by IR speckle, but the companion to KW 265 is not seen in either Palomar speckle or IRTF shift-and-add observations, despite the relatively wide separation of 0".425 measured by optical speckle. An additional 3 of the 48 stars—KW 385, KW 224, and KW 232—are binaries based on this IR survey, but not by the optical speckle measurements. KW 385 is outside the optical speckle field-of-view and the remaining two systems have ΔK values that correspond to ΔV values fainter than the optical speckle $\Delta V = 3$ mag detection limit (Mason 1993). An additional 15 visual binaries are compiled in the ADS, IDS, and WDS catalogues (Aitken 1932, Jeffers et al. 1963) and the projected separations range from 1".4 to 99".68. The two systems with separations less than 3".4 (the IRTF camera field-of-view)—KW 224 and KW 385—are both resolved by the IR speckle measurements. The remaining 13 systems are not considered physically associated since even the closest pair has a separation as large as ~21". Excluding the 13 widest pairs results in a lower limit to the overall single:binary:triple:quadruple ratio of 67:28:3:0 and a total *CSF* of $0.38 \pm 0.06$ for the observed Praesepe speckle sample. Figure A1 plots a color-magnitude diagram for the Praesepe sample with the binaries detected by any technique indicated.

α Persei

Although α Persei has not been as extensively surveyed as Praesepe, a number of spectroscopic and visual binaries are known in the cluster. In a spectroscopic survey of early spectral type members, Morrell & Abt (1992) observed 23 stars in this IR sample and discovered 4 binaries among them—HE 423, HE 817, HE 774, and HE 775. An additional 3 systems—HE 868, HE 955, and HE 965—are categorized as probable SB1 stars (Morrell & Abt 1992). Because HE 965 also has a speckle companion, this system is a triple. Another two spectroscopic binaries in the speckle sample – HE 848 and HE 143—are reported in Mermilliod (1991). Membership surveys also detect spectroscopic binaries, and one system—HE 314—is listed as a binary, while another system—HE 285— is listed as a probable spectroscopic binary in Prosser (1992). The HE 285 system is also resolved by speckle, and is counted as a binary rather than a triple.

A total of 8 visual binaries are listed in α Persei (Aitken 1932, Jeffers et al. 1963) with separations ranging from 0".68 to 26".91. HE 835 and HE 1082 have separations within the IRTF camera field-of-view. The visual binary HE 835 is also a seen in the IR data, but HE 1082 was not detected by this survey; HE 1082 has a reported separation of 2”.05 in 1881, but no subsequent measurements. The remaining 6 of these systems—HE 665, HD 18537/HD 18538, HE 955, HE 490, HE 799, and HE 828—have projected separations greater than 3”.4. The closest 5 of these 6 visual binaries are considered physically associated since the probability that these systems are chance projections is less than 50% (based on the same estimation as in § 5.3.1). The widest system – HE 955 – is not counted as a binary because the probability it is a background object exceeds 50% and successive measurements show that the two stars do not share a common proper motion. This accounting results in a lower limit single:binary:triple:quadruple ratio of 112:28:3:0 and a *CSF* of 0.24 ± 0.04 for the α Persei sample. Figure A2 plots a color-magnitude diagram for the α Persei sample with the binaries detected by any technique indicated.

Appendix 2: Magnitudes and Masses

The ground-based data for α Persei and Praesepe only measure ΔK for the binaries and $\Delta K_{lim}$ for the singles, but absolute $M_K$ magnitudes for each star are needed to determine mass ratios and mass ratio limits. The single star or binary system K magnitudes, the stellar distances and the extinction are required to convert the ΔK and $\Delta K_{lim}$ into binary $M_{K1}$, $M_{K2}$, and single $M_K$ and $M_{klim}$ from which masses are determined with mass-magnitude relations. The K-band mass-magnitude relations are taken from Henry & McCarthy (1993) for the solar-mass and lower targets, and an extension to higher mass stars listed in Patience et al. (1998) is used for the early-type stars. The relations are repeated here

$\log(M/M_{Sun}) = -0.159M_K + 0.49$ $\quad M_K < 3.07$

$-0.1048\, M_K + 0.3217$ $\quad M_K = 3.07\text{-}5.94$

$-0.2521\, M_K + 1.1965$ $\quad M_K = 5.94\text{-}7.70$

$-0.1668\, M_K + 0.5395$ $\quad M_K = 7.70\text{-}9.81$

For the majority of the cluster stars, $K_s$ measurements are available from the 2MASS database (ref). Since the mass-magnitude relations are based on K rather than $K_s$, the 2MASS $K_s$ values are converted to K values based on the transformation

$K = K_{CIT} = (K_s)_{2MASS} + 0.024$

given in Carpenter (2001). The CIT system is chosen since it is closest to that used in the photomety for constructing the mass-magnitude relations.

The current 2MASS data release covers the majority of the sources considered in this work. Of the 544 sources considered in this work, only 26 α Persei targets, 13 Praesepe targets, 7 Pleiades binaries, and 7 Hyades binaries are not included in the 2MASS point source catalogue. For the α Persei and Praesepe samples, the stars measured by 2MASS are used to construct a $V\text{-}(K_s)_{2MASS}$ relation from which the K for each star without a 2MASS measurement is estimated. The empirical $V\text{-}(K_s)_{2MASS}$ fit is

$(K_s)_{2MASS} = -3.34 + 12.14\log(V)$

for α Persei and

$(K_s)_{2MASS} = -5.69 + 14.49\log(V)$

for Praesepe. System K magnitudes for the Pleiades binaries not in the 2MASS sample are taken from measurements given in Bouvier et al. (1997). For the Hyades binaries not covered by 2MASS, the system K magnitude is estimated by converting the (B-V) color into a (V-K) color based on the relation given in Patience et al. (1998). These alternate methods, which affect only a small portion of the sample, were checked for stars with 2MASS measurements and give magnitudes within 10% of the 2MASS values. Once the absolute system magnitude is determined by

$M_{Ksystem} = K_{system} - 5\log(D/10) - A_K$

the primary and secondary absolute magnitdes are given by

$M_{Kprim} = M_{Ksystem} + 2.5\log[1 + 10\text{^}(-\Delta K/2.5)]$

$M_{Ksec} = M_{Kprim} + \Delta K$

and these values are substituted into the appropriate mass-magnitude relation. The distances to $\alpha$ Persei, Praesepe, and the Pleiades are taken from Pinsonneault et al. (1998) and are given in the sections 2 and 4.3.1. The distances to the Hyades members are determined individually by scaling values from Schwan (1991) or Reid (1992) as described in Patience et al. (1998). There is no reddening correction for Praesepe or the Hyades, but $A_K$ is 0.035 for $\alpha$ Persei and 0.014 for the Pleiades based on the E(B-V) listed in Pinsonneault et al. 1998 and the relations given in Rieke & Lebovsky (1985).

For the $\alpha$ Persei stars observed with HST, a similar procedure was used with J magnitudes instead of K. In this case, the data provide $\Delta$F140W and $\Delta F140W_{lim}$, so an extra step of changing the F140W into J is required. The single stars with measured 2MASS $J_{2MASS}$ magnitudes were converted to $J_{CIT}$ with the relation from Carpenter (2001)

$$J_{CIT} = 0.947J_{2MASS} + 0.053K_{2MASS} + 0.36.$$

As with the K-band data, the 4 stars that were not observed with 2MASS were assigned estimated J2MASS magnitudes based on their V magnitudes with the relation

$$J_{2MASS} = -5.72 + 14.8\log(V)$$

After switching the measured or estimated 2MASS magnitudes to the CIT system, the single stars were used to construct a F140W - J relation (the zeropoint is arbitrary)

$$J_{CIT} = -3.42 + 1.03(F140W)$$

Since the slope is not exactly unity, the observed $\Delta$F140W values are multiplied by the slope 1.03 to obtain $\Delta$J. The extinction $A_J$ for $\alpha$ Persei is 0.087. With the absolute $M_J$ system determined, the component absolute magnitudes are solved for and the masses are estimated with the Henry & McCarthy relations (1993)

$$\log(M/M_{Sun}) = -0.0863M_J + 0.3007 \qquad M_J = 3.48\text{-}6.97$$

$$-0.2791\, M_J + 1.6440 \qquad M_J = 6.97\text{-}8.56$$

$$-0.1593\, M_J + 0.6177 \qquad M_J = 8.56\text{-}10.77.$$

Appendix 3: CSF Distribution Details

The tables and explanations in this appendix are designed to help reconstruct the *CSF* distributions of open cluster stars, field G-dwarfs, nearby T Tauri stars, and T Tauri stars in Orion that are plotted in Figures 4a-d. Tables A1-A3 list the binaries included in each separation bin and give references for the surveys from which the binaries were drawn. Whenever possible, the binaries counted from each survey are limited to the binaries with separations in which the entire bin is covered by the survey. The one exception to this is the log(sep[AU])=0.95-1.4 bin in the nearby T Tauri distribution; the closest speckle systems resolved with the Hale 5-m are included, although this dataset is not as sensitive to all separations covered by this bin.For the open clusters and the nearest star forming regions, some additional binaries are known but not included. Some binaries exceed the ΔK=4mag limit and others have separations placing them in bins not entirely covered by the survey.

Because the original G-dwarf distribution (Duquennoy & Mayor 1991) includes corrections, it is not possible to generate an analogous table indicating which binaries are included in the individual bins. Instead, the distribution shown in Figures 4b and 5 is a scaled version of the original Figure 7 plot in Duquennoy & Mayor (1991). Two scalings are required: one reduction to account for the sensitivity difference and a second decrease in the number of companions due to the smaller bin size of the distributions in Figure 4a-d. The log(sep[AU]) bins are 2/3 the width of the G-dwarf log(P[dys]) bins, so the Figure 4b incorporates the bin size scaling and Figure 5 includes the additional sensitivity scaling. For example, the log(sep[AU])=0.95-1.4 bin is 67% of the ΔlogP[dys]=4-5 bin which includes 18 companions including correction in the original G-dwarf distribution; the corresponding number of companions in Figure 4b is 12 or a *CSF* of 0.074. In Figure 5, the G-dwarf curve is further scaled down by 16% as discussed in section 4.3.3 to account for the difference in sensitivity.

Figure 1: Binaries and Detection Limits

The detected binaries in the α Persei and Praesepe samples are plotted with a different symbol depending upon the telescope used; HST companions are further differentiated based on the companion mass. Open circles connected by solid lines delineate the median detection limits of the sample, with the error bars indicating the standard deviation of the limits. Dashed lines mark the bounds of separation range considered for the complete sample. Since the diffraction-limit of the IRTF is 0".15, this separation is taken as the lower limit for the complete range, although a number of binaries are resolved to separations as little as 0".027 by Keck, Palomar, and HST. The outer limit is 3".3 for α Persei and 3".4 for Praesepe, with the slight difference due to 5 pc variation in the cluster distances.

Figure 2: Age Dependence of the $CSF_{26\text{-}581AU}$

The *CSF* over the separation range 26 to 581 AU is plotted as a function of age for the 2 samples surveyed for this study and for 3 comparison samples; values from this work are noted by filled circles (α Persei at 90 Myr and Praesepe at 660 Myr), while open circles mark the values from previous surveys (nearby T Tauri stars at 2 Myr, the Pleiades at 125 Myr and field G-dwarfs at 5 Gyr). Since the α Persei and Praesepe data are sensitive to companions with $\Delta K < 4$ mag, the data from the comparison surveys are trimmed to this sensitivity level before calculating the *CSF;* details of this procedure are given in section in section 4.3. The result from the Hyades speckle survey is not included because the separation range coverage is different and the age is the same as that of Praesepe. All the samples older than the T Tauri stars have consistent, significantly lower multiplicities than the T Tauri stars, suggesting that the *CSF* does not decline with age on timescales measureable with the cluster data.

Figures 3 a-d: *CSF* Distributions of the Open Cluster Samples

Idividual *CSF* distributions for each of the four open clusters are plotted over the range of separations covered by large scale multiplicity surveys. The data for (a) the α Persei distribution is taken from the current survey, while (b) the Praesepe plot includes spectroscopic binaries in addition to the speckle systems reported in the current survey. The comparison samples of (c) the Pleiades and (d) the Hyades are also shown; the Pleiades binaries were detected by an AO survey and the Hyades represents a combination of spectroscopic and speckle systems. The references for the cluster surveys are given in sections 2 and 4.3.1 and Table A1 lists the specific binaries that are included in each bin of the distributions.

Figures 4 a-d: *CSF* Distributions of Four Samples

Comprehensive *CSF* distributions spanning several orders of magnitude in separation are constructed for (a) cluster stars, (b) nearby G-dwarfs, and (c) T Tauri stars. The less complete data available for Orion are also shown in (d); the x's denote that the surveys are incomplete at the extremes of the bins. Compared to the original plot (Duquennoy & Mayor 1991) of the G-dwarf distribution in terms of log(P[dys]), each of the 0.45

log(sep[AU]) bins is 0.675 in logP[dys]. The specifics of the G-dwarf rescaling are given in section 4.3 which also describes the selection criteria for the T Tauri binary sample. The peak of the cluster distribution occurs at log(sep[AU]) = 0.6 ± 0.1 ($4^{+1}_{-1.5}$ AU), which is a significantly smaller value than the location of the G-dwarf peak value log(sep[AU]) = 1.6 ± 0.2 ($40^{+23}_{-15}$ AU) and the T Tauri distribution peak of log(sep) = 1.8 ± 0.2 ($62^{+38}_{-22}$ AU); the peak positions of the youngest and oldest samples are consistent with each other. Gaussian curve fits are included for the first 3 distributions, however, there is too little data for a fit to Orion.

Figure 5: Gaussian Fits to the *CSF* Distributions

The Gaussian fits to the (long-dashed line) scaled G-dwarf sample, (dotted line) T Tauri data, and (thin solid line) Hyades/α Persei/Praesepe/Pleiades cluster sample are plotted along with the (thick solid line) combination of cluster and T Tauri distributions that best matches the shape of the G-dwarf curve. Based on this very simple population synthesis, the data suggest that the G-dwarf binary population is a combination of approximately one-third binaries formed in loose T associations and two-thirds binaries formed in higher stellar density regions.

Figure 6: Cluster *q* Distribution

The mass ratio distribution (*q*) of the combined cluster sample is shown. In order to avoid any detection bias, the binaries only include systems with mass ratios of 0.40 or greater and with specific separations. For the Hyades, the separation range is restricted to 5 to 50 AU, while the range considered for α Persei, Praesepe, and the Pleiades is 26 to 581 AU.

Figures 7 a-d: Cluster $(B – V)_0$ Color Histograms

Histograms of the $(B\text{-}V)_o$ color are displayed for the stars in the current surveys of (a) α Persei and (b) Praesepe. Two comparison samples observed in previous high resolution surveys and included in much of the analysis are also plotted: (c) the Pleiades and (d) the Hyades. Considering all four surveys, the spectral types observed cover B to early M, which corresponds to $(B\text{-}V)_o$ colors from -0.2 to 1.7 and masses between 5 $M_{SUN}$ and 0.5 $M_{SUN}$. The Praesepe sample is limited to early types, the α Persei and Hyades samples include both early-type stars, and the Pleiades sample is composed almost exclusively of solar-type stars. The range of stellar masses observed is important for testing mass-dependent predictions of binary formation models.

Figure 8: Mass Dependence of the *q* Distribution

The normalized mass ratio distributions for bluer (higher mass) binaries (dashed line) and redder (lower mass) binaries (solid line) are plotted. At a dividing $(B\text{-}V)_0$ color of 0.5, which corresponds to a spectral type of ~F7, the distributions are significantly different, with the lower mass systems showing fewer low mass ratio companions. This deficit of low-mass companions occurs above the stellar limit.

Figure 9: Mass Dependence of the *CSF*

The *CSF* calculated for four ranges of target star color is plotted as a function of the median color; the 545 star sample is split into four subsets with ~136 stars in each color range. Only binaries with $\Delta K < 4$ mag and separations of 26-581AU (for α Persei, Praesepe, and the Pleiades) or 5-50AU (for the Hyades) are included in the calculation. Since the mass ratio limit is comparable, the redder color bins have progressively lower companion mass limits. Over the range of stellar colors (masses) considered for this survey, the companion star fraction (*CSF*) rises toward redder colors or smaller masses. The upward slope of $0.13 \pm 0.04$ is 3 σ different from a flat line. An decrease in *CSF* with mass is inconsistent with both the diskless small-N capture and the scale-free fragmentation formation scenarios.

Figure 10: α Persei *vsini* vs. Separation

The rotational velocity is plotted as a function of binary separation for the solar-type stars in α Persei. Since the largest rotational evolution occurs for solar-type stars, this sample is limited to binaries with spectral types later than F7, or colors of $0.50 \geq (B\text{-}V)_0 \geq 1.6$. Filled circles mark the binaries with separations of 26 to 581 AU and mass ratios exceeding 0.25. The open circles at small separations represent binaries separated by less than 26 AU that could only be detected with the higher resolving power of the Keck telescope, while the open circles at larger separations represent binaries with small mass ratios (but with stellar companions) that could only be observed with the greater dynamic range of HST. Systems separated by less than 60 AU have significantly higher velocities than the wider pairs, suggesting that these stars lost their disk-braking mechanism earlier, possibly due to disk-disruption by the companion star.

Figure A1: Praesepe Sample Color-Magnitude Diagram

The V-(V-K) color-magnitude diagram is plotted for the observed Praesepe stars excluding the stars determined to be nonmembers and the 3 G and K giants in the sample that were saturated in 2MASS photometry. Single stars are labelled with open diamonds, while binary stars are denoted with a filled-in diamonds and account for most of the stars above the main sequence. These stars have been well-surveyed by multiple techniques, and the binaries detected by any method are included in the figure. A detailed listing of the binaries is given in Tables 5 and 7.

Figure A2: α Persei Sample Color-Magnitude Diagram

The V-(V-K) color magnitude diagram is plotted for all stars in the observed α Persei sample. Single stars are marked with an open diamond and any known binary stars are distinguished by filled diamonds. The faint potential substellar companions in the HST dataset are not included, as those systems require additional confirmation. The α Persei cluster has not been as comprehensively searched for binaries as Praesepe, however, and a significant number of multiple systems are probably not yet identified. The binaries included in this plot are listed in Tables 4a and 6.

**Table1: α Persei Sample**

| Object | BD | HD | RA[a] | 2000 | | DEC[a] | 2000 | | V[a] | B-V[a] | SpTy[a] | K | LogLx[c] | LogLx[d] | vsini | Tel |
|---|---|---|---|---|---|---|---|---|---|---|---|---|---|---|---|---|
| HD 18537 | +51 665 | | 3 | 0 | 52.0 | 52 | 21 | 7 | 5.00 | -0.04 | B7V | 5.468 | | | 220 | I |
| HD 18538 | +51 665B | | 3 | 0 | 53.3 | 52 | 21 | 8 | 6.74 | 0 | B9V | 6.691 | | | | I |
| HE 12 | +48 851 | | 3 | 7 | 50.3 | 49 | 6 | 30 | 10.09 | 0.51 | F6V | 8.797 | | | 49 | P, I |
| HE 56 | | | 3 | 9 | 51.9 | 48 | 28 | 16 | 10.84 | 0.81 | G3V | 9.007 | | | <10 | P, I |
| HE 93 | | | 3 | 10 | 41.6 | 50 | 31 | 32 | 11.09 | 0.7 | G4 | 9.275 | | | 12 | I |
| HE 92 | | | 3 | 10 | 44.2 | 50 | 20 | 47 | 11.06 | 0.65 | | 9.788 | | | 27 | I |
| HE 94 | | | 3 | 11 | 16.6 | 48 | 10 | 37 | 10.42 | 0.64 | F9V | 8.728 | | | | K |
| HE 104 | +47 776 | 19655 | 3 | 11 | 40.9 | 48 | 3 | 16 | 8.60 | 0.34 | F2Vn | 7.778 | | | >200 | P, I |
| HE 135 | +49 868 | | 3 | 11 | 49.8 | 50 | 22 | 48 | 9.71 | 0.49 | F5V | 8.556 | | | 16 | I |
| HE 143 | +49 870 | 19767 | 3 | 12 | 2.9 | 50 | 23 | 32 | 10.47 | 0.71 | F8IV-V | 8.738 | | | <10 | P, I |
| HE 151 | +47 780 | 19805 | 3 | 12 | 42.5 | 47 | 50 | 20 | 8.97 | 0.32 | F0Vn | 8.156 | | | 140 | I |
| HE 167 | +48 862 | 19893 | 3 | 13 | 5.1 | 49 | 0 | 35 | 7.94 | 0.12 | B9.5V | 7.679 | | | <20 | P, I |
| HE 212 | +49 876 | 19954 | 3 | 13 | 50.2 | 49 | 34 | 9 | 7.15 | 0.04 | B9V | 7.070 | | | 280 | P, I |
| HE 220 | +48 865 | | 3 | 14 | 16.5 | 48 | 34 | 41 | 9.14 | 0.33 | A3 | 8.328 | | | 85 | I |
| HE 270 | +48 871 | | 3 | 15 | 23.4 | 49 | 26 | 26 | 10.11 | 0.51 | F7V | 8.861 | | | 33 | I |
| HE 299 | | | 3 | 15 | 58.6 | 50 | 24 | 19 | 11.19 | 0.64 | F7 | 9.582 | | | 17 | I |
| HE 285 | +47 792 | 20135 | 3 | 16 | 1.8 | 48 | 1 | 41 | 8.09 | 0.21 | A0p | 7.400 | | | 35 | P, I |
| HE 314 | +50 728 | 20122 | 3 | 16 | 13.4 | 51 | 25 | 45 | 9.25 | 0.43 | F2V | 8.164 | | | 110 | I |
| HE 309 | | | 3 | 16 | 23.0 | 49 | 37 | 34 | 9.96 | 0.49 | F5V | 8.805 | | | 65 | K |
| HE 334 | +49 892 | | 3 | 16 | 48.9 | 51 | 13 | 7 | 7.19 | 0.03 | B9 | 7.028 | | | 19 | P, I |
| HE 333 | +50 731 | 20191 | 3 | 16 | 59.2 | 49 | 55 | 37 | 10.37 | 0.55 | F7V | 9.038 | | | 230 | P, I |
| HE 338 | +48 876 | | 3 | 17 | 20.1 | 49 | 30 | 9 | 9.93 | 0.56 | F7V | 8.621 | | | 56 | P, I |
| HE 350 | | | 3 | 17 | 36.7 | 48 | 50 | 9 | 11.13 | 0.71 | G3 | 9.268 | | 29.81 | 47 | P, I |
| AP 121 | | | 3 | 17 | 42.1 | 49 | 1 | 48 | 11.89 | 0.79 | G5 | 9.865 | <29.69 | | <10 | I |
| HE 361 | +49 896 | | 3 | 18 | 1.6 | 49 | 38 | 40 | 9.68 | 0.44 | F4V | 8.614 | | | 30 | I |
| HE 365 | +49 897 | | 3 | 18 | 5.0 | 49 | 54 | 22 | 9.90 | 0.5 | F6V | 8.716 | | | 108 | I |
| HE 373 | | | 3 | 18 | 27.3 | 47 | 21 | 17 | 11.50 | 0.77 | G3 | 9.421 | | | 140 | I |
| HE 383 | +49 899 | 20365 | 3 | 18 | 37.5 | 50 | 13 | 21 | 5.15 | -0.06 | B3V | 5.312 | | | 145 | P, I |
| HE 386 | +49 900 | 20391 | 3 | 18 | 44.6 | 49 | 46 | 13 | 7.93 | 0.12 | A2V | 7.663 | | | 260 | I |
| AP 90 | | | 3 | 18 | 49.9 | 49 | 43 | 53 | 11.17 | 0.67 | F9 | 9.523 | | | <10 | I |
| AP 92 | | | 3 | 19 | 2.0 | 49 | 33 | 38 | 15.68 | 1.5 | | 11.234 | <29.82 | | 90 | H |
| AP 93 | | | 3 | 19 | 2.3 | 48 | 10 | 57 | 11.99 | 0.93 | | 9.433 | | 30.18 | 75 | K |
| HE 401 | +49 902 | 20418 | 3 | 19 | 7.5 | 50 | 5 | 43 | 5.04 | -0.08 | B5V | 5.203 | <29.63 | | 320 | I |
| HE 407 | | | 3 | 19 | 19.4 | 50 | 44 | 49 | 11.18 | 0.64 | A2 | 9.499 | | | 28 | P, I |
| HE 421 | +48 885 | 20475 | 3 | 19 | 41.5 | 48 | 54 | 50 | 9.23 | 0.45 | F2V | 8.010 | 30.1 | 30.02 | 90 | I |
| AP 125 | | | 3 | 19 | 45.6 | 50 | 8 | 36 | 12.16 | | | 10.100 | <29.48 | | 48 | H |
| HE 423 | +48 886 | 20487 | 3 | 19 | 47.1 | 48 | 37 | 42 | 7.64 | 0.07 | A0Vn | 7.481 | <29.36 | <29.28 | 280 | I |
| AP 95 | | | 3 | 19 | 57.3 | 49 | 52 | 7 | 12.28 | 0.88 | | 9.934 | 29.88 | <29.91 | 140 | H |
| AP 127 | | | 3 | 20 | 1.4 | 46 | 53 | 8 | 12.57 | 1.12 | | 9.894 | | | 80 | H |
| AP 97 | | | 3 | 20 | 41.4 | 48 | 24 | 36 | 12.08 | 0.87 | G6.5 | 9.909 | <29.53 | 29.52 | <10 | I |
| AP 98 | | | 3 | 21 | 6.1 | 48 | 26 | 11 | 12.80 | 1 | G9 | 10.325 | <29.39 | 29.11 | <10 | H |
| AP 100 | | | 3 | 21 | 15.6 | 48 | 35 | 6 | 12.80 | 1.13 | | 9.599 | 29.87 | 29.72 | 205 | H |
| AP 102 | | | 3 | 21 | 19.9 | 48 | 45 | 27 | 11.96 | 0.8 | | 9.849 | 29.68 | 29.7 | 11 | I |
| AP 101 | | | 3 | 21 | 22.1 | 49 | 57 | 3 | 13.89 | 1.25 | K6 | 10.436 | 29.37 | <29.99 | <10 | H |
| HE 481 | +47 808 | | 3 | 21 | 30.1 | 48 | 29 | 39 | 9.18 | 0.36 | F1IVn | 8.219 | <29.27 | <28.85 | 180 | I |
| HE 490 | +48 892 | | 3 | 21 | 40.1 | 49 | 7 | 13 | 9.59 | 0.43 | F3IV-V | 8.489 | 28.85 | <29.11 | 15 | I |
| AP 104 | | | 3 | 22 | 4.8 | 48 | 49 | 36 | 12.06 | 0.78 | G3.5 | 10.193 | <28.95 | <28.90 | <10 | I |
| AP 139 | | | 3 | 22 | 6.8 | 47 | 34 | 8 | 12.00 | | | 9.782 [b] | | 29.86 | >200 | K |
| HE 520 | | | 3 | 22 | 21.8 | 49 | 8 | 29 | 11.69 | 0.79 | G3 | 9.651 | 30.38 | 30.29 | 87 | I |
| AP 106 | | | 3 | 22 | 40.6 | 49 | 40 | 42 | 12.94 | 1.01 | | 10.460 | 29.61 | 29.52 | <10 | H |
| HE 557 | +48 899 | 20809 | 3 | 23 | 13.0 | 49 | 12 | 49 | 5.26 | -0.08 | B5V | 5.558 | <28.96 | <28.90 | 250 | P |
| AP 108 | | | 3 | 23 | 36.3 | 48 | 58 | 53 | 12.92 | 1.03 | | 10.340 | 29.8 | 29.85 | 14 | H |
| AP 6 | | | 3 | 23 | 42.4 | 49 | 10 | 32 | 15.53 | 1.56 | | 11.327 | <28.95 | | <15 | H |
| HE 575 | +51 728 | 20842 | 3 | 23 | 43.0 | 51 | 46 | 15 | 7.85 | 0.1 | A0V | 7.554 | | | 85 | P, I |
| HE 581 | +48 903 | 20863 | 3 | 23 | 47.2 | 48 | 36 | 17 | 6.99 | 0.01 | B9V | 6.902 | 30.49 | 30.35 | 200 | P |
| HE 588 | +49 914 | | 3 | 23 | 54.9 | 50 | 18 | 25 | 10.01 | 0.54 | F5V | 8.485 | 29.8 | | 120 | P, I |
| AP 14 | | | 3 | 24 | 19.8 | 48 | 47 | 20 | 11.94 | 0.83 | G4 | 9.971 | 29.73 | 29.62 | <10 | I |
| HE 609 | +49 918 | | 3 | 24 | 24.9 | 50 | 19 | 35 | 9.23 | 0.42 | F0V | 7.996 | 30.07 | | 175 | I |
| AP 15 | | | 3 | 24 | 25.0 | 48 | 48 | 22 | 14.12 | 1.29 | | 10.744 | 29.86 | 29.58 | 52 | H |
| HE 612 | +48 906 | 20931 | 3 | 24 | 29.8 | 49 | 8 | 25 | 7.87 | 0.09 | A1V | 7.712 | <28.96 | <29.04 | 85 | P, I |
| AP 17 | | | 3 | 24 | 32.1 | 49 | 18 | 29 | 15.27 | 1.55 | | 10.947 | 29.76 | 29.3 | >60 | H |
| HE 621 | +47 816 | | 3 | 24 | 46.9 | 48 | 24 | 43 | 9.86 | 0.49 | F4V | 8.724 | <29.30 | <28.95 | 28 | I |
| AP 149 | | | 3 | 24 | 48.3 | 48 | 53 | 21 | 11.71 | 0.82 | | 9.502 | 30.33 | 30.23 | 117 | H |
| AP 19 | | | 3 | 24 | 49.6 | 48 | 52 | 20 | 11.62 | 0.79 | | 9.624 | 30.39 | 30.44 | 61 | H |
| HE 625 | +47 817 | 20961 | 3 | 24 | 51.9 | 47 | 54 | 56 | 7.63 | 0.11 | B9.5V | 7.338 | | <29.04 | 25 | P, I |
| AP 20 | | | 3 | 24 | 52.4 | 49 | 4 | 16 | 15.66 | 1.55 | | 10.906 | 29.7 | 29.11 | 70 | H |
| HE 632 | +46 745 | | 3 | 24 | 54.9 | 47 | 24 | 55 | 9.73 | 0.46 | F4V | 8.676[b] | | <29.63 | 160 | I |
| AP 110 | | | 3 | 24 | 55.7 | 50 | 1 | 52 | 12.27 | 0.92 | G8 | 10.058 | 29.48 | <30.00 | <10 | K |
| AP 21 | | | 3 | 25 | 1.2 | 49 | 2 | 6 | 15.56 | 1.6 | | 11.103 | 29.58 | 29.2 | 25 | H |
| HE 639 | +48 907 | 20986 | 3 | 25 | 9.8 | 49 | 15 | 7 | 8.15 | 0.12 | A3Vn | 7.882 | <29.13 | <28.9 | 210 | P, I |
| AP 25 | | | 3 | 25 | 16.1 | 48 | 22 | 24 | 12.25 | 0.88 | | 10.008 | 29.73 | 30.12 | 12 | H |
| HE 651 | +48 909 | 21005 | 3 | 25 | 20.6 | 49 | 18 | 59 | 8.42 | 0.19 | A5Vn | 8.029 | <29.16 | <28.95 | 250 | I |
| AP 112 | | | 3 | 25 | 31.8 | 48 | 30 | 9 | 13.72 | 1.13 | | 10.610 | 29.57 | 29.9 | 13 | H |
| HE 665 | +46 748 | 21046 | 3 | 25 | 37.5 | 47 | 1 | 16 | 8.64 | 0.29 | A7V | 8.050[b] | | | 70 | P |
| AP 28 | | | 3 | 25 | 53.8 | 48 | 31 | 9 | 13.09 | 1.05 | | 10.432 | 29.59 | 29.8 | 12 | H |
| HE 675 | +48 913 | 21071 | 3 | 25 | 57.2 | 49 | 7 | 16 | 6.06 | -0.08 | B7V | 6.315 | <29.20 | <28.90 | 70 | P, I |
| HE 684 | | | 3 | 26 | 4.1 | 48 | 48 | 9 | 10.59 | 0.57 | F9V | 9.201 | 30.2 | 30.12 | 71 | P, I |
| AP 37 | | | 3 | 26 | 16.3 | 48 | 50 | 29 | 12.61 | 0.96 | | 10.295 | 29.91 | 29.96 | 29 | H |
| HE 696 | | | 3 | 26 | 19.2 | 49 | 13 | 34 | 11.61 | 0.76 | G3 | 9.727 | 29.81 | 29.81 | 10 | I |
| HE 699 | | | 3 | 26 | 22.1 | 49 | 25 | 39 | 11.27 | 0.71 | G1V | 9.416 | 30.18 | 30.36 | 90 | I |
| AP 156* | | | 3 | 26 | 22.6 | 47 | 16 | 10 | 11.89 | | G6 | 9.733[b] | | | <10 | I |
| AP 41 | | | 3 | 26 | 25.3 | 48 | 20 | 7 | 12.03 | 0.85 | G5 | 9.894 | <29.56 | 29.45 | 10 | H |
| AP 43 | | | 3 | 26 | 27.5 | 49 | 2 | 13 | 12.84 | 0.97 | | 10.129 | 30.16 | 29.99 | 72 | H |
| AP 158 | | | 3 | 26 | 33.7 | 50 | 13 | 55 | 11.93 | 0.85 | | 9.736 | 29.4 | | 13 | I |
| HE 715 | | | 3 | 26 | 40.5 | 48 | 46 | 38 | 9.75 | 0.48 | F4V | 8.687[b] | 30.03 | 29.15 | 110 | I |
| HE 709 | | | 3 | 26 | 43.8 | 49 | 54 | 35 | 10.95 | 0.68 | G0V | 9.299[b] | 30.46 | 30.26 | 59 | P, I |
| HE 727 | | | 3 | 26 | 49.9 | 48 | 47 | 33 | 10.32 | 0.56 | F7V | 8.987[b] | 30.18 | 30.09 | 70 | P, I |
| HE 735 | +47 828 | 21185 | 3 | 27 | 5.1 | 48 | 12 | 21 | 6.83 | -0.02 | BVn.. | 6.811[b] | | <28.85 | 345 | P, I |
| AP 56 | | | 3 | 27 | 23.3 | 48 | 22 | 25 | 13.00 | 1 | | 10.305 | <29.64 | 29.72 | 110 | H |
| HE 750 | | | 3 | 27 | 37.6 | 48 | 59 | 30 | 10.59 | 0.59 | F9V | 9.206 | <29.55 | 29.18 | 26 | P, I |
| AP 60 | | | 3 | 27 | 38.5 | 48 | 25 | 0 | 15.74 | 1.7 | | 10.865 | <29.77 | 29.4 | 105 | H |
| AP 63 | | | 3 | 27 | 50.9 | 49 | 12 | 11 | 12.29 | 0.92 | | 9.919 | <29.56 | 29.81 | | K |
| HE 767 | | | 3 | 27 | 54.9 | 49 | 45 | 38 | 10.69 | 0.62 | F9V | 9.204 | <29.36 | <29.38 | 10 | P, I |
| HE 775 | +47 831 | 21279 | 3 | 27 | 55.6 | 47 | 44 | 10 | 7.26 | 0.05 | B8.5V | 7.133[b] | | <29.26 | 200 | I |
| HE 774 | +48 920 | 21278 | 3 | 28 | 2.9 | 49 | 3 | 48 | 4.97 | -0.09 | B5V | 5.243 | <29.57 | <28.90 | 65 | I |
| HE 780 | | | 3 | 28 | 18.4 | 49 | 57 | 11 | 8.09 | 0.17 | A1Vn | 7.547 | 30.12 | 29.74 | 230 | K |
| AP 70 | | | 3 | 28 | 18.6 | 48 | 39 | 49 | 12.83 | 1 | K0 | 10.274 | <29.66 | 29.68 | <10 | H |
| AP 72 | | | 3 | 28 | 22.5 | 49 | 14 | 30 | 12.78 | 0.99 | K0 | 10.443 | <29.63 | 28.7 | <10 | H |
| HE 799 | +48 923 | | 3 | 28 | 31.4 | 48 | 56 | 28 | 9.66 | 0.45 | F4V | 8.636 | <29.67 | 29 | 20 | I |
| AP 75 | | | 3 | 28 | 47.3 | 49 | 16 | 28 | 13.82 | 1.27 | | 10.296 | <29.56 | 29.43 | 11 | H |
| AP 117 | | | 3 | 28 | 48.3 | 49 | 11 | 54 | 13.05 | 0.95 | | 10.367 | <29.83 | 29.89 | 83 | H |
| HE 810 | +49 944 | 21362 | 3 | 28 | 52.2 | 49 | 50 | 55 | 5.58 | -0.04 | B6Vn | 5.687 | <29.36 | <29.28 | 385 | P, I |
| HE 817 | +48 927 | 21375 | 3 | 28 | 53.5 | 49 | 4 | 14 | 7.46 | 0.11 | A1V | 7.153 | <29.78 | 29.45 | 270 | P, I |
| HE 828 | | | 3 | 28 | 59.5 | 48 | 14 | 10 | 11.62 | 0.71 | F8 | 9.748 | | 29.2 | 10 | I |
| HE 831 | +47 835 | 21398 | 3 | 29 | 7.5 | 48 | 18 | 12 | 7.36 | 0.01 | B9V | 7.354 | | <28.85 | 135 | P, I |
| HE 833 | | | 3 | 29 | 8.1 | 48 | 10 | 52 | 10.05 | 0.42 | F6V | 8.806 | | <29.18 | 30 | I |
| AP 173 | | | 3 | 29 | 14.1 | 49 | 41 | 18 | 12.26 | | K3.2 | 9.784 | | <29.38 | <10 | K |
| HE 835 | +49 945 | 21428 | 3 | 29 | 21.9 | 49 | 30 | 33 | 4.66 | -0.1 | B3V | 4.881 | 30.15 | 30.05 | 190 | I |
| HE 841 | | | 3 | 29 | 24.8 | 48 | 57 | 46 | 10.29 | 0.54 | F7V | 8.944 | | 30.02 | 65 | P, I |
| HE 848 | | | 3 | 29 | 26.1 | 48 | 12 | 13 | 10.00 | 0.6 | | 8.537 | | 29.53 | <20 | P, I |
| AP 78 | | | 3 | 29 | 26.5 | 49 | 20 | 36 | 13.06 | 1.02 | | 10.475 | <29.56 | 29.72 | 13 | H |
| HE 862 | +48 930 | 21480 | 3 | 29 | 46.9 | 49 | 9 | 15 | 8.52 | 0.31 | A7V | 7.809 | <29.56 | <28.85 | 50 | P, I |
| HE 868 | +48 933 | 21479 | 3 | 29 | 51.7 | 49 | 12 | 50 | 7.28 | 0.09 | A1IVn | 7.034 | <29.56 | <28.78 | 180 | P, I |
| HE 885 | +48 934 | 21527 | 3 | 30 | 19.1 | 48 | 29 | 59 | 8.79 | 0.28 | A7IV | 8.192 | | 29.2 | 80 | I |
| AP 86 | | | 3 | 30 | 22.4 | 48 | 24 | 42 | 14.31 | 1.32 | | 10.780 | | 29.3 | 140 | H |
| HE 906 | +47 842 | 21553 | 3 | 30 | 33.8 | 47 | 37 | 43 | 8.78 | 0.28 | A6Vn | 8.135 [b] | | <29.15 | 150 | I |
| HE 904 | +47 844 | 21551 | 3 | 30 | 36.8 | 48 | 6 | 14 | 5.82 | -0.04 | B8V | 5.936 | | <28.90 | 380 | I |
| HE 917 | | | 3 | 30 | 47.5 | 47 | 53 | 23 | 10.93 | 0.66 | F4 | 9.256 | | 30.21 | 40 | P, I |
| HE 921 | +49 953 | 21600 | 3 | 31 | 14.5 | 49 | 42 | 24 | 8.59 | 0.2 | A6Vn | 8.158 | <28.99 | <29.30 | 200 | I |
| HE 935 | | | 3 | 31 | 28.8 | 48 | 59 | 30 | 10.05 | 0.62 | | 8.847 [b] | | 30.62 | 78 | P, I |
| HE 931 | +49 954 | 21619 | 3 | 31 | 30.1 | 49 | 54 | 9 | 8.75 | 0.26 | A6V | 8.180 | <28.90 | <29.76 | 90 | I |
| HE 955 | +47 380 | 21641 | 3 | 31 | 32.9 | 47 | 51 | 46 | 6.75 | -0.02 | B8.5V | 6.796 | | <29.08 | 215 | P, I |
| HE 944 | +49 957 | | 3 | 31 | 44.4 | 49 | 32 | 13 | 9.62 | 0.43 | F3V | 8.616 [b] | | 29.59 | 56 | I |
| HE 965 | +48 943 | 21672 | 3 | 31 | 53.8 | 48 | 44 | 8 | 6.62 | -0.03 | B8V | 6.646 [b] | | 29.48 | 225 | P, I |
| HE 968 | | | 3 | 31 | 54.1 | 48 | 31 | 40 | 10.41 | 0.57 | F8V | 9.032 [b] | | 30.21 | 38 | P, I |
| HE 970 | +48 944 | | 3 | 31 | 55.6 | 48 | 35 | 3 | 8.19 | 0.19 | A4V | 7.768 [b] | | <28.90 | 120 | P, I |
| HE 985 | +47 847 | 21699 | 3 | 32 | 8.5 | 48 | 1 | 26 | 5.46 | -0.1 | B8IIImnp | 5.631 [b] | | <28.95 | 50 | I |
| AP 193 | | | 3 | 32 | 10.2 | 49 | 8 | 29 | 12.28 | 0.85 | | 9.903 [b] | 29.75 | 30.3 | 64 | H |
| AP 194 | | | 3 | 32 | 14.9 | 46 | 39 | 23 | 12.02 | 0.74 | | 10.129 | | | <10 | I |
| AP 118 | | | 3 | 32 | 30.6 | 49 | 10 | 35 | 12.06 | 0.81 | | 9.808 [b] | 30.01 | 30.14 | 160 | H |
| AP 199 | | | 3 | 32 | 44.4 | 47 | 41 | 36 | 12.10 | 0.98 | | 9.602 | | 29.99 | 23 | I |
| AP 201 | | | 3 | 32 | 51.1 | 49 | 50 | 44 | 13.08 | | K5 | 10.236 [b] | 29.85 | <29.84 | 12 | H |
| HE 1082 | +48 949 | 21931 | 3 | 34 | 13.0 | 48 | 37 | 3 | 7.37 | 0.02 | B9V | 7.212 [b] | | <29.43 | 205 | P, I |
| HE 1100 | | | 3 | 34 | 28.6 | 47 | 4 | 25 | 11.50 | 0.4 | | 9.557 [b] | | | <10 | I |
| AP 213 | | | 3 | 34 | 39.4 | 48 | 18 | 44 | 11.55 | 0.83 | G6 | 9.580 [b] | | <29.35 | 0 | P, I |
| HE 1101 | | | 3 | 35 | 8.6 | 49 | 44 | 40 | 11.25 | 0.69 | G4 | 9.441 [b] | 30.56 | | 35 | P, I |
| HE 1153 | +46 773 | 22136 | 3 | 35 | 58.5 | 47 | 5 | 28 | 6.878 | -0.023 | B8V | 6.952 | | | 25 | I |
| AP 225 | | | 3 | 36 | 22.0 | 49 | 9 | 21 | 11.83 | 0.78 | | 9.706 [b] | 30.61 | | 138 | I |
| HE 1185 | | | 3 | 36 | 57.6 | 48 | 44 | 47 | 11.19 | 0.723 | F7 | 9.405 | | | <10 | P, I |
| HE 1259 | +47 865 | 22401 | 3 | 38 | 15.6 | 47 | 34 | 37 | 7.46 | 0.01 | A0V | 7.415 | | | 45 | I |
| AP 256 | | | 3 | 43 | 38.5 | 46 | 3 | 48 | 11.79 | 0.81 | | 9.689 [b] | | | 10 | I |
| AP 264 | | | 3 | 50 | 27.8 | 47 | 49 | 6 | 12.12 | 1.01 | | 9.834 [b] | | | 14 | I |

[a] Prosser & Stauffer (ftp://cfa-ftp.harvard.edu/pub/stauffer/clusters/) or SIMBAD, [b] not in 2MASS database, [c] Prosser et al. 1996, [d] Randich et al. 1996

## Table 2: Praesepe Sample

| name | BD | HD | RA [a] 2000 | | | DEC [a] 2000 | | | V[a] | B-V[a] | SpTy[a] | K | LogLx[c] | Tel |
|---|---|---|---|---|---|---|---|---|---|---|---|---|---|---|
| II490 | | | 8 | 27 | 59 | 22 | 6 | 8 | 9.67 | 0.45 | | 8.693 | | I |
| II582 | | | 8 | 29 | 54 | 22 | 26 | 31 | 9.88 | 0.49 | | 8.823 | | I |
| A70 | | | 8 | 31 | 13 | 18 | 9 | 21 | 9.86 | 0.46 | | 8.767 | | I |
| I563 | | | 8 | 33 | 27 | 16 | 35 | 19 | 10.11 | 0.5 | | 8.849 | | I |
| A365 | | | 8 | 35 | 17 | 20 | 33 | 49 | 8.42 | 0.26 | | 7.779 | | I |
| JS88 | +20 2119 | | 8 | 35 | 28.1 | 20 | 11 | 47 | 10.34 | 0.59 | | 8.924 | | I |
| KW534 | +20 2125 | 72942 | 8 | 36 | 17.47 | 20 | 20 | 29.43 | 7.53 | 0.22 | Am | 7.185 | | I |
| KW536 | +19 2045 | | 8 | 36 | 29.898 | 18 | 57 | 56.77 | 9.42 | 0.51 | F6V | 8.329 | | I |
| KW538 | +19 2047 | 73045 | 8 | 36 | 48.049 | 18 | 52 | 57.88 | 8.62 | 0.38 | Am | 7.991 | | I |
| KW16 | +20 2128 | 73081 | 8 | 37 | 2.074 | 19 | 36 | 16.98 | 9.16 | 0.51 | F6V | 8.059 | 28.90 | I |
| KW38 | +20 2131 | 73161 | 8 | 37 | 33.866 | 20 | 0 | 48.99 | 8.69 | 0.33 | F0Vn | 7.944[b] | | I |
| KW40 | +20 2132 | 73174 | 8 | 37 | 37.031 | 19 | 43 | 58.19 | 7.79 | 0.22 | Am | 7.255[b] | 29.15 | I |
| KW45 | +20 2133 | 73175 | 8 | 37 | 40.743 | 19 | 31 | 5.97 | 8.25 | 0.23 | F0Vn | 7.617[b] | | P, I |
| KW47 | +19 2052 | | 8 | 37 | 42.408 | 19 | 8 | 1.34 | 9.82 | 0.5 | F4V | 8.713[b] | | I |
| KW50 | +19 2053 | 73210 | 8 | 37 | 46.799 | 19 | 16 | 1.73 | 6.75 | 0.19 | A5V | 6.354[b] | | P, I |
| A609 | | | 8 | 38 | 8 | 17 | 3 | 3 | 9.77 | 0.44 | | 8.720 | | I |
| KW100 | | | 8 | 38 | 24.351 | 20 | 6 | 21.47 | 10.55 | 0.58 | G0V | 9.164[b] | | K |
| KW114 | +20 2138 | 73345 | 8 | 38 | 37.909 | 19 | 59 | 22.97 | 8.14 | 0.21 | F0V | 7.653 | | P, I |
| KW124 | +20 2139 | 73397 | 8 | 38 | 46.976 | 19 | 30 | 2.93 | 9 | 0.32 | F4V | 8.273 | | P, I |
| KW142 | +20 2140 | | 8 | 39 | 2.876 | 19 | 43 | 28.63 | 9.31 | 0.49 | F7V | 8.135 | 29.30 | P, I |
| KW143 | +20 2141 | 73430 | 8 | 39 | 3.628 | 19 | 59 | 58.92 | 8.31 | 0.23 | A9V | 7.780 | | P, I |
| KW146 | +20 2142 | 73429 | 8 | 39 | 5.279 | 20 | 7 | 1.5 | 9.39 | 0.4 | F5V | 8.432 | | I |
| KW150 | +20 2143 | 73449 | 8 | 39 | 6.148 | 19 | 40 | 36.18 | 7.45 | 0.25 | A9Vn | 6.728 | | P, I |
| KW154 | +20 2144 | 73450 | 8 | 39 | 9.131 | 19 | 35 | 32.25 | 8.5 | 0.25 | A9V | 7.889 | | P, I |
| KW155 | +20 22143 | | 8 | 39 | 10.182 | 19 | 40 | 42.01 | 9.42 | 0.41 | F6 | 8.449 | | I |
| KW181 | +19 2061 | | 8 | 39 | 25.017 | 19 | 27 | 33.34 | 10.47 | 0.59 | F7V | 9.039 | 28.90 | K |
| KW182 | +20 2146 | | 8 | 39 | 30.455 | 20 | 4 | 8.3 | 10.31 | 0.68 | F8V | 8.827 | 28.78 | I |
| KW203 | +20 2148 | 73574 | 8 | 39 | 42.835 | 20 | 5 | 10.28 | 7.73 | 0.22 | A5V | 7.190 | | P, I |
| KW253 | +20 2158 | 73665 | 8 | 39 | 36.574 | 20 | 54 | 36.07 | 6.39 | 0.98 | G8III | | | P |
| KW204 | +20 2149 | 73575 | 8 | 39 | 42.715 | 19 | 46 | 42.3 | 6.67 | 0.25 | F0III | 6.027 | | P, I |
| KW207 | | | | 8 | 39 | 44.696 | 19 | 16 | 30.38 | 7.67 | 0.2 | A7Vn | 7.127 | | P, I |
| KW212 | +20 | 2150 | 73598 | 8 | 39 | 50.762 | 19 | 32 | 26.76 | 6.59 | 0.96 | K0III | | 29.90 | P |
| KW217 | | | | 8 | 39 | 52.369 | 19 | 18 | 45.15 | 10.23 | 0.51 | F5V | 9.086 | 29.08 | K |
| KW218 | +21 | 1882 | 73597 | 8 | 39 | 54.374 | 20 | 33 | 36.9 | 9.36 | 0.44 | F6V | 8.436 | | I |
| KW222 | +20 | 2151 | | 8 | 39 | 55.109 | 20 | 3 | 53.73 | 10.11 | 0.49 | F4V | 8.985 | 29.00 | K |
| KW224 | +20 | 2152 | 73618 | 8 | 39 | 56.527 | 19 | 33 | 10.5 | 7.32 | 0.19 | Am | 6.809 | 29.94 | P, I |
| KW229 | +20 | 2153 | 73619 | 8 | 39 | 57.811 | 19 | 32 | 29.1 | 7.54 | 0.25 | Am | 7.030 | 29.94 | P, I |
| KW227 | +19 | 2066 | 73641 | 8 | 39 | 58.103 | 19 | 12 | 5.48 | 9.49 | 0.41 | F6V | 8.533 | | I |
| KW226 | +20 | 2154 | 73616 | 8 | 39 | 58.422 | 20 | 9 | 29.36 | 8.89 | 0.32 | F2V | 8.113 | | P, I |
| KW232 | +20 | 2155 | 73617 | 8 | 39 | 59.119 | 20 | 1 | 52.87 | 9.23 | 0.39 | F5V | 8.248 | | I |
| KW238 | | | | 8 | 40 | 0.66 | 19 | 48 | 23.17 | 10.3 | 0.51 | F6V | 9.091 | 29.00 | K |
| KW239 | +20 | 2156 | 73640 | 8 | 40 | 1.333 | 20 | 8 | 7.91 | 9.67 | 0.44 | F6V | 8.599 | 29.00 | I |
| KW244 | +19 | 2068 | | 8 | 40 | 1.742 | 18 | 59 | 59.05 | 9.98 | 0.62 | F8V | 8.704 | 30.05 | I |
| KW250 | +20 | 2157 | | 8 | 40 | 4.942 | 19 | 43 | 44.94 | 9.79 | 0.47 | F6V | 8.693 | | I |
| KW258 | | | | 8 | 40 | 6.308 | 19 | 27 | 14.64 | 10.24 | 0.57 | F6V | 8.893 | | I |
| KW265 | +20 | 2159 | 73666 | 8 | 40 | 11.506 | 19 | 58 | 16.1 | 6.61 | 0.01 | A1V | 6.543 | | P, I |
| KW268 | +20 | 2160 | | 8 | 40 | 12.354 | 19 | 38 | 21.92 | 9.89 | 0.48 | F5V | 8.698 | | I |
| KW271 | +20 | 2161 | | 8 | 40 | 15.396 | 19 | 59 | 39.13 | 8.81 | 0.32 | F2V | 8.064 | | I |
| KW275 | +20 | 2162 | | 8 | 40 | 17.658 | 19 | 47 | 14.86 | 9.96 | 0.58 | F8V | 8.628 | | I |
| KW276 | +20 | 2163 | 73711 | 8 | 40 | 18.135 | 19 | 31 | 54.93 | 7.54 | 0.16 | F0III | 7.187 | 29.04 | P, I |
| KW284 | +19 | 2069 | 73712 | 8 | 40 | 20.178 | 19 | 20 | 56.07 | 6.78 | 0.26 | A9V | 6.063 | | P, I |
| KW279 | +20 | 2165 | 73709 | 8 | 40 | 20.796 | 19 | 41 | 11.97 | 7.7 | 0.2 | F2III | 7.290 | | P, I |
| KW283 | +20 | 2166 | 73710 | 8 | 40 | 21 | 19 | 40 | 25 | 6.44 | 1.02 | G9III | | 29.04 | P |
| KW282 | +20 | 2164 | | 8 | 40 | 22.355 | 20 | 6 | 23.97 | 10.08 | 0.51 | F2III | 8.837 | | K |
| KW286 | +20 | 2168 | 73730 | 8 | 40 | 23.508 | 19 | 50 | 5.67 | 8.02 | 0.19 | F2III | 7.599 | | P, I |
| KW293 | | | | 8 | 40 | 25.575 | 19 | 28 | 32.44 | 9.85 | 0.47 | | 8.796 | | I |
| KW295 | +20 | 2170 | | 8 | 40 | 26.18 | 19 | 41 | 10.96 | 9.37 | 0.42 | F6V | 8.381 | 28.90 | I |
| KW292 | +20 | 2169 | 73729 | 8 | 40 | 26.791 | 20 | 10 | 55 | 8.18 | 0.3 | F2Vn | 7.426 | | P, I |
| KW300 | +20 | 2171 | 73731 | 8 | 40 | 27.058 | 19 | 32 | 41.31 | 6.3 | 0.17 | A5m | 5.880 | 28.70 | P, I |
| KW318 | +19 | 2072 | 73746 | 8 | 40 | 32.994 | 19 | 11 | 39.29 | 8.65 | 0.29 | F0V | 7.965 | | P |
| KW323 | +19 | 2073 | 73763 | 8 | 40 | 39.277 | 19 | 13 | 41.48 | 7.8 | 0.22 | A9V | 7.249 | | P |
| KW328 | +20 | 2172 | 73785 | 8 | 40 | 43.26 | 19 | 43 | 9.45 | 6.85 | 0.2 | A9III | 6.348 | 28.90 | P, I |
| KW332 | +19 | 2074 | | 8 | 40 | 46.116 | 19 | 18 | 34.26 | 9.55 | 0.43 | F6V | 8.566 | | I |

| | | | | | | | | | | | | | | |
|---|---|---|---|---|---|---|---|---|---|---|---|---|---|---|
| KW340 | +20 2173 | 73798 | 8 | 40 | 52.523 | 20 | 15 | 59.19 | 8.48 | 0.26 | F0Vn | 7.822 | | P |
| KW341 | | | 8 | 40 | 52.564 | 19 | 28 | 59.18 | 10.3 | 0.52 | F7V | 9.067 | | K |
| KW348 | +20 2175 | 73819 | 8 | 40 | 56.337 | 19 | 34 | 48.96 | 6.78 | 0.17 | A6Vn | 6.293 | | P, I |
| KW350 | +20 2174 | 73818 | 8 | 40 | 56.962 | 19 | 56 | 5.51 | 8.71 | 0.32 | Am | 8.062 | | I |
| KW365 | +19 2076 | | 8 | 41 | 7.399 | 19 | 4 | 16.06 | 10.18 | 0.64 | G0 | 8.689 | | I |
| KW371 | +20 2176 | | 8 | 41 | 10.053 | 19 | 30 | 31.72 | 10.11 | 0.5 | F6V | 8.952 | | P |
| KW370 | +20 2177 | 73854 | 8 | 41 | 10.708 | 19 | 49 | 46.04 | 9.04 | 0.36 | F5V | 8.212 | 29.04 | P |
| KW375 | +20 2179 | 73872 | 8 | 41 | 13.816 | 19 | 55 | 18.9 | 8.33 | 0.2 | A5V | 7.796 | 29.66 | P, I |
| KW385 | +19 2078 | 73890 | 8 | 41 | 18.437 | 19 | 15 | 38.97 | 7.92 | 0.24 | A7Vn | 7.328 | | P, I |
| KW396 | +20 2180 | | 8 | 41 | 26.996 | 19 | 32 | 32.46 | 9.83 | 0.46 | F4V | 8.751 | | I |
| KW411 | +19 2080 | 73937 | 8 | 41 | 36.241 | 19 | 8 | 33.28 | 9.32 | 0.39 | F4V | 8.373 | | I |
| KW416 | +20 2183 | | 8 | 41 | 42.335 | 19 | 39 | 37.65 | 9.59 | 0.41 | F6V | 8.512 | | I |
| KW418 | +20 2184 | | 8 | 41 | 43.853 | 20 | 13 | 36.26 | 10.47 | 0.56 | F7 | 9.194 | | K |
| KW421 | +19 2081 | | 8 | 41 | 45.535 | 19 | 16 | 1.91 | 10.17 | 0.52 | | 8.958 | | K |
| KW429 | +20 2186 | 73993 | 8 | 41 | 53.177 | 20 | 9 | 33.59 | 8.53 | 0.3 | F2Vn | 7.807 | | P, I |
| KW439 | +19 2082 | 73994 | 8 | 41 | 57.857 | 18 | 54 | 41.84 | 9.45 | 0.39 | F5V | 8.484 | | I |
| KW445 | +19 2083 | 74028 | 8 | 42 | 6.54 | 19 | 24 | 40.34 | 7.96 | 0.21 | A7V | 7.428 | | P, I |
| KW449 | +19 2084 | 74050 | 8 | 42 | 10.848 | 18 | 56 | 3.42 | 7.91 | 0.21 | A7Vn | 7.359 | | I |
| KW454 | | | 8 | 42 | 15.528 | 19 | 41 | 15.22 | 9.88 | 0.46 | | 8.782 | | K |
| KW458 | +20 2189 | | 8 | 42 | 20.166 | 20 | 2 | 11.31 | 9.71 | 0.55 | F6V | 8.497 | | I |
| KW459 | +20 2191 | 74058 | 8 | 42 | 21.646 | 20 | 10 | 53.42 | 9.23 | 0.39 | F3Vn | 8.284 | 29.51 | P |
| JS495 | +18 2020 | | 8 | 42 | 36.8 | 18 | 23 | 20 | 10.13 | 0.52 | | 8.994 | | I |
| KW472 | +20 2192 | | 8 | 42 | 40.762 | 19 | 32 | 35.13 | 9.77 | 0.45 | F2III | 8.758 | | I |
| KW478 | +20 2193 | | 8 | 42 | 44.442 | 19 | 34 | 47.5 | 9.68 | 0.43 | F6V | 8.664 | | I |
| A1196 | | | 8 | 42 | 53 | 20 | 49 | 11 | 8.84 | 0.33 | | 8.106 | | I |
| KW496 | +19 2088 | 74186 | 8 | 43 | 7.091 | 19 | 4 | 6.11 | 9.56 | 0.52 | F8V | 8.376 | | I |
| JS532 | | | 8 | 43 | 22.7 | 21 | 40 | 18 | 10.59 | 0.71 | | 9.027 | 29.04 | I |
| KW515 | +20 2196 | | 8 | 43 | 35.601 | 20 | 11 | 22.07 | 10.13 | 0.51 | F6V | 8.914 | | I |
| KW549 | +19 2089 | | 8 | 43 | 48.197 | 18 | 48 | 2.8 | 10.11 | 0.51 | F8 | 8.965 | | I |
| KW553 | +20 2198 | | 8 | 44 | 7.369 | 20 | 4 | 36.5 | 10.15 | 0.45 | | 9.081 | | K |
| JS589 | | | 8 | 45 | 14.7 | 20 | 59 | 51 | 9.46 | 0.42 | | 8.498 | | I |
| A1480 | | | 8 | 45 | 18 | 18 | 53 | 27 | 9.51 | 0.46 | | 8.625 | | I |
| A1501 | | | 8 | 45 | 29 | 20 | 23 | 38 | 8.51 | 0.28 | | 7.888 | | I |

| | | | | | | | | | | |
|---|---|---|---|---|---|---|---|---|---|---|
| JS600 | 8 | 45 | 30.5 | 20 | 35 | 24 | 9.82 | 0.46 | 8.797 | I |
| A1528 | 8 | 45 | 47 | 19 | 3 | 1 | 8.03 | 0.24 | 7.546 | I |
| A1565 | 8 | 46 | 11 | 18 | 10 | 42 | 8.27 | 0.23 | 7.677 | I |
| A1583 | 8 | 46 | 15 | 19 | 42 | 30 | 8.39 | 0.24 | 7.813 | I |

[a] Prosser & Stauffer (ftp://cfa-ftp.harvard.edu/pub/stauffer/clusters/) or SIMBAD, [b] not in 2MASS database, [c] Randich & Schmitt 1994, Stern et al. 1995

## Table 3a: $\alpha$ Persei Observing Summary

| Telescope | Date | # obs. | Technique |
|---|---|---|---|
| Palomar | 1997 Oct 8 | 45 | Speckle |
| IRTF | 1997 Dec 4-9 | 43/3 | Speckle + Shift&Add/ Shift&Add only |
| | 1998 Feb 18-20 | 14/33 | Speckle + Shift&Add/ Shift&Add only |
| HST | 1998 Feb 28; Jul 25; Aug 8, 23, 26-28, 31 | 33 | Imaging |
| Keck | 1999 Jan 6&7 | 7 | Speckle + Shift&Add |

## Table 3b: Praesepe Observing Summary

| Telescope | Date | # obs. | Technique |
|---|---|---|---|
| Palomar | 1995 Nov 12&13 | 8 | Speckle |
| | 1996 Jan 8&9 | 26 | Speckle |
| IRTF | 1998 Feb 18-20 | 20/23 | Speckle + Shift&Add/ Shift&Add only |
| | 1999 Feb 13-15 | 35 | Speckle + Shift&Add |
| Keck | 1999 Jan 6&7 | 11 | Speckle + Shift&Add |

## Table 4a: α Persei Stellar Binaries

| Object | Tel | Date | Separation | Position Angle | ΔK/ΔF140W | M1 | M2 | q | # comp. | Notes [a] |
|---|---|---|---|---|---|---|---|---|---|---|
| AP 60 | HST | 1998 Aug 27 | 0".053 ± 0".004 | 350 ± 3 | 0.12 ± 0.02 | 0.59 | 0.57 | 0.98 | 1 | new |
| HE 935 | Pal | 1997 Oct 08 | 0".056 ± 0".006 | 6 ± 9 | 0.3 ± 0.3 | 0.97 | 0.91 | 0.94 | 1 | new |
| HE 965 | Pal | 1997 Oct 08 | 0".117 ± 0".004 | 350 ± 6 | 2.5 ± 0.2 | 2.61 | 1.04 | 0.40 | 2 | new, ? (M&A) |
| HE 581 | Pal | 1997 Oct 08 | 0".120 ± 0".003 | 17 ± 4 | 2.8 ± 0.2 | 2.40 | 0.90 | 0.38 | 1 | new, S (M&A) |
| AP 139 | Keck | 1999 Jan 7 | 0".19 ± 0".01 | 72 ± 4 | 3.5 ± 0.1 | 0.89 | 0.26 | 0.30 | 1 | new |
| HE 285 | Pal | 1997 Oct 08 | 0".214 ± 0".005 | 320 ± 1 | 1.25 ± 0.03 | 1.84 | 1.17 | 0.63 | 1 | ?B (P) |
| AP 149AB | HST | 1998 Aug 26 | 0".327 ± 0".003 | 97.7 ± 0.4 | 0.956 ± 0.009 | 0.88 | 0.72 | 0.82 | 1 | new |
| AP 98 | HST | 1998 Feb 28 | 0".332 ± 0".008 | 298 ± 1 | 2.93 ± 0.08 | 0.79 | 0.31 | 0.39 | 1 | new |
| HE696 | IRTF | 1997 Dec 08 | 0".43 ± 0".01 | 184 ± 4 | 2.5 ± 0.1 | 0.89 | 0.47 | 0.53 | 1 | new |
| AP 41 | HST | 1998 Aug 26 | 0".532 ± 0".007 | 315.26 ± 0.09 | 2.770 ± 0.007 | 0.88 | 0.48 | 0.55 | 1 | new |
| AP 201 | HST | 1998 Aug 30 | 0".544 ± 0".002 | 337.4 ± 0.3 | 1.40 ± 0.03 | 0.77 | 0.58 | 0.75 | 1 | new |
| HE 835 | IRTF | 1997 Dec 04 | 0".67 ± 0".01 | 34 ± 1 | 1.96 ± 0.03 | 4.87 | 2.37 | 0.49 | 1 | S (M&A), B (ADS) 0".68 |
| AP 17 | HST | 1998 Aug 23 | 0".969 ± 0".001 | 8.15 ± 0.09 | 2.78 ± 0.03 | 0.66 | 0.19 | 0.29 | 1 | new |
| HE 780 | Keck | 1999 Jan 7 | 1".38 ± 0".03 | 317 ± 1 | 2.88 ± 0.01 | 1.90 | 0.76 | 0.40 | 1 | new |
| AP 121 | IRTF | 1997 Dec 06 | 1".44 ± 0".03 | 31.2 ± 2 | 3.5 ± 0.2 | 0.86 | 0.24 | 0.28 | 1 | new |
| HE 828 | IRTF | 1997 Dec 06 | 1".63 ± 0.03 | 234 ± 1 | 2.60 ± 0.06 | 0.89 | 0.44 | 0.50 | 2 | new, B (P) 10".5 |
| AP 75 | HST | 1998 Aug 27 | 1".910 ± 0".005 | 79.0 ± 0.3 | 5.107 ± 0.008 | 0.77 | 0.10 | 0.13 | 1 | new |
| AP 193 | HST | 1998 Aug 28 | 2".102 ± 0".005 | 50.9 ± 0.4 | 5.221 ± 0.008 | 0.88 | 0.12 | 0.14 | 1 | new |
| AP 6 | HST | 1998 Aug 23 | 2".876 ± 0".001 | 263.47 ± 0.08 | 1.51 ± 0.09 | 0.60 | 0.32 | 0.54 | 1 | new |
| AP 106 | HST | 1998 Aug 23 | 4".571 ± 0".002 | 17.06 ± 0.08 | 2.34 ± 0.02 | 0.76 | 0.41 | 0.54 | 1 | Prosser |
| AP 108 | HST | 1998 Aug 31 | 4".947 ± 0".004 | 50.61 ± 0.06 | 4.76 ± 0.01 | 0.80 | 0.13 | 0.16 | 1 | new |

[a] M&A - Morrell & Abt 1992, P - Prosser 1992, ADS - Aitken 1932

**Table 4b: α Persei Faint Doubles**

| Object | Tel | Date | Separation | Position Angle | ΔF140W | M1 | M2 | q | # comp. |
|---|---|---|---|---|---|---|---|---|---|
| AP 72 | HST | 1998 Jul 25 | 3".045 ± 0".003 | 35.4 ± 0.2 | 8.471 ± 0.006 | 0.78 | 0.03 | 0.037 | 1 |
| AP 15 | HST | 1998 Aug 31 | 3".512 ± 0".001 | 197.8 ± 0.3 | 7.70 ± 0.04 | 0.73 | 0.03 | 0.047 | 1 |
| AP 112 | HST | 1998 Aug 27 | 4".402 ± 0".006 | 337.42 ± 0.03 | 6.557 ± 0.007 | 0.73 | 0.05 | 0.073 | 1 |
| AP 101 | HST | 1998 Aug 23 | 4".999 ± 0".001 | 160.62 ± 0.06 | 5.68 ± 0.09 | 0.75 | 0.08 | 0.104 | 1 |
| AP 149AC | HST | 1998 Aug 26 | 5".05 ± 0".01 | 160.2 ± 0.1 | 7.46 ± 0.01 | 0.94 | 0.06 | 0.064 | 3 |
| AP 149 AD | HST | 1998 Aug 26 | 5".381 ± 0".002 | 7.842 ± 0.002 | 8.46 ± 0.06 | 0.94 | 0.04 | 0.044 | |
| AP 86 | HST | 1998 Aug 27 | 5".412 ± 0".002 | 66.3 ± 0.1 | 7.34 ± 0.08 | 0.70 | 0.04 | 0.052 | 1 |
| AP 95 | HST | 1998 Feb 28 | 6".087 ± 0".002 | 188.52 ± 0.05 | 7.186 ± 0.005 | 0.88 | 0.06 | 0.067 | 1 |
| AP 19 | HST | 1998 Aug 26 | 7".287 ± 0".003 | 49.88 ± 0.02 | 6.77 ± 0.02 | 0.94 | 0.08 | 0.083 | 1 |

**Table 5: Praesepe Binaries**

| Object | Tel | Date | Separation | Position Angle | ΔK | M1 | M2 | q | # comp. | Notes [a] |
|---|---|---|---|---|---|---|---|---|---|---|
| KW 284 | Palomar | 1995 Nov 12 | 0".104 ± 0".002 | 159 ± 3 | 1.7 ± 0.1 | 3.00 | 1.61 | 0.54 | 1 | B (M) 0".054, B (AW99) |
| KW 275 | IRTF | 1998 Feb 20 | 0".198 ± 0".004 | 91 ± 1 | 0.06 ± 0.06 | 0.98 | 0.96 | 0.99 | 1 | S (B91), B (MM99) |
| KW 232 | IRTF | 1998 Feb 18 | 0".27 ± 0".05 | 22 ± 5 | 2.5 ± 0.4 | 1.40 | 0.68 | 0.49 | 1 | new, S (M) |
| KW 458 | IRTF | 1998 Feb 19 | 0".279 ± 0".005 | 167 ± 1 | 0.66 ± 0.04 | 1.12 | 0.91 | 0.82 | 1 | S (B91), B (MM99) |
| A1565 | IRTF | 1999 Feb 13 | 0".330 ± 0".008 | 352 ± 2 | 1.16 ± 0.04 | 1.59 | 1.04 | 0.65 | 1 | new |
| KW 365 | IRTF | 1998 Feb 20 | 0".340 ± 0".006 | 72 ± 1 | 0.80 ± 0.06 | 1.06 | 0.85 | 0.81 | 2 | B (B91) 21dy, T (MM99) |
| KW 282 | Keck | 1999 Jan 6 | 0".383 ± 0".008 | 153 ± 1 | 4.0 ± 0.1 | 1.16 | 0.33 | 0.28 | 1 | new |
| KW 203 | Palomar | 1996 Jan 8 | 0".499 ± 0".009 | 40 ± 2 | 0.80 ± 0.03 | 1.83 | 1.37 | 0.75 | 1 | B (P&W) 0".326, B (M) 0".550, S (AW99) |
| KW 212 | Palomar | 1995 Nov 13 | 0".80 ± 0".05 | 233 ± 1 | 4.5 ± 0.3 | giant | | | 1 | B (P&W) 0".346, B (M) 0".492, S (AW99) |
| KW 250 | IRTF | 1998 Feb 20 | 0".90 ± 0".06 | 208 ± 1 | 2.90 ± 0.04 | 1.20 | 0.56 | 0.46 | 1 | new, S (MM99) |
| KW 224 | IRTF | 1998 Feb 19 | 0".96 ± 0.02 | 250 ± 1 | 2.30 ± 0.01 | 2.35 | 1.01 | 0.43 | 2 | B (IDS) 1".4, S (P&W), S (M), B (B&C), B (AW99) |
| KW 385 | IRTF | 1998 Feb 20 | 2".14 ± 0".04 | 14 ± 1 | 1.51 ± 0.03 | 1.86 | 1.07 | 0.58 | 1 | B (ADS) 2".4, S (M) |

[a] M — Mason et al. 1993, B91 — Bolte 1991, IDS — Jeffers et al. 1963, P & W — Peterson & White 1984, B & C — Burkhart & Coupry 1989

## Table 6: α Persei Single Stars

| Name | Telescope | Date | ΔK/ΔF140W | qlim | # comp | Notes [a] |
|---|---|---|---|---|---|---|
| HD 18537 | IRTF | 1998 Feb 18 | 2.87 | 0.34 | 0 | 12".18 (ADS) |
| HD 18538 | IRTF | 1998 Feb 18 | 2.03 | 0.45 | 0 | |
| HE 12 | Palomar | 1997 Oct 08 | 4.2 | 0.26 | 0 | |
| | IRTF | 1998 Feb 18 | | | | |
| HE 56 | Palomar | 1997 Oct 08 | 3.76 | 0.31 | 0 | |
| | IRTF | 1998 Feb 18 | | | | |
| HE 92 | IRTF | 1997 Dec 06 | 2.49 | 0.51 | 0 | |
| HE 93 | IRTF | 1997 Dec 06 | 1.98 | 0.60 | 0 | |
| HE 94 | Keck | 1999 Jan 7 | 4.23 | 0.26 | 0 | |
| HE 104 | Palomar | 1997 Oct 08 | 4.29 | 0.29 | 0 | |
| | IRTF | 1998 Feb 18 | | | | |
| HE 1082 | Palomar | 1997 Oct 08 | 4.25 | 0.27 | 0 | S (M&A) |
| | IRTF | 1998 Feb 19 | | | | |
| HE 1100 | IRTF | 1997 Dec 06 | 4.18 | 0.22 | 0 | |
| HE 1101 | Palomar | 1997 Oct 08 | 4.18 | 0.22 | 0 | |
| | IRTF | 1998 Feb 20 | | | | |
| HE 1153 | IRTF | 1997 Dec 05 | 3.54 | 0.31 | 0 | S (M&A) |
| HE 1185 | Palomar | 1997 Oct 08 | 4.27 | 0.22 | 0 | |
| | IRTF | 1998 Feb 20 | | | | |
| HE 1259 | IRTF | 1997 Dec 05 | 2.7 | 0.40 | 0 | S (M&A) |
| HE 135 | IRTF | 1997 Dec 08 | 1.95 | 0.54 | 0 | |
| HE 143 | Palomar | 1997 Oct 08 | 4.45 | 0.22 | 1 | B (M-P) |
| | IRTF | 1998 Feb 18 | | | | |
| HE 151 | IRTF | 1997 Dec 08 | 2.6 | 0.45 | 0 | |
| HE 167 | Palomar | 1997 Oct 08 | 3.73 | 0.33 | 0 | |
| | IRTF | 1997 Dec 08 | | | | |
| HE 212 | Palomar | 1997 Oct 08 | 3.54 | 0.32 | 0 | |
| | IRTF | 1998 Feb 18 | | | | |
| HE 220 | IRTF | 1997 Dec 08 | 3.92 | 0.33 | 0 | |
| HE 270 | IRTF | 1997 Dec 06 | 2.44 | 0.51 | 0 | |
| HE 299 | IRTF | 1997 Dec 06 | 3.08 | 0.40 | 0 | |
| HE 309 | Keck | 1999 Jan 7 | 4.87 | 0.17 | 0 | |
| HE 314 | IRTF | 1998 Feb 18 | 2.07 | 0.50 | 1 | B (P) |
| HE 333 | Palomar | 1997 Oct 08 | 3.92 | 0.28 | 0 | |
| | IRTF | 1998 Feb 18 | | | | |
| HE 334 | Palomar | 1997 Oct 08 | 4.01 | 0.28 | 0 | |
| | IRTF | 1998 Feb 18 | | | | |
| HE 338 | Palomar | 1997 Oct 08 | 3.89 | 0.32 | 0 | |
| | IRTF | 1998 Feb 18 | | | | |
| HE 350 | Palomar | 1997 Oct 08 | 3.62 | 0.33 | 0 | |
| | IRTF | 1998 Feb 20 | | | | |
| HE 361 | IRTF | 1998 Feb 19 | 2.13 | 0.52 | 0 | |
| HE 365 | IRTF | 1998 Feb 19 | 2.94 | 0.44 | 0 | |
| HE 373 | IRTF | 1997 Dec 09 | 3.73 | 0.29 | 0 | |
| HE 383 | Palomar | 1997 Oct 08 | 3.98 | 0.23 | 0 | S (M&A) |
| | IRTF | 1997 Dec 04 | | | | |
| HE 386 | IRTF | 1997 Dec 07 | 2.78 | 0.40 | 0 | |
| HE 401 | IRTF | 1997 Dec 04 | 3.08 | 0.32 | 0 | |
| HE 407 | Palomar | 1997 Oct 08 | 3.25 | 0.37 | 0 | |
| | IRTF | 1998 Feb 20 | | | | |
| HE 421 | IRTF | 1997 Dec 09 | 2.6 | 0.44 | 0 | |
| HE 423 | IRTF | 1997 Dec 07 | 3.4 | 0.34 | 1 | |
| HE 481 | IRTF | 1997 Dec 09 | 2.78 | 0.43 | 0 | |
| HE 490 | IRTF | 1997 Dec 09 | 1.89 | 0.54 | 0 | |
| HE 520 | IRTF | 1997 Dec 08 | 2.28 | 0.56 | 0 | |
| HE 557 | Palomar | 1997 Oct 08 | 4.45 | 0.21 | 0 | |
| HE 575 | Palomar | 1997 Oct 08 | 4.39 | 0.27 | 0 | S (M&A) |
| | IRTF | 1998 Feb 20 | | | | |
| HE 588 | Palomar | 1997 Oct 08 | 3.45 | 0.38 | 0 | |
| | IRTF | 1998 Feb 20 | | | | |
| HE 609 | IRTF | 1997 Dec 07 | 2.87 | 0.41 | 0 | |
| HE 612 | Palomar | 1997 Oct 08 | 4.33 | 0.28 | 0 | S (M&A) |
| | IRTF | 1998 Feb 20 | | | | |
| HE 621 | IRTF | 1997 Dec 06 | 3.86 | 0.32 | 0 | |
| HE 625 | Palomar | 1997 Oct 08 | 3.76 | 0.31 | 0 | S (M&A) |
| | IRTF | 1998 Feb 20 | | | | |
| HE 632 | IRTF | 1997 Dec 06 | 2.5 | 0.49 | 0 | |
| HE 639 | Palomar | 1997 Oct 08 | 3.73 | 0.33 | 0 | |
| | IRTF | 1998 Feb 20 | | | | |
| HE 651 | IRTF | 1997 Dec 07 | 2.87 | 0.41 | 0 | |
| HE 665 | Palomar | 1997 Oct 08 | 3.25 | 0.38 | 0 | 10".6 (IDS) |
| HE 675 | Palomar | 1997 Oct 08 | 3.83 | 0.27 | 0 | S (M&A) |
| | IRTF | 1998 Feb 19 | | | | |
| HE 684 | Palomar | 1997 Oct 08 | 4.29 | 0.22 | 0 | |
| | IRTF | 1998 Feb 19 | | | | |
| HE 699 | IRTF | 1997 Dec 08 | 2.94 | 0.45 | 0 | |
| HE 709 | Palomar | 1997 Oct 08 | 3.73 | 0.30 | 0 | |
| | IRTF | 1998 Feb 20 | | | | |
| HE 715 | IRTF | 1997 Dec 06 | 3.01 | 0.44 | 0 | |
| HE 727 | Palomar | 1997 Oct 08 | 3.8 | 0.31 | 0 | |
| | IRTF | 1998 Feb 20 | | | | |
| HE 735 | Palomar | 1997 Oct 08 | 3.73 | 0.29 | 0 | S (M&A) |
| | IRTF | 1998 Feb 18 | | | | |
| HE 750 | Palomar | 1997 Oct 08 | 4.13 | 0.24 | 0 | |
| | IRTF | 1998 Feb 18 | | | | |
| HE 767 | Palomar | 1997 Oct 08 | 3.69 | 0.31 | 0 | |
| | IRTF | 1998 Feb 20 | | | | |
| HE 774 | IRTF | 1997 Dec 04 | 2.17 | 0.43 | 1 | B (M&A) 21.7dy |
| HE 775 | IRTF | 1997 Dec 05 | 2.94 | 0.36 | 1 | B (M&A) 21.2dy |
| HE 799 | IRTF | 1998 Feb 18 | 3.45 | 0.39 | 0 | 6".68 (ADS) |
| HE 810 | Palomar | 1997 Oct 08 | 3.76 | 0.25 | 0 | S (M&A) |
| | IRTF | 1998 Feb 18 | | | | |
| HE 817 | Palomar | 1997 Oct 08 | 3.62 | 0.31 | 1 | B (M&A) 30.9dy |
| | IRTF | 1998 Feb 18 | | | | |
| HE 831 | Palomar | 1997 Oct 08 | 4.03 | 0.29 | 0 | S (M&A) |
| | IRTF | 1997 Dec 05 | | | | |
| HE 833 | IRTF | 1998 Feb 18 | 2.6 | 0.49 | 0 | |
| HE 841 | Palomar | 1997 Oct 08 | 3.08 | 0.44 | 0 | |
| | IRTF | 1998 Feb 20 | | | | |
| HE 848 | Palomar | 1997 Oct 08 | 3.45 | 0.39 | 1 | B (M-P) |
| | IRTF | 1998 Feb 20 | | | | |
| HE 862 | Palomar | 1997 Oct 08 | 4.98 | 0.20 | 0 | |
| | IRTF | 1998 Feb 20 | | | | |
| HE 868 | Palomar | 1997 Oct 08 | 4.03 | 0.28 | 1 | ? (M&A) |
| | IRTF | 1998 Feb 20 | | | | |
| HE 885 | IRTF | 1997 Dec 06 | 2.17 | 0.49 | 0 | |
| HE 904 | IRTF | 1997 Dec 05 | 3.98 | 0.25 | 0 | S (M&A) |
| HE 906 | IRTF | 1997 Dec 05 | 4.08 | 0.32 | 0 | |
| HE 917 | Palomar | 1997 Oct 08 | 3.92 | 0.27 | 0 | |
| | IRTF | 1998 Feb 19 | | | | |
| HE 921 | IRTF | 1997 Dec 07 | 4.41 | 0.26 | 0 | |
| HE 931 | IRTF | 1997 Dec 07 | 3.49 | 0.37 | 0 | |
| HE 944 | IRTF | 1997 Dec 06 | 4.06 | 0.29 | 0 | |
| HE 955 | Palomar | 1997 Oct 08 | 4.11 | 0.27 | 1 | ? (M&A) |
| | IRTF | 1997 Dec 05 | | | | |
| HE 968 | Palomar | 1997 Oct 08 | 4.2 | 0.24 | 0 | |
| | IRTF | 1998 Feb 19 | | | | |
| HE 970 | Palomar | 1997 Oct 08 | 4.35 | 0.28 | 0 | |
| | IRTF | 1998 Feb 20 | | | | |
| HE 985 | IRTF | 1997 Dec 05 | 3.98 | 0.24 | 0 | S (M&A) |
| AP 14 | IRTF | 1997 Dec 09 | 2.6 | 0.45 | 0 | |
| AP 20 | HST | 1998 Aug 27 | 1.06 | 0.80 | 0 | |
| AP 21 | HST | 1998 Aug 27 | 1.04 | 0.78 | 0 | |
| AP 25 | HST | 1998 Aug 26 | 1.06 | 0.80 | 0 | |
| AP 28 | HST | 1998 Aug 27 | 1.07 | 0.80 | 0 | |
| AP 37 | HST | 1998 Aug 27 | 1.1 | 0.80 | 0 | |
| AP 43 | HST | 1998 Jul 25 | 1.07 | 0.80 | 0 | |
| AP 56 | HST | 1998 Aug 27 | 1.07 | 0.80 | 0 | |

| | | | | | |
|---|---|---|---|---|---|
| AP 63 | Keck | 1999 Jan 7 | 3.25 | 0.32 | 0 |
| AP 70 | HST | 1998 Aug 27 | 1.06 | 0.80 | 0 |
| AP 78 | HST | 1998 Aug 27 | 1.05 | 0.80 | 0 |
| AP 90 | IRTF | 1997 Dec 06 | 3.2 | 0.38 | 0 |
| AP 92 | HST | 1998 Feb 28 | 1.06 | 0.71 | 0 |
| AP 97 | IRTF | 1997 Dec 09 | 2.48 | 0.49 | 0 |
| AP 100 | HST | 1998 Aug 08 | 1.06 | 0.80 | 0 |
| AP 102 | IRTF | 1998 Feb 20 | 2.78 | 0.43 | 0 |
| AP 104 | IRTF | 1998 Feb 20 | 1.98 | 0.57 | 0 |
| AP 110 | Keck | 1999 Jan 7 | 3.73 | 0.23 | 0 |
| AP 117 | HST | 1998 Aug 27 | 1.07 | 0.80 | 0 |
| AP 118 | HST | 1998 Aug 28 | 1.06 | 0.80 | 0 |
| AP 125 | HST | 1998 Feb 28 | 1.07 | 0.80 | 0 |
| AP 127 | HST | 1998 Jul 25 | 1.07 | 0.80 | 0 |
| AP 156* | IRTF | 1998 Feb 19 | 2.17 | 0.57 | 0 |
| AP 158 | IRTF | 1997 Dec 08 | 3.8 | 0.25 | 0 |
| AP 173 | Keck | 1999 Jan 7 | 4.57 | 0.17 | 0 |
| AP 194 | IRTF | 1998 Feb 19 | 3.08 | 0.33 | 0 |
| AP 199 | IRTF | 1998 Feb 19 | 2.23 | 0.57 | 0 |
| AP 213 | Palomar | 1997 Oct 08 | 3.49 | 0.32 | 0 |
| | IRTF | 1998 Feb 20 | | | |
| AP 225 | IRTF | 1997 Dec 08 | 3.2 | 0.36 | 0 |
| AP 256 | IRTF | 1997 Dec 08 | 3.83 | 0.25 | 0 |
| AP 264 | IRTF | 1997 Dec 08 | 2.05 | 0.59 | 0 |

[a] M&A - Morrell & Abt 1992, P - Prosser 1992, ADS - Aitken 1932, IDS — Jeffers et al. 1963

## Table 7: Praesepe Single Stars

| Object | Telescope | Date | ΔK 0".15 | q lim. | # comp | Notes [a] |
|---|---|---|---|---|---|---|
| A 70 | IRTF | 1999 Feb 15 | 3.14 | 0.44 | 0 | |
| A 365 | IRTF | 1999 Feb 13 | 2.35 | 0.47 | 0 | |
| A 609 | IRTF | 1999 Feb 15 | 3.4 | 0.41 | 0 | |
| A 1196 | IRTF | 1999 Feb 14 | 0 | 1.00 | 0 | |
| A 1480 | IRTF | 1999 Feb 14 | 3.73 | 0.35 | 0 | |
| A 1501 | IRTF | 1999 Feb 13 | 1.87 | 0.54 | 0 | |
| A 1528 | IRTF | 1999 Feb 13 | 2.78 | 0.41 | 0 | |
| A 1583 | IRTF | 1999 Feb 13 | 2.32 | 0.48 | 0 | |
| I 563 | IRTF | 1999 Feb 15 | 3.69 | 0.34 | 0 | |
| II 490 | IRTF | 1999 Feb 14 | 4.39 | 0.23 | 0 | |
| II 582 | IRTF | 1999 Feb 15 | 2.6 | 0.51 | 0 | |
| JS 88 | IRTF | 1999 Feb 15 | 3.54 | 0.36 | 0 | |
| JS 495 | IRTF | 1999 Feb 15 | 3.66 | 0.33 | 0 | |
| JS 532 | IRTF | 1999 Feb 15 | 3.08 | 0.46 | 0 | |
| JS 589 | IRTF | 1999 Feb 14 | 4.08 | 0.29 | 0 | |
| JS 600 | IRTF | 1999 Feb 15 | 3.2 | 0.44 | 0 | |
| KW 16 | IRTF | 1999 Feb 14 | 0 | 1.00 | 1 | ? (B91), S (M), B (MM99) |
| KW 38 | IRTF | 1999 Feb 14 | 0 | 1.00 | 0 | S (M), S (P&W) |
| KW 40 | IRTF | 1999 Feb 13 | 2.6 | 0.42 | 2 | S (M), S (P&W), B (B&C), T (AW99) |
| KW 45 | Palomar | 1996 Jan 9 | 4.16 | 0.30 | 0 | S (M) |
| | IRTF | 1998 Feb 18 | | | | |
| KW 47 | IRTF | 1999 Feb 14 | 4.58 | 0.21 | 1 | B (MM99) |
| KW 50 | Palomar | 1996 Jan 9 | 4.25 | 0.25 | 0 | S (M), B (AW99) |
| | IRTF | 1998 Feb 18 | | | | |
| KW 100 | Keck | 1999 Jan 7 | 3.86 | 0.29 | 0 | S (MM99) |
| KW 114 | Palomar | 1996 Jan 9 | 4.67 | 0.25 | 0 | S (M), S (P&W) |
| | IRTF | 1998 Feb 18 | | | | |
| KW 124 | Palomar | 1996 Jan 9 | 4.13 | 0.30 | 0 | S (M), S (P&W) |
| | IRTF | 1998 Feb 18 | | | | |
| KW 142 | Palomar | 1996 Jan 9 | 3.36 | 0.39 | 1 | S (M), B (MM99) |

| | | | | | |
|---|---|---|---|---|---|
| | IRTF | 1998 Feb 18 | | | |
| KW 143 | Palomar | 1996 Jan 9 | 4.46 | 0.27 | 1 S (M), B (P&W) 0".0441 |
| | IRTF | 1998 Feb 18 | | | |
| KW 146 | IRTF | 1998 Feb 19 | 3.95 | 0.32 | 0 S (P&W) |
| KW 150 | Palomar | 1996 Jan 9 | 4.86 | 0.22 | 0 S (M), S (P&W), S (AW99) |
| | IRTF | 1998 Feb 18 | | | |
| KW 154 | Palomar | 1996 Jan 9 | 3.89 | 0.33 | 0 S (M), S (P&W) |
| | IRTF | 1998 Feb 18 | | | |
| KW 155 | IRTF | 1998 Feb 19 | 2.7 | 0.47 | 0 S (MM99) |
| KW 181 | Keck | 1999 Jan 7 | 3.2 | 0.43 | 1 B (MM99) |
| KW 182 | IRTF | 1999 Feb 15 | 4.39 | 0.23 | 1 S (B91), B (MM99) |
| KW 204 | Palomar | 1996 Jan 8 | 4.89 | 0.20 | 0 S (M), S (P&W), S (AW99) |
| | IRTF | 1998 Feb 18 | | | |
| KW 207 | Palomar | 1996 Jan 8 | 4.9 | 0.23 | 0 S (M), S (AW99) |
| | IRTF | 1998 Feb 18 | | | |
| KW 217 | Keck | 1999 Jan 6 | 4.53 | 0.20 | 0 S (B91), S (MM99) |
| KW 218 | IRTF | 1999 Feb 14 | 3.92 | 0.32 | 0 S (M), S (MM99) |
| KW 222 | Keck | 1999 Jan 6 | 4.89 | 0.16 | 0 S (MM99) |
| KW 226 | Palomar | 1996 Jan 9 | 3.36 | 0.38 | 0 S (M), S (P&W) |
| | IRTF | 1998 Feb 19 | | | |
| KW 227 | IRTF | 1998 Feb 18 | 2.39 | 0.52 | 0 S (B91), S (MM99) |
| KW 229 | Palomar | 1996 Jan 9 | 3.89 | 0.29 | 0 S (M), B (AW99) |
| | IRTF | 1998 Feb 18 | | | |
| KW 238 | Keck | 1999 Jan 7 | 3.14 | 0.44 | 0 S (MM99) |
| KW 239 | IRTF | 1998 Feb 20 | 3.49 | 0.40 | 0 S (MM99) |
| KW 244 | IRTF | 1999 Feb 14 | 4.48 | 0.22 | 1 B (B91) 0.4dy |
| KW 253 | Palomar | 1995 Nov 13 | 3.54 | giant | 0 S (B91), S (M), S (AW99) |
| KW 258 | IRTF | 1999 Feb 15 | 3.58 | 0.36 | 0 |
| KW 265 | Palomar | 1995 Nov 12 | 3.73 | 0.29 | 1 B (M) 0".425, S (P&W), S (B&C), S (AW99) |
| | IRTF | 1998 Feb 19 | | | |
| KW 268 | IRTF | 1998 Feb 18 | 1.87 | 0.60 | 1 B (MM99) |
| KW 271 | IRTF | 1998 Feb 18 | 2.39 | 0.49 | 0 S (M) |
| KW 276 | Palomar | 1995 Nov 12 | 3.8 | 0.31 | 1 S (M), S (P&W), B (B&C), S (AW99) |
| | IRTF | 1998 Feb 18 | | | |
| KW 279 | Palomar | 1995 Nov 12 | 2.49 | 0.43 | 1 S (M), S (P&W), B (B&C), B (AW99) |
| | IRTF | 1998 Feb 18 | | | |
| KW 283 | Palomar | 1995 Nov 12 | 2.7 | giant | 0 S (B91), S (M), S (P&W), S (AW99) |
| KW 286 | Palomar | 1996 Jan 9 | 4.13 | 0.30 | 0 S (M), S (P&W), S (B&C) |
| | IRTF | 1998 Feb 19 | | | |
| KW 292 | Palomar | 1996 Jan 9 | 4.08 | 0.30 | 1 B (B91), S(M) |
| | IRTF | 1998 Feb 19 | | | |
| KW 293 | IRTF | 1999 Feb 15 | 3.49 | 0.38 | 0 S (MM99) |
| KW 295 | IRTF | 1998 Feb 18 | 2.6 | 0.48 | 0 |
| KW 300 | Palomar | 1995 Nov 13 | 3.45 | 0.28 | 1 S (M), S (P&W), B (B&C), B (AW99) |
| | IRTF | 1998 Feb 18 | | | |
| KW 318 | Palomar | 1996 Jan 8 | 5.05 | 0.19 | 0 S (M) |
| KW 323 | Palomar | 1996 Jan 8 | 4.6 | 0.25 | 0 S (M) |
| KW 328 | Palomar | 1996 Jan 8 | 5.07 | 0.20 | 0 S (M), S (P&W), S (AW99) |
| | IRTF | 1998 Feb 19 | | | |
| KW 332 | IRTF | 1998 Feb 19 | 1.85 | 0.60 | 0 S (MM99) |
| KW 340 | Palomar | 1996 Jan 8 | 4.57 | 0.25 | 0 S (M) |
| KW 341 | Keck | 1999 Jan 6 | 4.74 | 0.18 | 1 B (MM99) |
| KW 348 | Palomar | 1996 Jan 9 | 3.58 | 0.29 | 0 S (M), S (P&W), S (AW99) |
| | IRTF | 1998 Feb 19 | | | |
| KW 350 | IRTF | 1998 Feb 18 | 2.87 | 0.43 | 1 S (M), B (B&C) |
| KW 370 | Palomar | 1996 Jan 8 | 5.16 | 0.17 | 0 S (M), S (P&W) |
| KW 371 | Palomar | 1996 Jan 8 | 4.57 | 0.20 | 1 B (MM99) |
| KW 375 | Palomar | 1996 Jan 8 | 5.11 | 0.19 | 0 S (M), S (P&W) |
| | IRTF | 1998 Feb 20 | | | |
| KW 396 | IRTF | 1999 Feb 15 | 3.76 | 0.33 | 0 S (MM99) |
| KW 411 | IRTF | 1998 Feb 19 | 2.6 | 0.48 | 0 S (M), S (MM99) |
| KW 416 | IRTF | 1998 Feb 19 | 1.89 | 0.58 | 1 B (MM99) |
| KW 418 | Keck | 1999 Jan 7 | 3.49 | 0.35 | 0 S (MM99) |
| KW 421 | Keck | 1999 Jan 6 | 4.57 | 0.20 | 0 S (MM99) |
| KW 429 | Palomar | 1996 Jan 9 | 4.01 | 0.32 | 0 S (M) |
| | IRTF | 1998 Feb 20 | | | |
| KW 439 | IRTF | 1999 Feb 14 | 4.89 | 0.18 | 0 S (M), B (MM99) |
| KW 445 | Palomar | 1996 Jan 9 | 4.25 | 0.28 | 0 S (M), S (P&W) |
| | IRTF | 1998 Feb 20 | | | |
| KW 449 | IRTF | 1999 Feb 13 | 2.5 | 0.43 | 0 S (M) |
| KW 454 | Keck | 1999 Jan 6 | 3.69 | 0.34 | 0 S (B91), S (MM99) |
| KW 459 | Palomar | 1996 Jan 9 | 4.11 | 0.30 | 0 S (M) |
| KW 472 | IRTF | 1998 Feb 20 | 1.83 | 0.61 | 0 |
| KW 478 | IRTF | 1999 Feb 15 | 2.78 | 0.48 | 0 |
| KW 496 | IRTF | 1999 Feb 14 | 3.98 | 0.32 | 1 S (B91), B (MM99) |
| KW 515 | IRTF | 1999 Feb 15 | 3.4 | 0.39 | 1 ? (B91) |
| KW 534 | IRTF | 1999 Feb 13 | 3.2 | 0.36 | 1 S (M), S (P&W), S (B&C), B (AW99) |
| KW 536 | IRTF | 1999 Feb 14 | 4.06 | 0.31 | 0 S (M) |
| KW 538 | IRTF | 1999 Feb 13 | 2.87 | 0.43 | 0 S (M) |
| KW 549 | IRTF | 1999 Feb 15 | 4.46 | 0.21 | 0 |
| KW 553 | Keck | 1999 Jan 6 | 2.6 | 0.52 | 0 NONMEMBER |

M — Mason et al. 1993, B91 — Bolte 1991, IDS — Jeffers et al. 1963, P & W — Peterson & White 1984, B & C — Burkhart & Coupry 1989, MM99 - Mermilliod et al. 1999, AW99 - Abt & Willmarth 1999

**Table 8a: Summary of Observations**

| Observable | Trend in Speckle/AO/HST Cluster data |
|---|---|
| q distribution | Flat or slightly rising toward lower q |
| q vs. separation | No dependence |
| q distribution vs. Mass | Steeper slope for higher Mass |
| CSF vs. Mass | Lower for higher Mass |

**Table 8b: Summary of Comparison with Formation Models**

| | q distribution | | q vs. separation | | q dist. vs. Mass | | CSF vs. Mass | |
|---|---|---|---|---|---|---|---|---|
| Model | Prediction | Observed? | Prediction | Observed? | Prediction | Observed? | Prediction | Observed? |
| small-N capture | increase toward q~1 | N | | | | | increase for higher Mass | N |
| small-N w/disks | more lower q | Y | | | | | increase for higher Mass | ? [b] |
| few-body decay | | | | | steeper for higher Mass | Y | increase for higher Mass | ? [b] |
| scale-free fragmentation | | | | | independent of Mass | N | independent of Mass | N |
| disk fragmentation | | | should show dependence | N [a] | | | | |
| accretion in association | | | more q~1 closer | N | more q~1 for higher Mass | N | | |
| accretion in cluster | | | | | more low q for higher Mass | Y | | |

[a] - may require larger separation range to properly test, [b] - prediction for fixed companion mass limit, not mass ratio limit, data cannot address

## Table 9: X-ray Detections and Binaries among Early-type Stars

| Sample | # X-ray detections | # bin | CSF $_{X-ray}$ | # X-ray Upper limits | # bin [a] | CSF $_{no\ X-ray}$ |
|---|---|---|---|---|---|---|
| Hyades | 9 | 6 | 0.7 ± 0.3 | 14 | 5 | 0.4 ± 0.2 |
| Praesepe | 5 | 3 | 0.6 ± 0.3 | 27 | 7(2) | 0.19 ± 0.08 |
| a Persei | 8 | 4 | 0.5 ± 0.3 | 15 | 2 | 0.13 ± 0.09 |
| 3 Clusters | 22 | 13 | 0.6 ± 0.2 | 56 | 14(2) | 0.21 ± 0.06 |

[a] numbers in parentheses indicate binaries with two early-type components

## Table A1: Cluster Binaries in Figure 3a

| log(sep[AU]) | sample size | # bin | clusters | Techniques | Refs. | Binaries |
|---|---|---|---|---|---|---|
| -1.525 | 162 | 3 | Hy | spect. | 1 | Lei 20, +22 669, vB 38 |
| -1.075 | 241 | 9 | Hy, Pr | spect. | 1, 2 | vB 34, vB 40, vB 22, vB 121, vB 45, vB 62, vB 117, KW 181, KW 127 |
| -0.625 | 262 | 13 | Hy, Pr | spect. | 1, 2, 3 | vB 112, vB 75, vB 69, vB 162, Lei 83, KW 416, KW 47, KW 495, KW 142, KW 184, KW 365, KW 368, KW 300, KW 284 |
| -0.175 | 262 | 12 | Hy, Pr | spect. | 1, 2, 3 | vB 83, vB 124, vB 130, vB 140, vB 77, vB 185, vB 182, KW 268, KW 534, KW 428, KW 40 |
| 0.275 | 241 | 15 | Hy, Pr | spect. | 1, 2 | vB 177, vB 95, vB 43, vB 102, vB 151, Lei 63, +10 568, vB 142, vB 115, vB 120, KW 439, KW 508, KW 325, KW 540, KW 367 |
| 0.725 | 162 | 18 | Hy | spect., speck. | 1, 4 | Lei 57, Lei 90, vB 57, vB 63, vB 96, vB 113, vB 81, vB 106, vB 91, vB 39, Lei 59, vB 141, vB 24, vB 103, vB 114, vB 50, vB 59, vB 58 |
| 1.175 | 162 | 11 | Hy | speck. | 1, 4 | Lei 20, Lei 83, vB 102, vB 131, vB 85, vB 122, vB 75, Lei 92, vB 29, vB 185, vB 124 |
| 1.625 | 383 | 15 | αP, Pl, Pr | speck., AO, SAA | 5, 6 | KW 275, HE 285, HII 2106, HII 1061, A1565, KW 232, KW 458, HII 2278, AP 149AB, KW 365, HII 738, HII 357, HII 2500 AC, AP 139, AP 98 |
| 2.075 | 383 | 16 | αP, Pl, Pr | speck., AO, SAA | 5, 6 | AP38, HII 3197 AC, KW 203, HII 2193, AP 201, AP 41, HII 97, HII 1100, HII 885, HE 835, KW 250, HII 1298, KW 224, HII 1355, AP 17, KW 282 |
| 2.525 | 383 | 9 | αP, Pl, Pr | speck., AO, SAA | 5, 6 | MT 61 AB, HII 303, HII 134, HE 780 AC, HE 828, KW 385, AP 6, HII 102, AP 121 |

1 - Griffin et al. 1988; 2 - Mermilliod & Mayor 1999; 3 - Abt & Willmarth 1999; 4 - Patience et al. 1998; 5 - Bouvier et al. 1997; 6 - Patience et al. 2001

## Table A2: T Tauri Binaries in Figure 3c

| log(sep[AU]) | sample size | # bin | regions | Techniques | Refs. | Binaries |
|---|---|---|---|---|---|---|
| -1.525 | 53 | 1 | Tau/CrA/Sco-Oph | spect. | 1 | 155913-2233 |
| -1.075 | 53 | 1 | Tau/CrA/Sco-Oph | spect. | 1 | 160905-1859 |
| -0.625 | 53 | 2 | Tau/CrA/Sco-Oph | spect. | 1 | 162819-24235, 162814-2427 |
| -0.175 | 53 | 1 | Tau/CrA/Sco-Oph | spect. | 1 | 160814-1857 |
| 0.275 | 82 | 5 | Tau/Oph | lun. occ. | 2 | V853 Oph, ROXs 43B, HP Tau, SR 1, HP Tau G3/G2 |
| 0.725 | 82 | 5 | Tau/Oph | lun. occ. | 2 | Elias 12, FF Tau, ZZ Tau, HV Tau, ROXs 42B |
| 1.175 | 101 | 10 | Tau/Oph | lun. occ., speck. | 2, 3 | SR 20, DF Tau, VSSG 14, FW Tau, ROX 42C, V773 Tau, DI Tau, V410 Tau, V928 Tau, FO Tau |
| 1.625 | 254 | 29 | Tau/Tau(X)/Lup/Cha/CrA | speck. | 2, 3, 4, 5, 6 | SR 12, ROXs 31, GN Tau, V853 Oph, LkCa 3, IW Tau, LkHa 331, CoKu LkHa 332/G2, GH Tau, XZ Tau, CZ Tau, Elias 12, CoKu LkHa 332/G1, GG Tau, LkHa 332, IS Tau, FS Tau, HN Lup, HM Anon, RXJ0444.4+1952 AB, RXJ0438.2+2023, RXJ0452.8+1621, HD 284135, RXJ0451.9+28 |
| 2.075 | 254 | 26 | Tau/Tau(X)/Lup/Cha/CrA | speck., direct im. | 3, 4, 5, 6 | NTTS 043230+1746, FQ Tau, Haro 6-28, DD Tau, FX Tau, UY Aur, VY Tau, NTTS 034903+2431, T Tau, C 7-11, S CrA, VW Cha, WX Cha, RXJ0457.2+1524, RXJ0447.9+2755, FV Tau +c, RXJ0451.8+1758, RXJ0453.1+3311 AB, LkCa 7, HD 285281, RXJ0415.8+3100, RW Aur A/B, GG Ta |
| 2.525 | 254 | 19 | Tau/Tau(X)/Lup/Cha/CrA | direct im. | 2, 4, 5, 6 | Sz 41, NTTS 040047+2603, RXJ0444.9+2717, Sz 77, Sz 81, DoAr 24 E, CoKu Tau/3, VV CraA, IT Tau, CHXR 32, RXJ0412.8+1937, Haro 6-37 /c, UX Tau A/B, Sz 120, VW Cha, DK Tau, Sz 68, V710 A/B, UZ Tau e/w |
| 2.975 | 240 | 19 | Tau/Tau(X)/Lup/Cha/CrA | direct im. | 4, 5, 6 | Sz 65, Sz 91, RXJ0437.4+185, RXJ0444.4+1952 AB-C, RXJ0420.8+3009AB-C, RXJ0409.1+2901, RXJ0457.5+2014, RXJ0453.1+3311AB-C, BD+26718B, RXJ0438.2+2302, HD285957AB, RXJ0444.3+2017, RXJ0431.3+1800, RXJ0435.9+2352AB-C, UX Tau A/b, NTTS040142+2150W+E, NTTS035120 |

1 - Mathieu et al. 1989; 2 - Simon et al. 1995; 3 - Ghez et al. 1993; 4 - Leinert et al. 1993; 5 - Koehler & Leinert 1998; 6 - Ghez et al. 1997

**Table A3: Orion Binaries in Figure 3c**

| log(sep[AU]) | sample size | # bin | Techniques | Refs. | Binaries |
|---|---|---|---|---|---|
| -1.525 | | | | | |
| -1.075 | | | | | |
| -0.625 | | | | | |
| -0.175 | | | | | |
| 0.275 | | | | | |
| 0.725 | | | | | |
| 1.175 | | | | | |
| 1.625 | 361 | 11 | HST im., speck. | 1, 2 | TCC 56, PC 11/12, PC 13/14, PC 80/79, PC 99/98, PC 126/125, PC 152/150, PC 173/172, PC 202/203, PC 208/206, PC 322/321 |
| 2.075 | 480 | 20 | HST im., speck., AO | 1, 2, 3 | TCC 45, TCC 77/75, TCC 101, PC 16/17, PC 33/34, PC 48/49, PC 68/67, PC 111/112, PC 144/143, PC 224/222, PC 315/314, AO 45/46, AO 102/104, AO 128/131, AO5/6, AO 16/18, AO 29/30, AO 149/147, AO 179/180, AO 183/185 |
| 2.525 | 480 | 16 | HST im., speck., AO | 1, 2, 3 | PC 37/40, PC 57/58, PC 80/77, PC 129/130, PC 219/221, PC 235/234, PC 282/279, PC 294/296, PC 323/324, AO 51/48, AO 74/78, AO 110/116, AO 117/110, AO 117/116, AO 184/189, AO 253/252 |

1 - Prosser et al. 1994; 2 - Petr et al. 1998; 3 - Simon et al. 1999

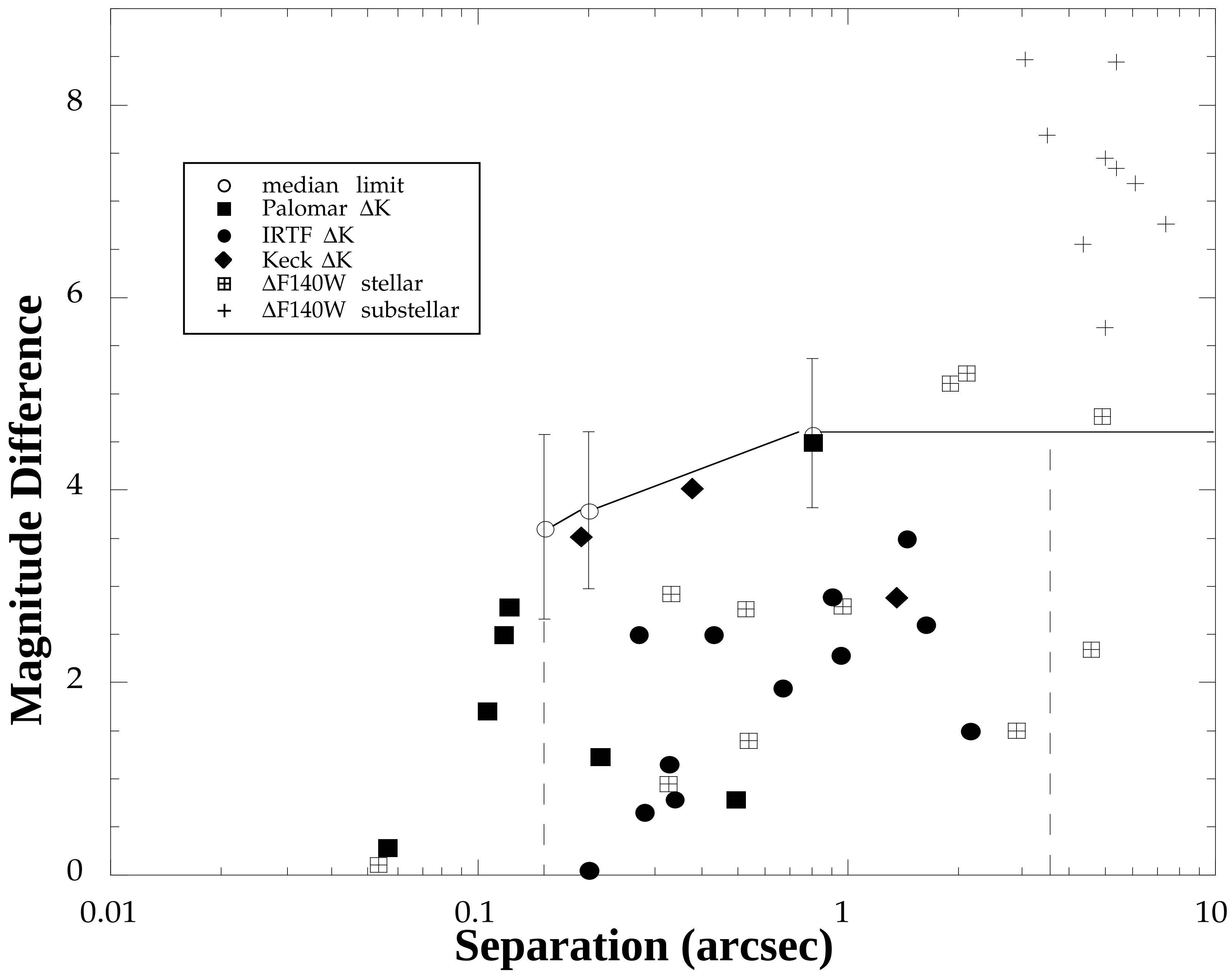

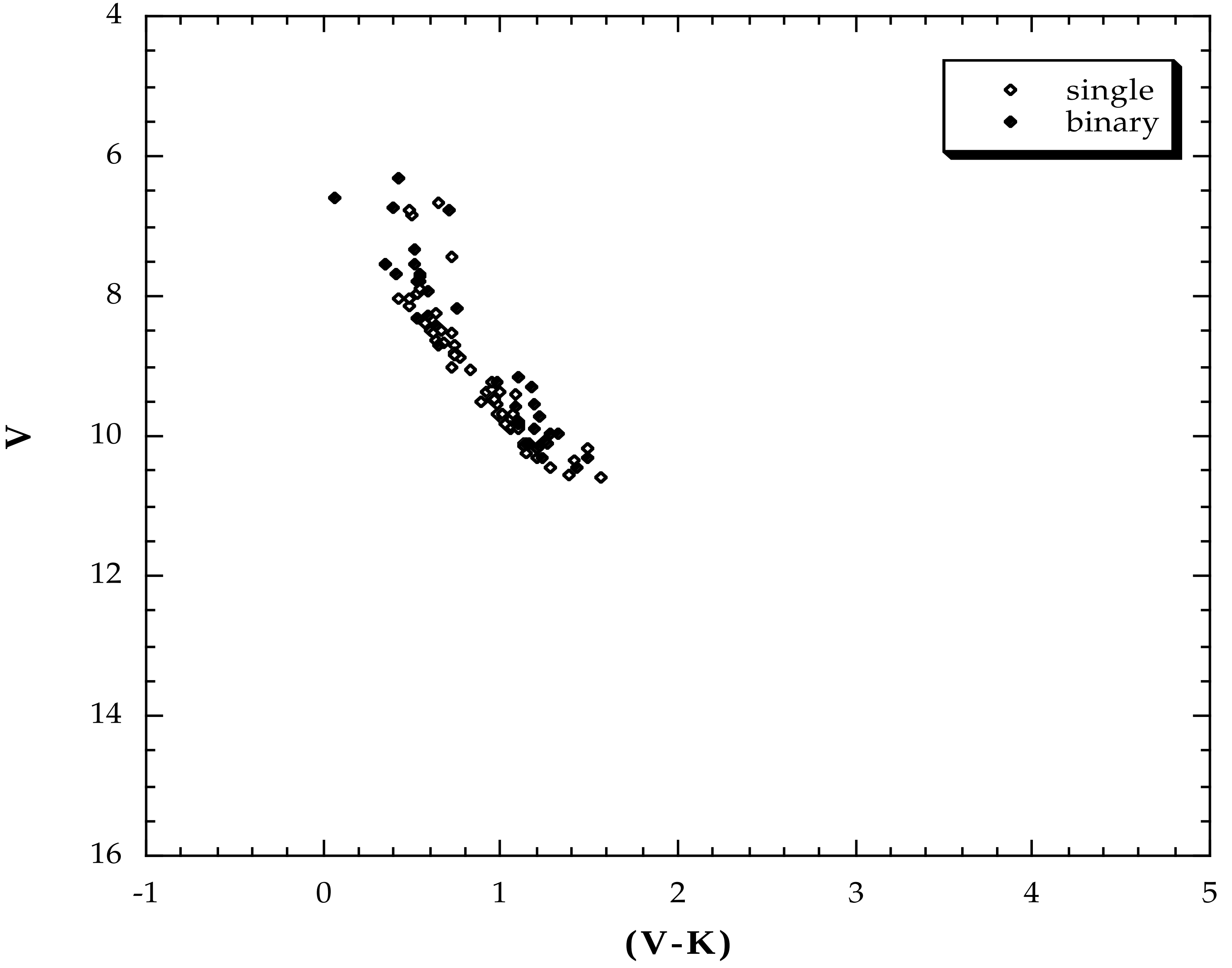

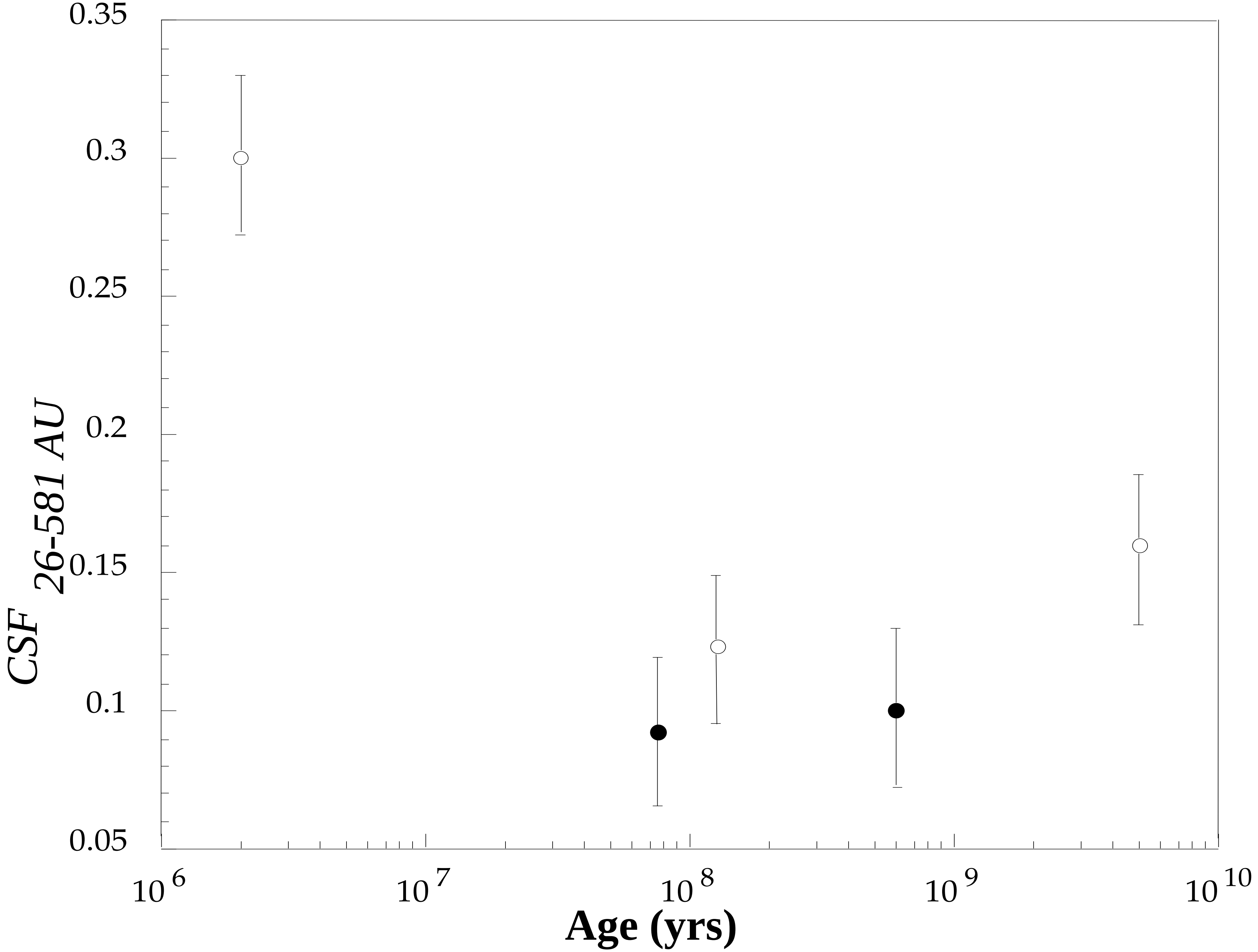

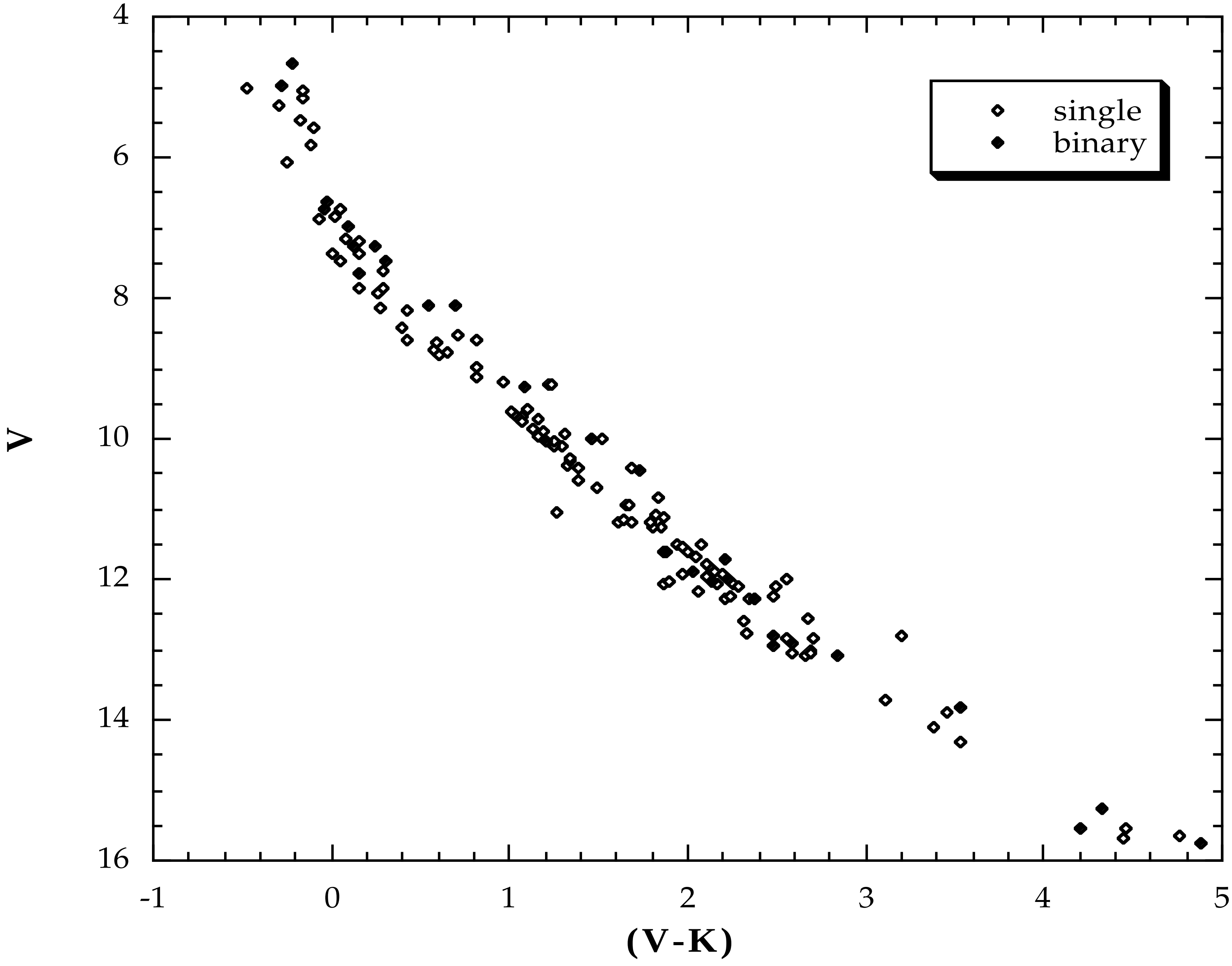

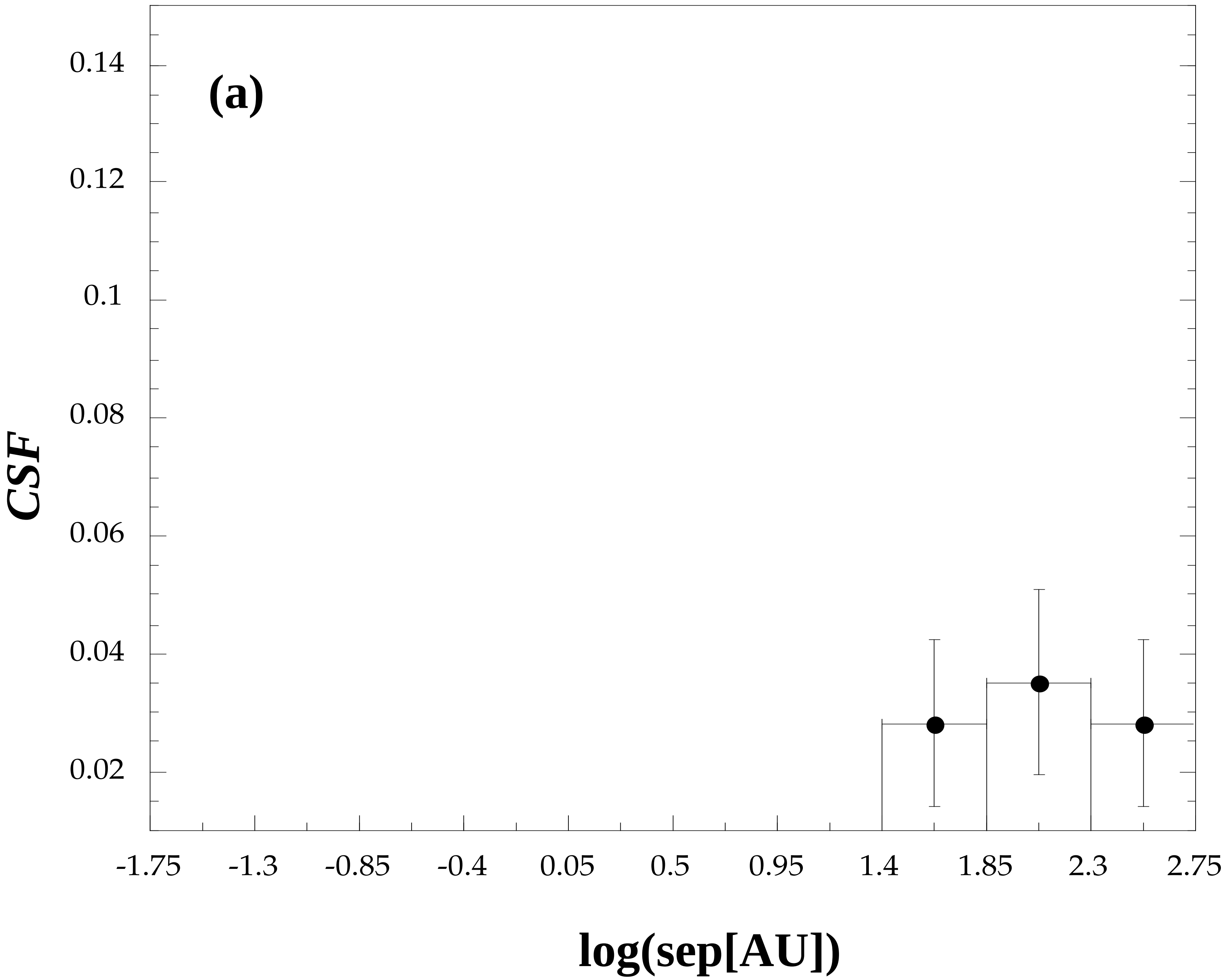

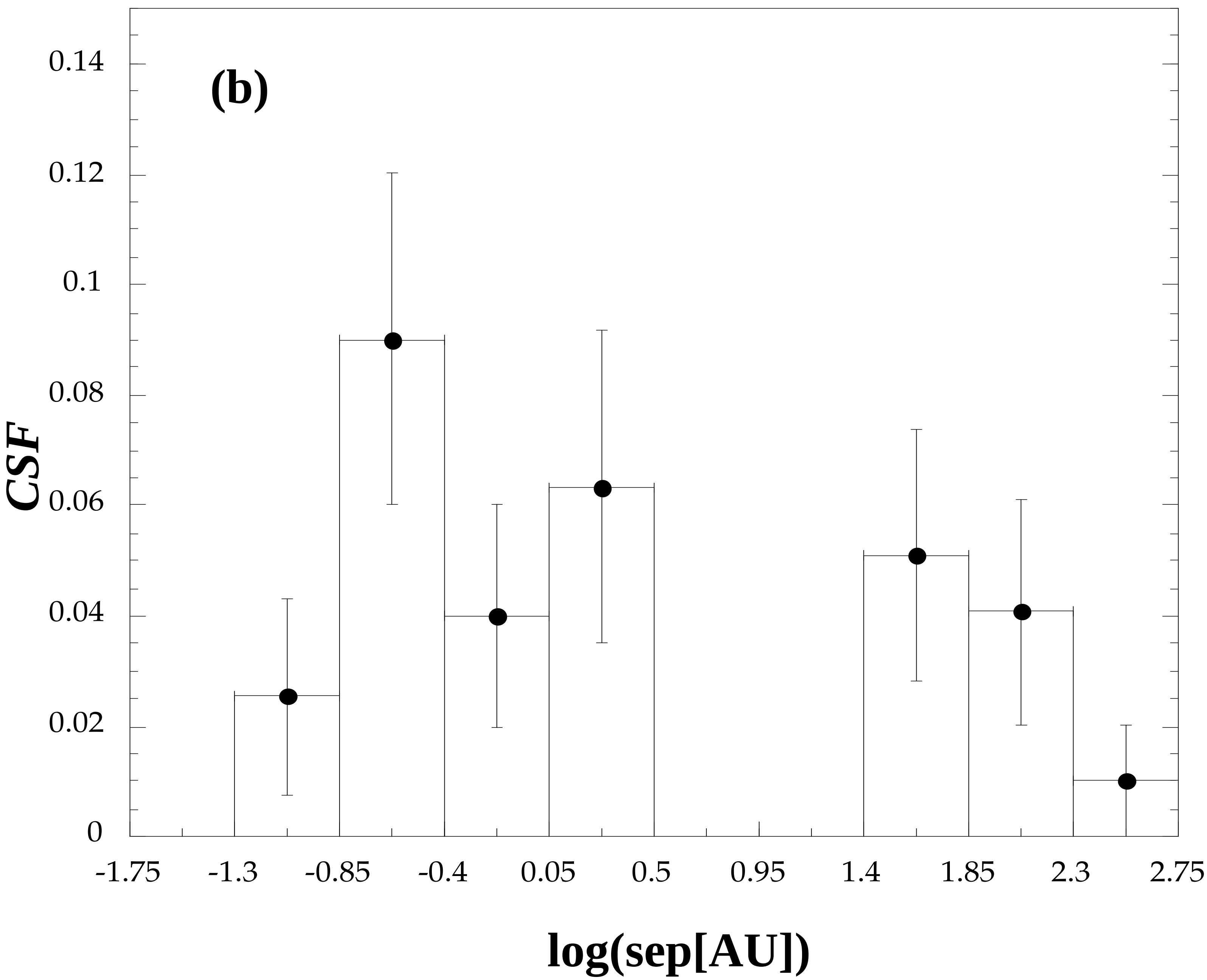

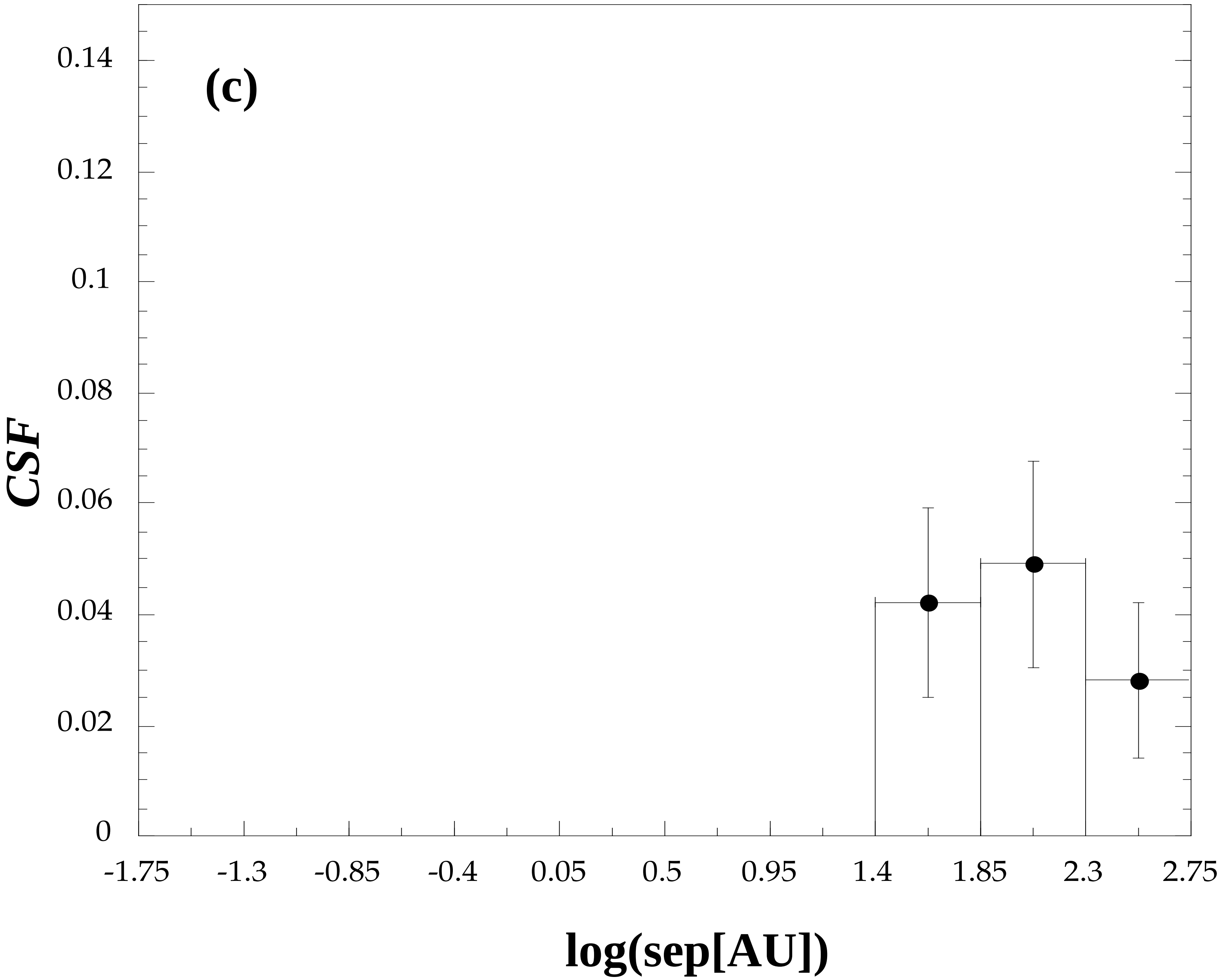

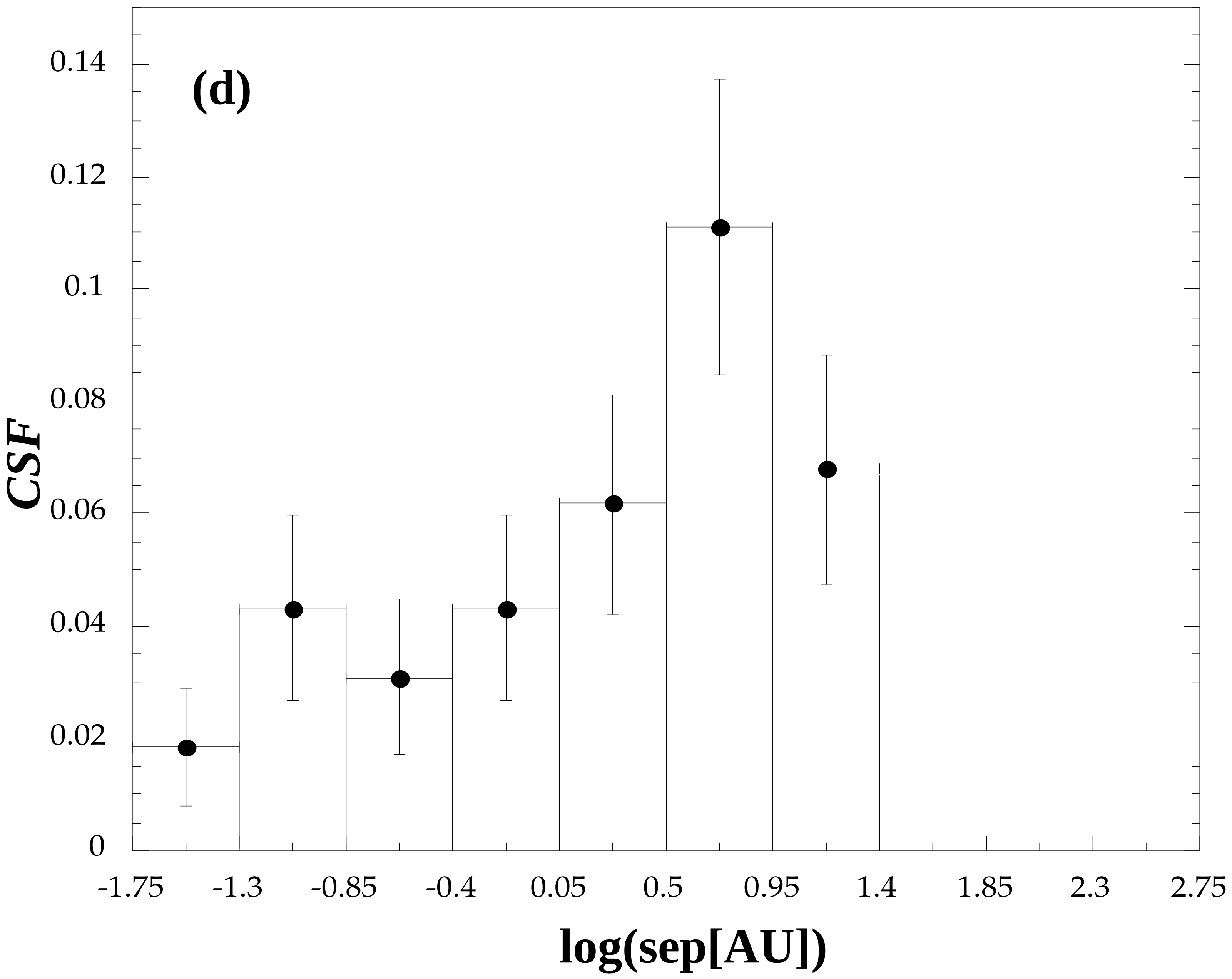

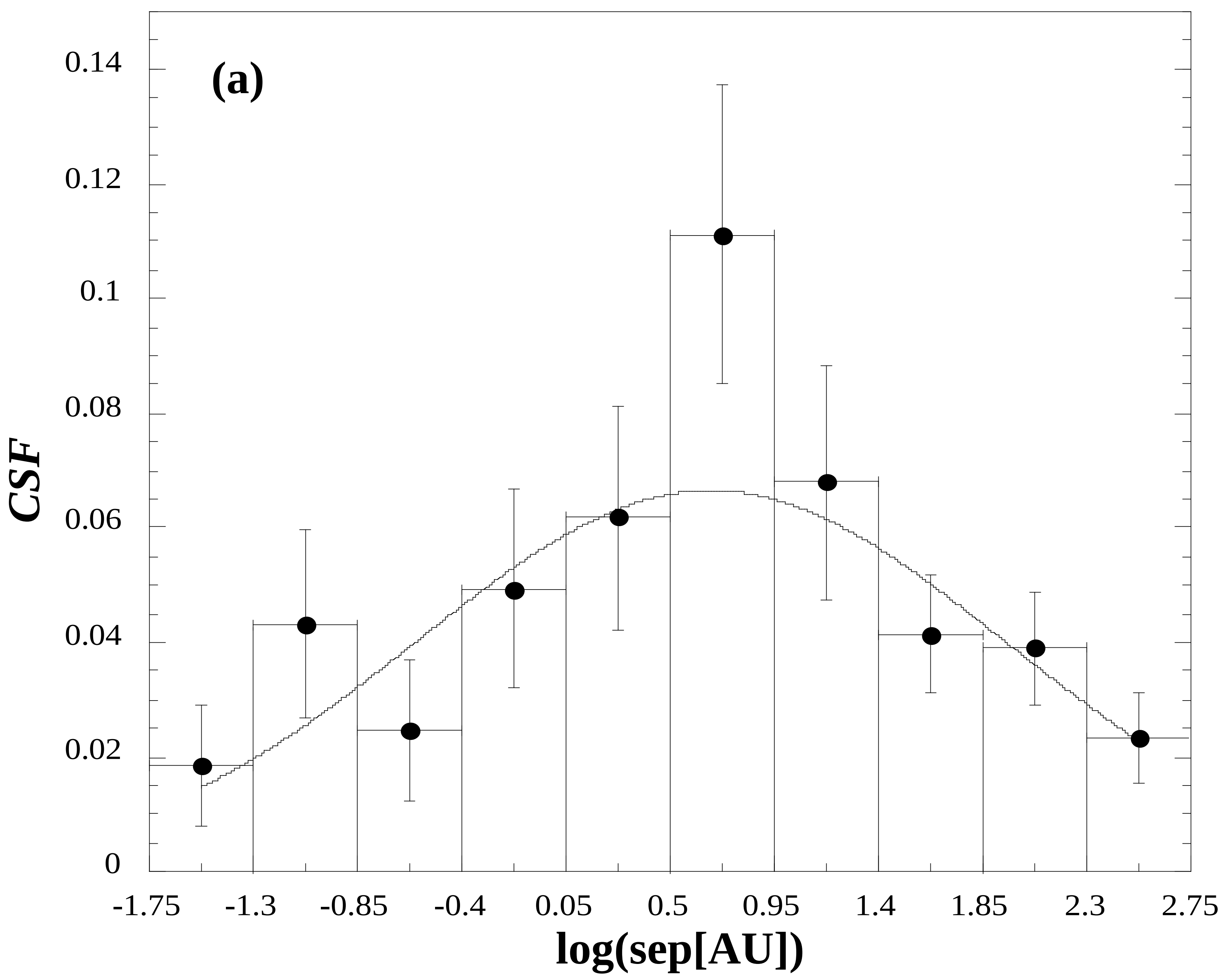

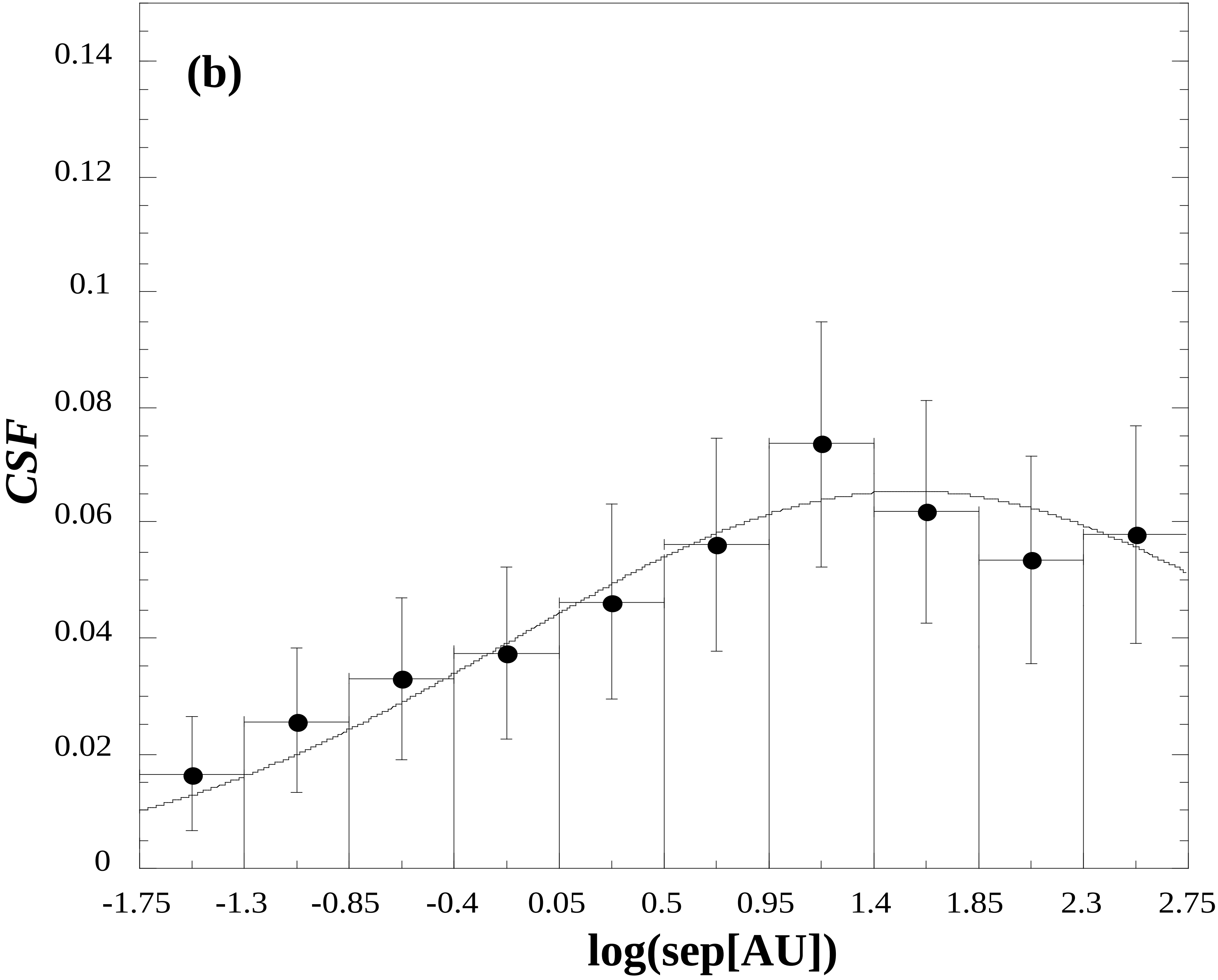

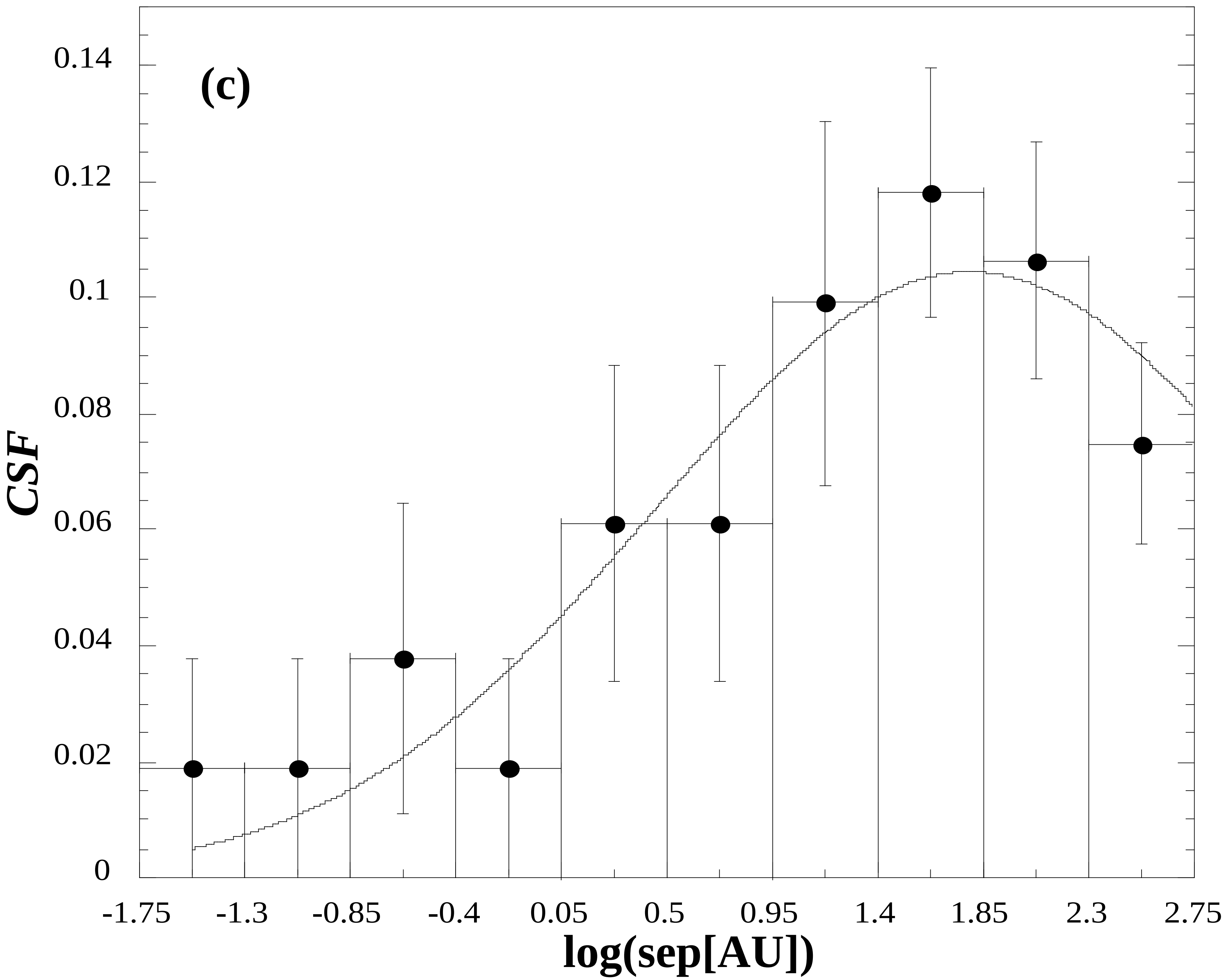

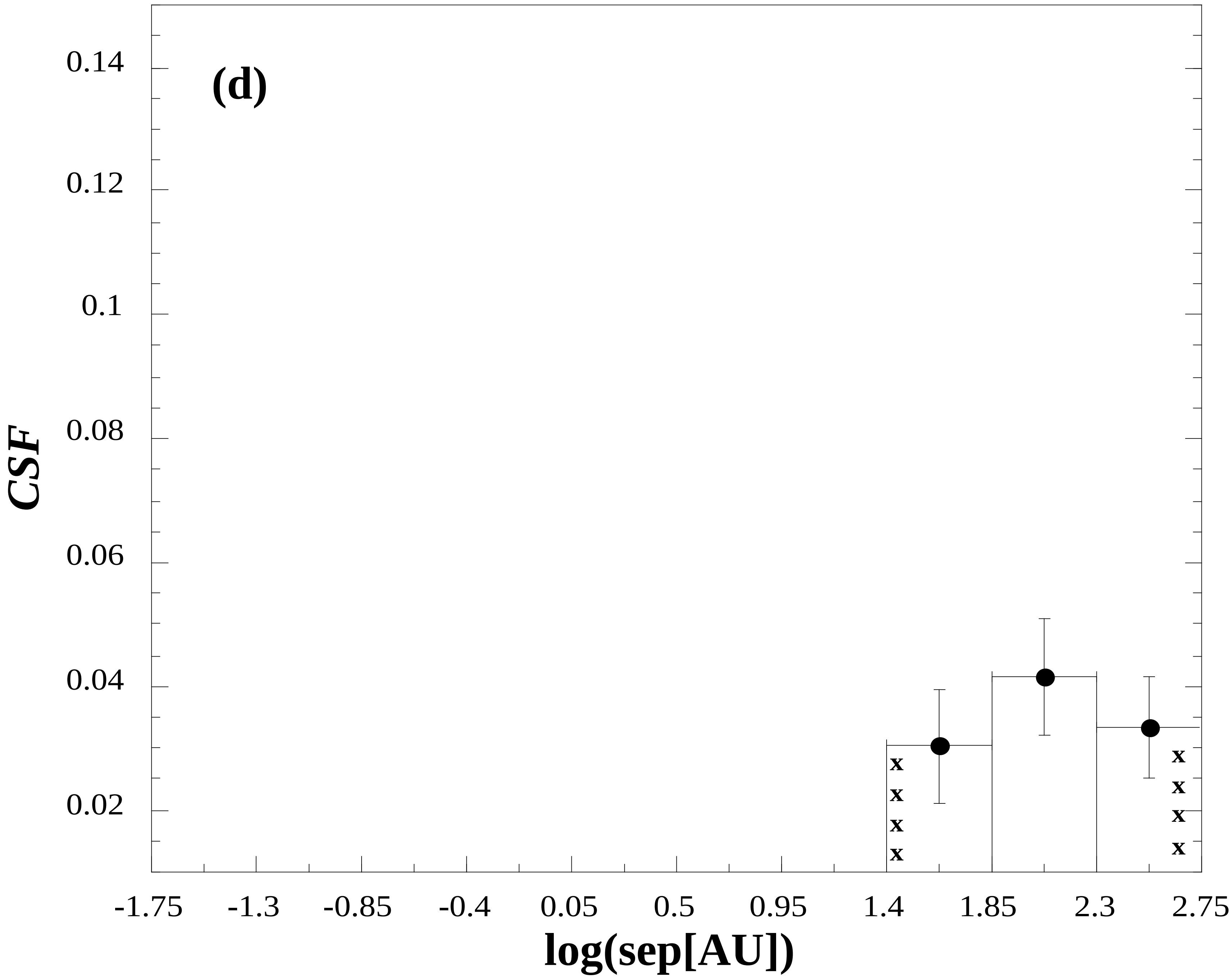

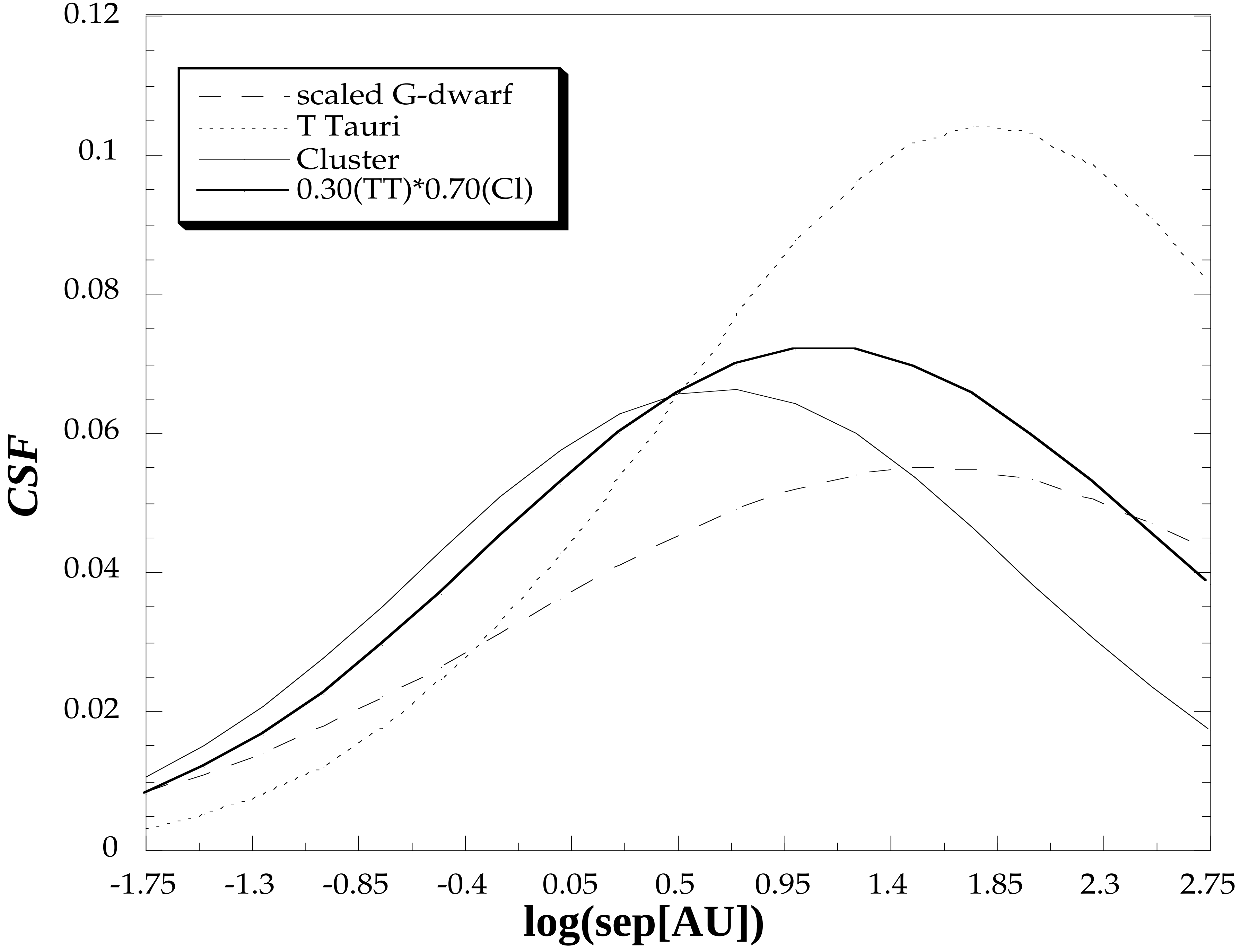

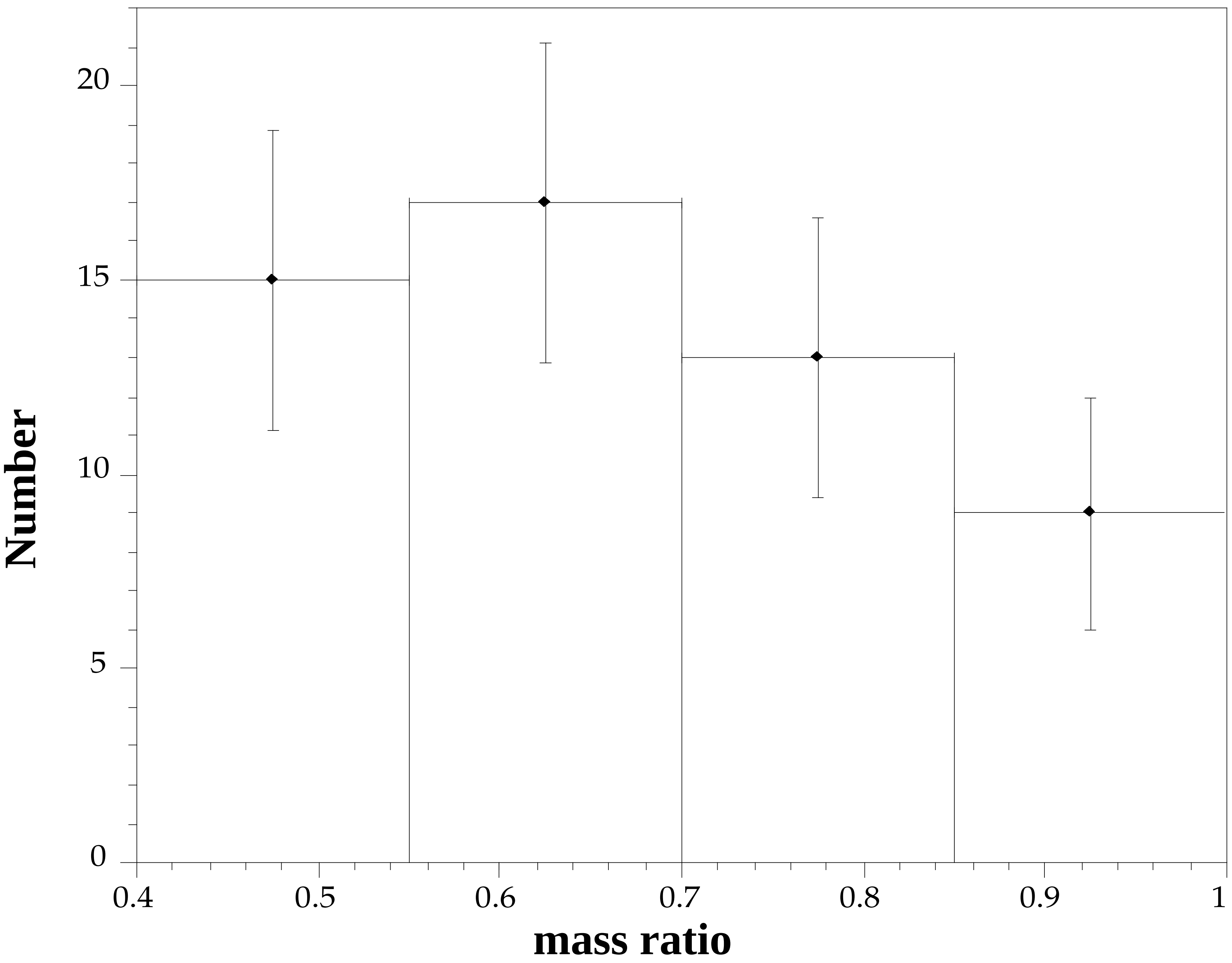

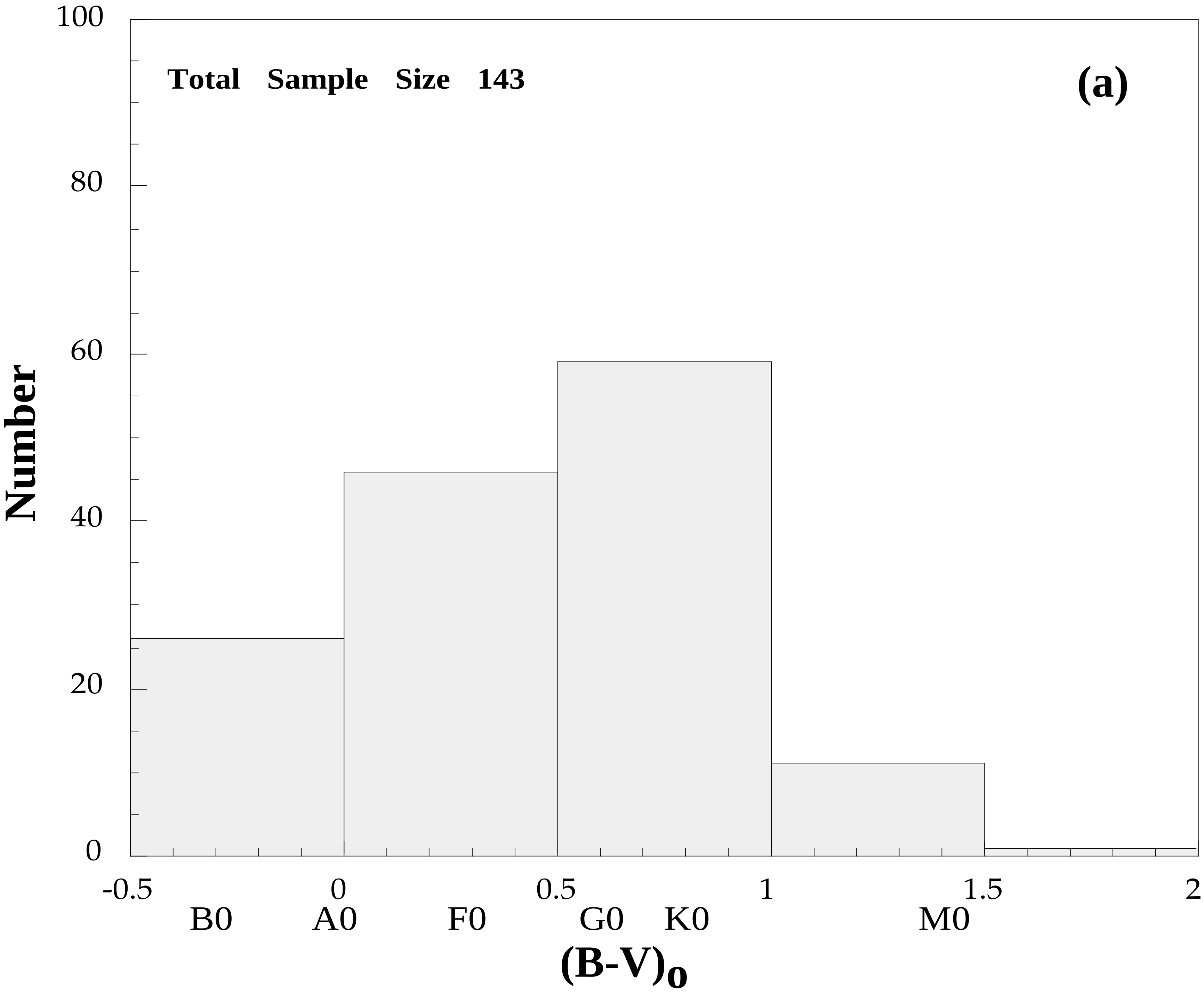

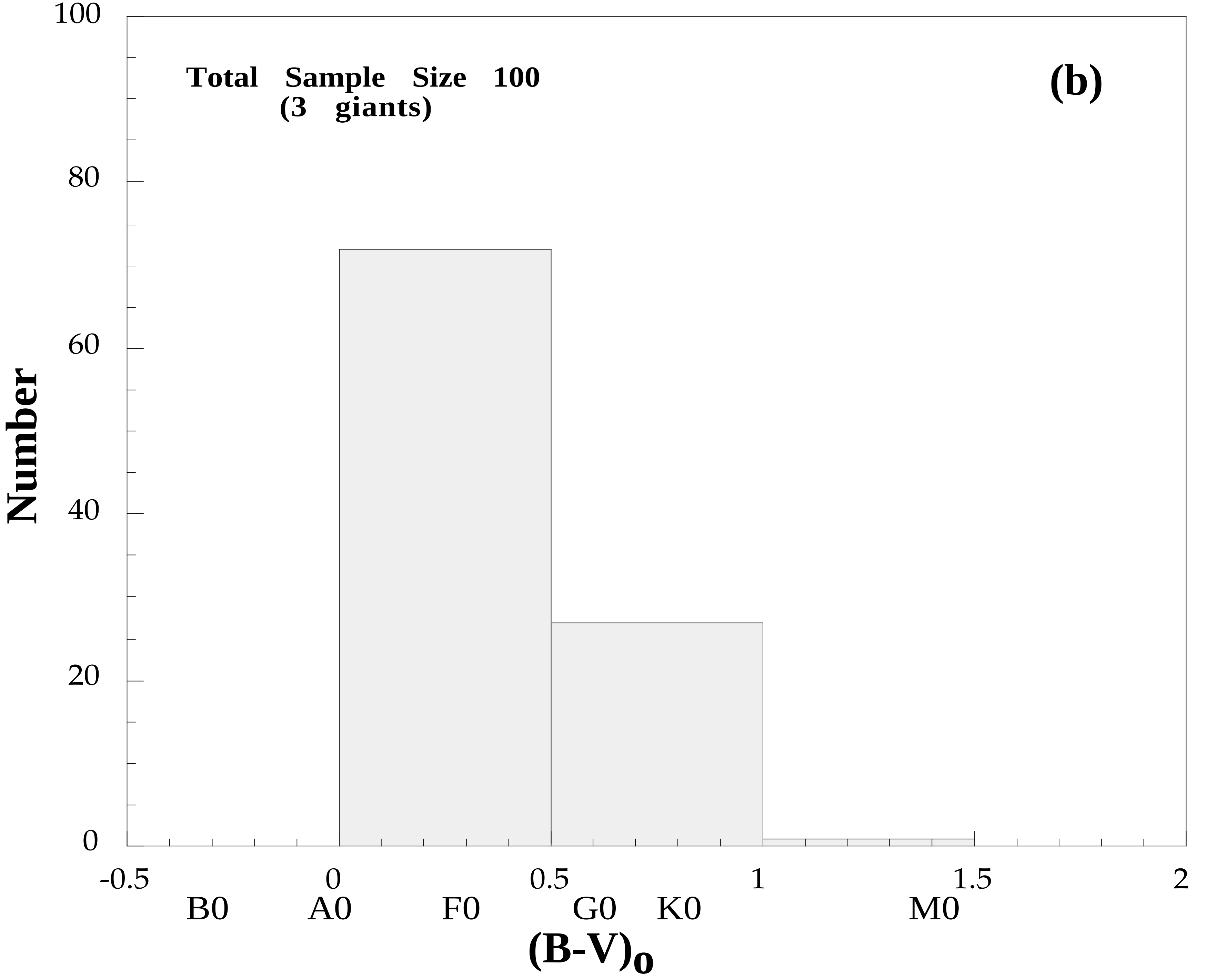

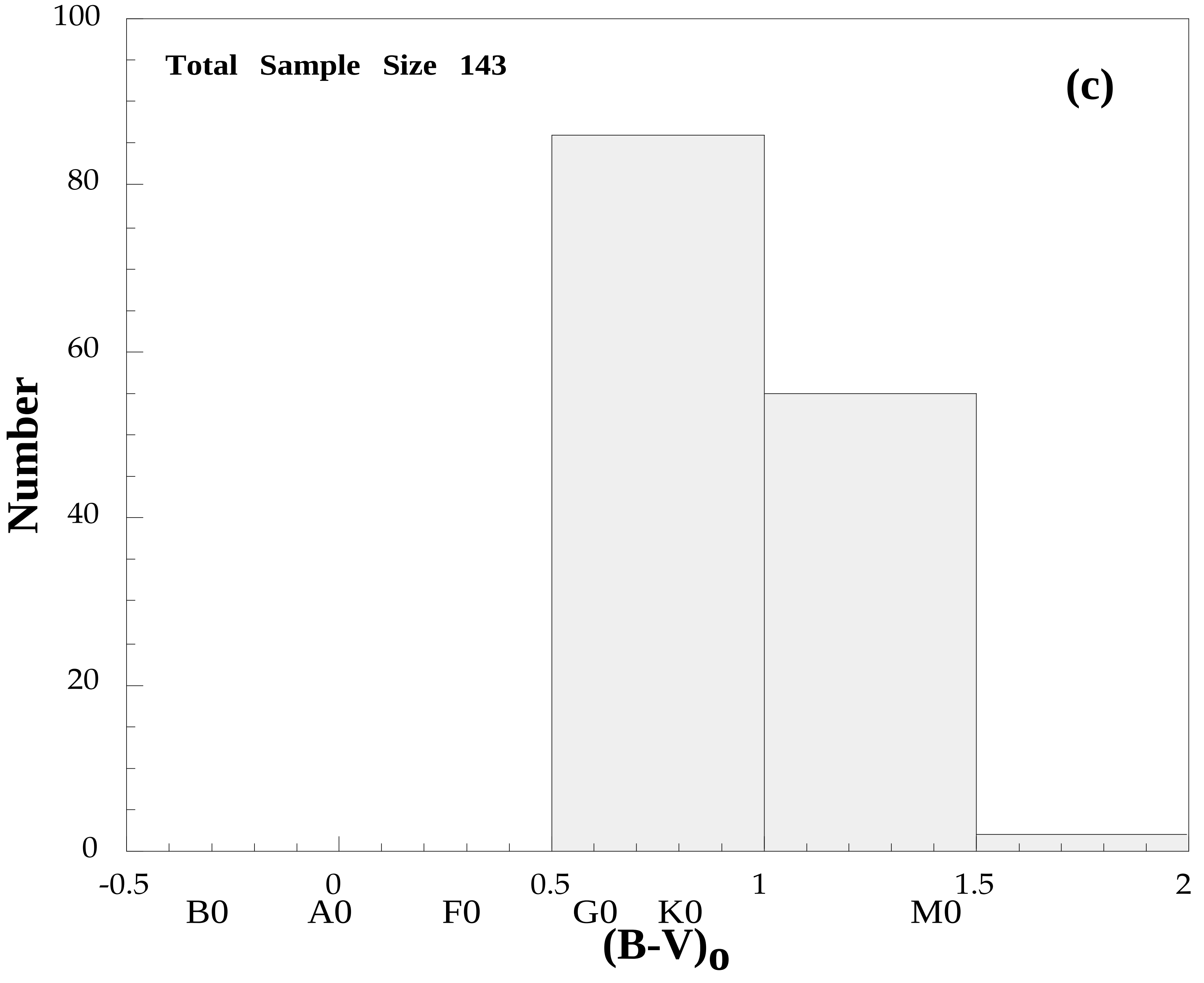

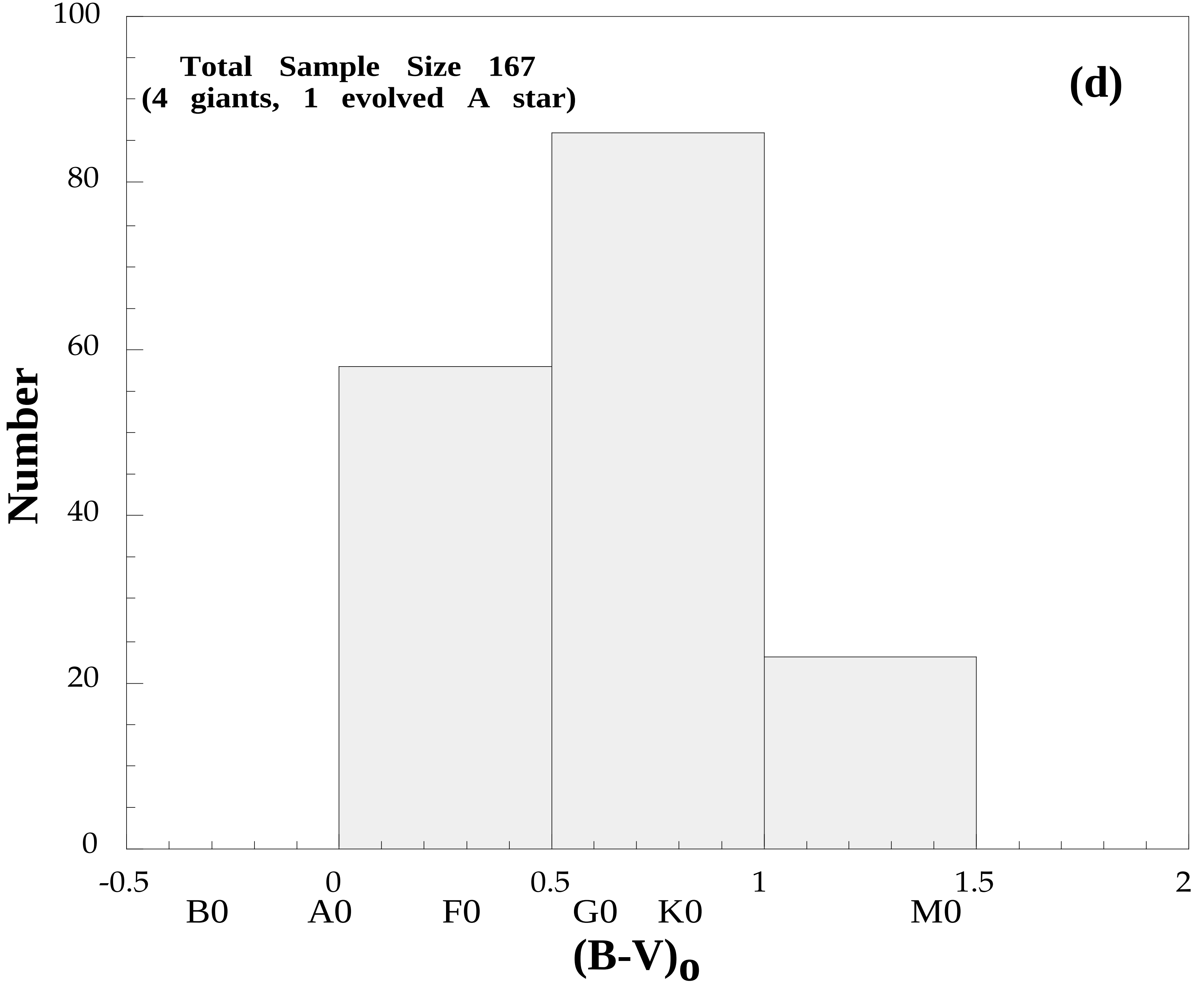

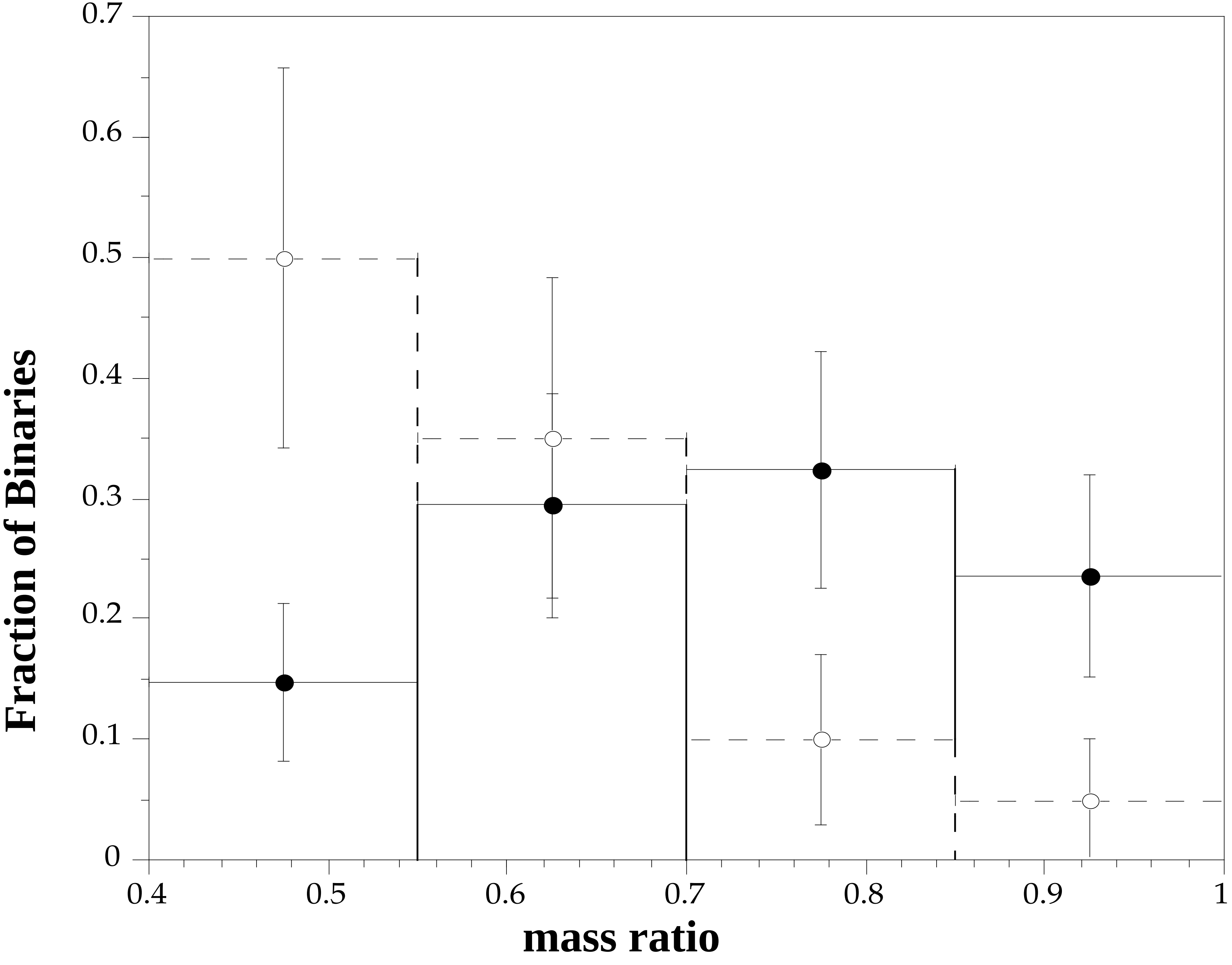

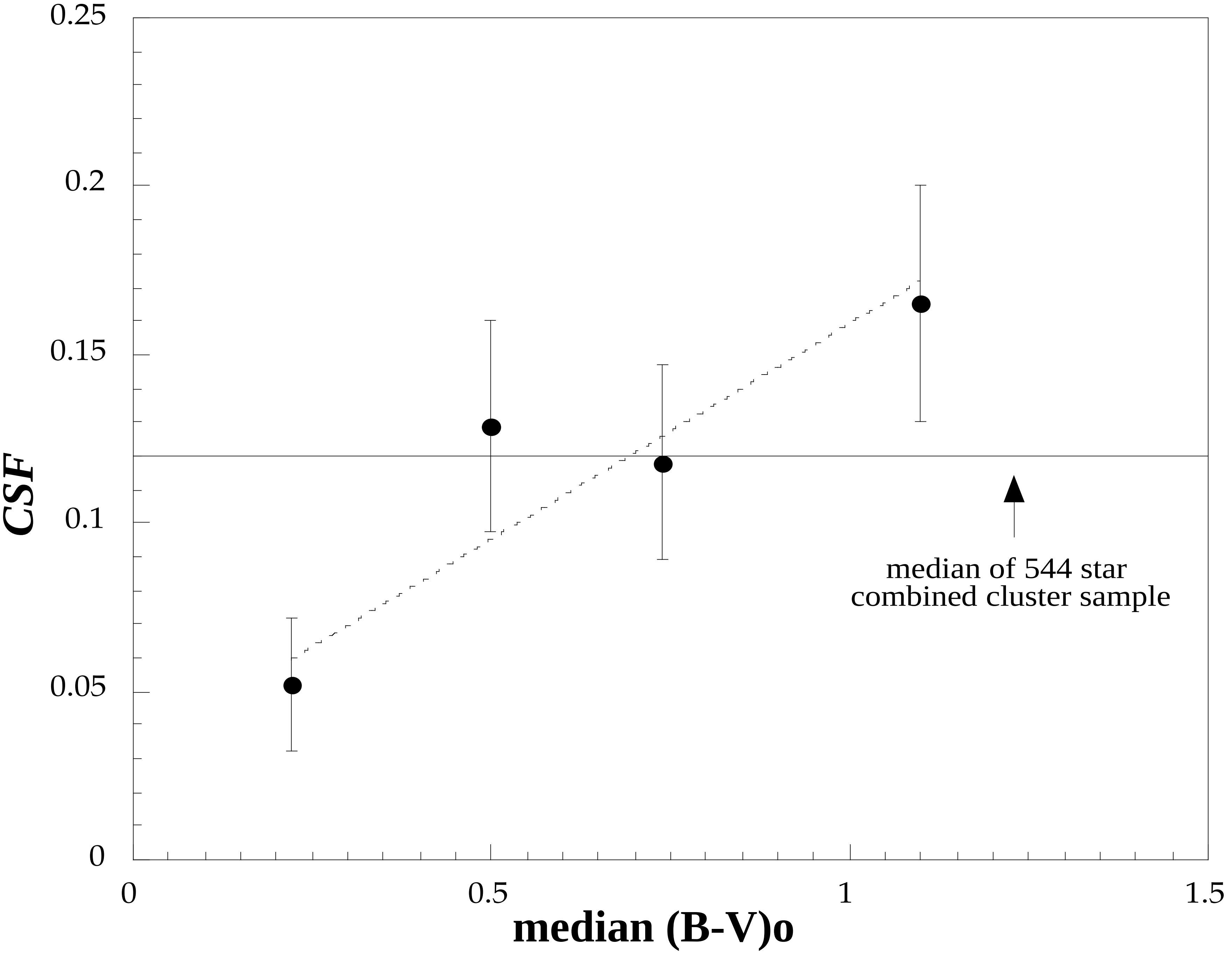

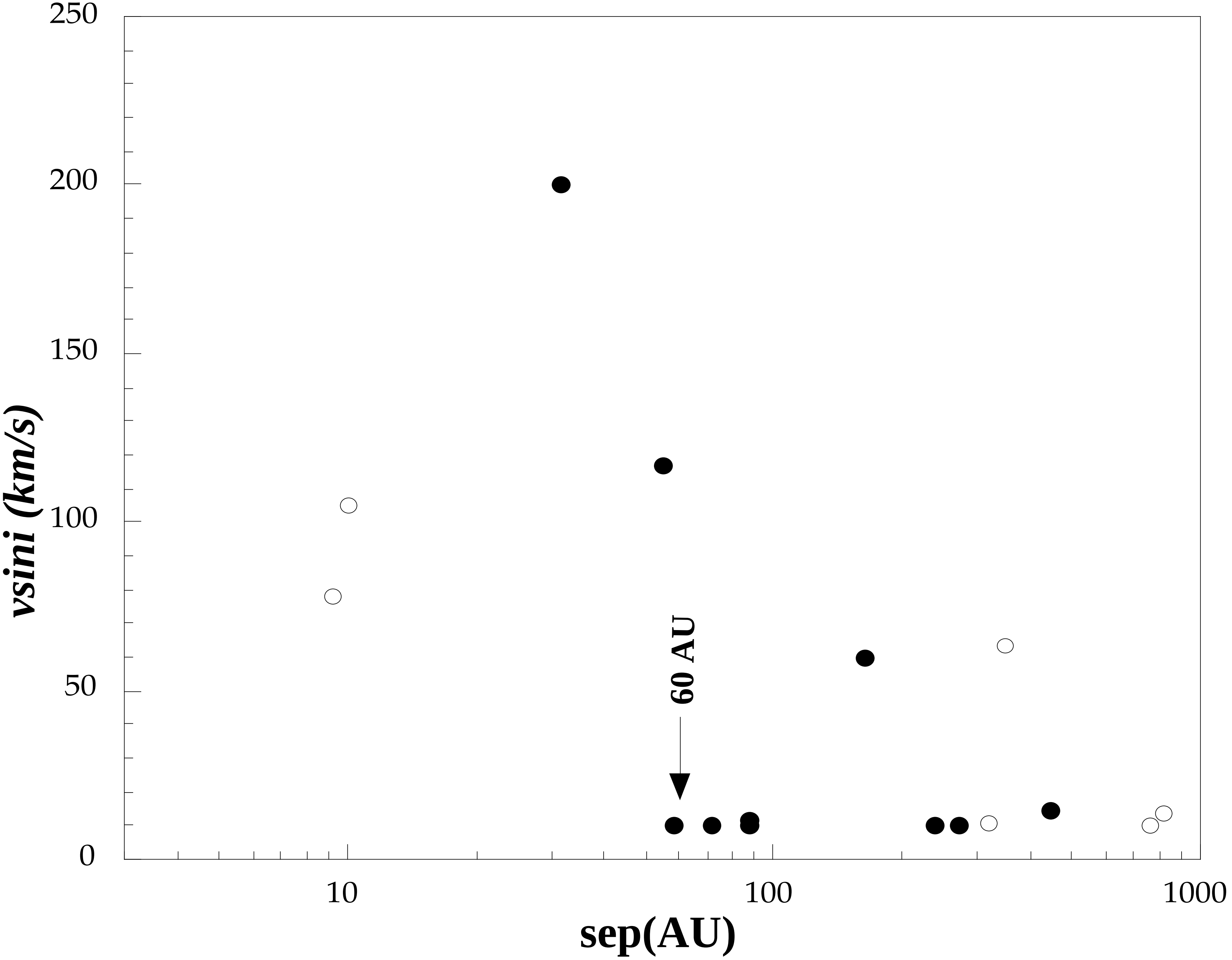